\documentclass[sigconf, noacm, dvipsnames]{acmart}

\usepackage{subcaption}
\usepackage{multirow}
\usepackage{amsmath}
\usepackage{xcolor}
\usepackage{colortbl}
\usepackage{adjustbox}
\usepackage{fancyhdr}
\usepackage{listings}
\usepackage{datetime}
\usepackage[shortlabels]{enumitem}

\usepackage{wrapfig}
\usepackage{graphicx}
\usepackage[normalem]{ulem}
\usepackage[labelfont=bf, textfont=bf, skip=2pt]{caption}
\usepackage[english]{babel}
\usepackage{booktabs}
\usepackage{url}
\usepackage{pifont}
\usepackage{footnote}
\usepackage{tikz}
\usepackage{ltablex}
\usepackage{xargs} 
\usepackage[first=0,last=9]{lcg}
\usepackage{color, colortbl}
\usepackage[colorinlistoftodos,prependcaption,textsize=tiny]{todonotes}
\usepackage{marginnote}
\usepackage{clipboard}
\usepackage{mathtools}
\usepackage[framemethod=tikz]{mdframed}
\usepackage{setspace}
\usepackage{blindtext}
\usepackage{imakeidx}
\usepackage{soul}
\usepackage[keeplastbox]{flushend}
\usepackage{hyperref}
\usepackage{blindtext}
\usepackage{imakeidx}
\usepackage{graphbox}

\newif\ifsubmission
\submissionfalse

\ifsubmission
    \newcommand\inlineedit[2]{#2}
    \newcommand\inlinecomment[3]{ }
\else
    \newcommand\inlineedit[2]{\noindent{\color{#1} {#2}}}
    \newcommand\inlinecomment[3]{\noindent{\color{#1} {\fbox{\bf{#2:}}\it#3}}}
\fi

\newcommand\ignore[1]{ }

\definecolor{dgreen}{rgb}{0.00, 0.75, 0.00}
\definecolor{ddgreen}{rgb}{0.00, 0.50, 0.00}
\definecolor{dblue}{rgb}{0.00, 0.00, 0.75}
\definecolor{ddblue}{rgb}{0.20, 0.30, 0.60}
\definecolor{dcyan}{rgb}{0.00, 0.50, 0.50}
\definecolor{dred}{rgb}{0.82, 0.10, 0.26}
\definecolor{dred}{rgb}{0.50, 0.00, 0.00}
\definecolor{dbrown}{rgb}{0.6, 0.38, 0.26}
\definecolor{electriccrimson}{rgb}{1.0, 0.0, 0.25}
\definecolor{airforceblue}{rgb}{0.36, 0.54, 0.66}
\definecolor{darkgoldenrod}{rgb}{0.72, 0.53, 0.04}

\newcommand\ie[1]{\inlinecomment{dcyan}{IE}{#1}}
\newcommand\jgl[1]{\noindent{\color{dred} {\fbox{\bf{JGL:}}\it#1}}}

\ignore{
\newcommand{\izzat}[1]{\inlineedit{dblue}{#1}}

\newcommand{\ivan}[1]{\textcolor{dbrown}{#1}}

}
\newcommand{\om}[1]{\inlineedit{black}{#1}}
\newcommand{\izzat}[1]{\inlineedit{black}{#1}}

\newcommand{\jieee}[1]{\textcolor{black}{#1}}
\newcommand{\ivan}[1]{\textcolor{black}{#1}}

\newcommand{\juancrr}[1]{\textcolor{black}{#1}}

\newcommand{\juanc}[1]{\textcolor{black}{#1}}

\newcommand{\juanfinal}[1]{\textcolor{black}{#1}}

\usepackage[most]{tcolorbox}
\usepackage{clipboard}

\newcommand{\boxbegin} {
	\begin{tcolorbox}[enhanced, frame hidden, colback=gray!50, breakable]
}
\newcommand{\boxend} {
	\end{tcolorbox}
}

\newcommand{\gboxbegin}[1] {
    \begin{tcolorbox}[colback=ddblue!5,colframe=ddblue!90!black,title=\textbf{\emph{KEY OBSERVATION {#1}}}]
}
\newcommand{\gboxend} {
	\end{tcolorbox}
}

\newcommand{\tboxbegin}[1] {
    \begin{tcolorbox}[colback=dred!5,colframe=dred!80!black,title=\textbf{\emph{KEY TAKEAWAY {#1}}}]
}
\newcommand{\tboxend} {
	\end{tcolorbox}
}

\newcommand{\pboxbegin}[1] {
    \begin{tcolorbox}[colback=ddgreen!5,colframe=ddgreen!80!black,title=\textbf{\emph{PROGRAMMING RECOMMENDATION {#1}}}]
}
\newcommand{\pboxend} {
	\end{tcolorbox}
}

\newcommand{\gpboxbegin}[1] {
    \begin{tcolorbox}[colback=dgreen!5,colframe=dgreen!80!black,title=\textbf{\emph{GENERAL PROGRAMMING  RECOMMENDATIONS}}]
}
\newcommand{\gpboxend} {
	\end{tcolorbox}
}

\newcommand{\yboxbegin} {
	\begin{tcolorbox}[enhanced, frame hidden, colback=yellow!50, breakable]
}

\newcommand{\yboxend} {
	\end{tcolorbox}
}

\usepackage{pifont}
\newcommand{\boldone}{\ding{202}}
\newcommand{\boldtwo}{\ding{203}}
\newcommand{\boldthree}{\ding{204}}
\newcommand{\boldfour}{\ding{205}}
\newcommand{\boldfive}{\ding{206}}
\newcommand{\boldsix}{\ding{207}}
\newcommand{\boldseven}{\ding{208}}

\definecolor{mygreen}{rgb}{0,0.6,0}
\definecolor{mygray}{rgb}{0.5,0.5,0.5}
\definecolor{mymauve}{rgb}{0.58,0,0.82}

\lstdefinestyle{myC}{
  language=Matlab,
  backgroundcolor=\color{white},  
  basicstyle=\footnotesize,        
  breakatwhitespace=false,  
  breaklines=true,       
  captionpos=b,                   
  commentstyle=\color{mygreen},    
  deletekeywords={...},           
  escapechar=\%,
  xleftmargin=0pt,
  xrightmargin=0pt,
  aboveskip=\medskipamount,
  belowskip=\medskipamount,
  extendedchars=true,            
  keepspaces=true,               
  keywordstyle=\color{blue},      
  language=C++,              
  morekeywords={move,lsl_add,lw,add,sw,jneq,mem_alloc, *,...},          
  numbers=left,                   
  numbersep=1pt,                   
  numberstyle=\tiny\color{mygray}, 
  rulecolor=\color{black},     
  showspaces=false,                
  showstringspaces=false,          
  showtabs=false,                  
  stepnumber=1,                    
  stringstyle=\color{mymauve},     
  tabsize=2,	                   
  title=\lstname                   
}

\definecolor{bluehl}{rgb}{0.8,0.874,1}
\definecolor{pinkhl}{rgb}{0.992156863,0.847058824,1}
\definecolor{macaroniandcheese}{rgb}{1.0, 0.74, 0.53}
\definecolor{mossgreen}{rgb}{0.68, 0.87, 0.68}
\definecolor{greenhl}{rgb}{0.835,0.996,0.939}
\definecolor{yellowhl}{rgb}{0.996,0.957,0.8}
\definecolor{palecerulean}{rgb}{0.61, 0.77, 0.89}

\newcommand{\HilightBlue}{\makebox[0pt][l]{\color{palecerulean}\rule[-4pt]{\linewidth}{10pt}}}
\newcommand{\HilightPink}{\makebox[0pt][l]{\color{macaroniandcheese}\rule[-4pt]{\linewidth}{10pt}}}
\newcommand{\HilightYellow}{\makebox[0pt][l]{\color{yellowhl}\rule[-4pt]{\linewidth}{10pt}}}
\newcommand{\HilightGreen}{\makebox[0pt][l]{\color{mossgreen}\rule[-4pt]{\linewidth}{10pt}}}

\AtBeginDocument{\DeclareCaptionSubType{lstlisting}}

\newcounter{rtaskno}
\DeclareRobustCommand{\rtask}[1]{%
   \refstepcounter{rtaskno}%
   \thertaskno\label{#1}}
\newcounter{ptaskno}
\DeclareRobustCommand{\ptask}[1]{%
   \refstepcounter{ptaskno}%
   \theptaskno\label{#1}}
\newcounter{gptaskno}
\DeclareRobustCommand{\gptask}[1]{%
   \refstepcounter{gptaskno}%
   \thegptaskno\label{#1}}
\newcounter{ttaskno}
\DeclareRobustCommand{\ttask}[1]{%
   \refstepcounter{ttaskno}%
   \thettaskno\label{#1}}

\newcommand{\astretch}[1]{\renewcommand{\arraystretch}{#1}}

\newcommand{\tsc}[1]{\textsuperscript{#1}} 
\newcommand{\affilETH}{\tsc{1}}
\newcommand{\affilAUB}{\tsc{2}}
\newcommand{\affilUMA}{\tsc{3}}
\newcommand{\affilNTUA}{\tsc{4}}

\AtBeginDocument{%
 \providecommand\BibTeX{{%
    \normalfont B\kern-0.5em{\scshape i\kern-0.25em b}\kern-0.8em\TeX}}}

\makeatletter
\def\bstctlcite{\@ifnextchar[{\@bstctlcite}{\@bstctlcite[@auxout]}}
\def\@bstctlcite[#1]#2{\@bsphack
  \@for\@citeb:=#2\do{%
    \edef\@citeb{\expandafter\@firstofone\@citeb}%
    \if@filesw\immediate\write\csname #1\endcsname{\string\citation{\@citeb}}\fi}%
  \@esphack}
\makeatother

\setcopyright{none}
\renewcommand\footnotetextcopyrightpermission[1]{}

\begin{document}
\bstctlcite{IEEEexample:BSTcontrol}

\title{Benchmarking a New Paradigm: An Experimental Analysis \\ of a Real Processing-in-Memory Architecture \vspace{-2mm}}




     
\author{
 {%
     Juan Gómez-Luna$^1$\quad 
     Izzat El Hajj$^2$\quad 
     Ivan Fernandez$^1$$^,$$^3$\quad 
     Christina Giannoula$^1$$^,$$^4$
 }
}
\author{
 {
     Geraldo F. Oliveira$^1$\quad
     Onur Mutlu$^1$
 }
}


\affiliation{
\institution{
      \vspace{5pt}
      \affilETH ETH Zürich \quad
      \affilAUB American University of Beirut \quad
      \affilUMA University of Malaga \quad
      \affilNTUA National Technical University of Athens
  }
}

\authorsaddresses{}
\renewcommand{\shortauthors}{}




\begin{abstract}

Many modern workloads, such as neural networks, databases, and graph processing, are fundamentally memory-bound.
For such workloads, the data movement between main memory and CPU cores imposes a significant overhead in terms of both latency and energy.
A major reason is that this communication happens through a narrow bus with high latency and limited bandwidth, and the low data reuse in memory-bound workloads is insufficient to amortize the cost of main memory access.
Fundamentally addressing this \emph{data movement bottleneck} requires a paradigm where the memory system assumes an active role in computing by integrating processing capabilities.
This paradigm is known as \emph{processing-in-memory} (\emph{PIM}).

Recent research explores different forms of PIM architectures, motivated by the emergence of new 3D-stacked memory technologies that integrate memory with a logic layer where processing elements can be easily placed. 
Past works evaluate these architectures in simulation or, at best, with simplified hardware prototypes.
In contrast, the UPMEM company has designed and manufactured the first publicly-available real-world PIM architecture. 
The UPMEM PIM architecture combines traditional DRAM memory arrays with general-purpose in-order cores, called \emph{DRAM Processing Units} (\emph{DPUs}), integrated in the same chip.

This paper provides the first comprehensive analysis of the first publicly-available real-world PIM architecture. 
We make two key contributions. 
First, we conduct an experimental characterization of the UPMEM-based PIM system using microbenchmarks to assess various architecture limits such as compute throughput and memory bandwidth, yielding new insights.
Second, we present \emph{PrIM} (\emph{\underline{Pr}ocessing-\underline{I}n-\underline{M}emory benchmarks}), a benchmark suite of 16 workloads from different application domains (e.g., dense/sparse linear algebra, databases, data analytics, graph processing, neural networks, bioinformatics, image processing), which we identify as memory-bound.
We evaluate the performance and scaling characteristics of PrIM benchmarks on the UPMEM PIM architecture, and compare their performance and energy consumption to their state-of-the-art CPU and GPU counterparts.
Our extensive evaluation conducted on two real UPMEM-based PIM systems with 640 and 2,556 DPUs provides new insights about suitability of different workloads to the PIM system, programming recommendations for software designers, and suggestions and hints for hardware and architecture designers of future PIM systems.

\end{abstract}

\keywords{processing-in-memory, near-data processing, memory systems, data movement bottleneck, DRAM, benchmarking, real-system characterization, workload characterization}

\maketitle
\pagestyle{plain}


\section{Introduction} 

In {modern} computing systems, a {large} fraction of the {execution time} and energy consumption of modern {data-intensive} workloads is spent moving data between memory and processor cores. 
This \emph{data movement bottleneck}~\cite{mutlu2019,mutlu2020modern,ghoseibm2019,ghose2019arxiv,ghose.bookchapter19.arxiv,mutlu.msttalk17,mutlu2019enabling} stems from the fact that, for decades, the performance of processor cores has been increasing at a faster rate than the memory {performance}.
The gap between an arithmetic operation and a memory access in terms of latency and energy keeps widening {and the memory access is becoming increasingly more expensive}.
As a result, {recent experimental studies report that} data movement accounts for 62\%~\cite{boroumand.asplos18} {(reported in 2018)}, 40\%~\cite{pandiyan.iiswc2014} {(reported in 2014)}, and 35\%~\cite{kestor.iiswc2013} {(reported in 2013)} of the total system energy in {various} consumer, scientific, and mobile applications, respectively. 

One promising way to alleviate the data movement bottleneck is \emph{processing-in-memory} (\emph{PIM}), which equips memory chips with processing capabilities{~\cite{mutlu2020modern}}. 
This paradigm has been explored for more than {50} years{~\cite{Kautz1969,stone1970logic,shaw1981non, kogge1994, gokhale1995processing, patterson1997case, oskin1998active, kang1999flexram, Mai:2000:SMM:339647.339673, Draper:2002:ADP:514191.514197,aga.hpca17,eckert2018neural,fujiki2019duality,kang.icassp14,seshadri.micro17,seshadri.arxiv16,seshadri2013rowclone,seshadri2018rowclone,angizi2019graphide,kim.hpca18,kim.hpca19,gao2020computedram,chang.hpca16,xin2020elp2im,li.micro17,deng.dac2018,hajinazarsimdram,rezaei2020nom,wang2020figaro,ali2019memory,li.dac16,angizi2018pima,angizi2018cmp,angizi2019dna,levy.microelec14,kvatinsky.tcasii14,shafiee2016isaac,kvatinsky.iccd11,kvatinsky.tvlsi14,gaillardon2016plim,bhattacharjee2017revamp,hamdioui2015memristor,xie2015fast,hamdioui2017myth,yu2018memristive,syncron,fernandez2020natsa,alser2020accelerating,cali2020genasm,kim.arxiv17,kim.bmc18,ahn.pei.isca15,ahn.tesseract.isca15,boroumand.asplos18,boroumand2019conda,boroumand2016pim,boroumand.arxiv17,singh2019napel,asghari-moghaddam.micro16,DBLP:conf/sigmod/BabarinsaI15,chi2016prime,farmahini2015nda,gao.pact15,DBLP:conf/hpca/GaoK16,gu.isca16,guo2014wondp,hashemi.isca16,cont-runahead,hsieh.isca16,kim.isca16,kim.sc17,DBLP:conf/IEEEpact/LeeSK15,liu-spaa17,morad.taco15,nai2017graphpim,pattnaik.pact16,pugsley2014ndc,zhang.hpdc14,zhu2013accelerating,DBLP:conf/isca/AkinFH15,gao2017tetris,drumond2017mondrian,dai2018graphh,zhang2018graphp,huang2020heterogeneous,zhuo2019graphq,santos2017operand,mutlu2019,mutlu2020modern,ghoseibm2019,ghose2019arxiv,wen2017rebooting,besta2021sisa,ferreira2021pluto,olgun2021quactrng,lloyd2015memory,elliott1999computational,zheng2016tcam,landgraf2021combining,rodrigues2016scattergather,lloyd2018dse,lloyd2017keyvalue,gokhale2015rearr,nair2015active,jacob2016compiling,sura2015data,nair2015evolution,balasubramonian2014near,xi2020memory,PIM_fall2021, CompArch_fall2021, Seminar_fall2021, yavits2021giraf, olgun2021pidram, lee2022isscc, ke2021near, kwon202125, lee2021hardware, asgarifafnir, herruzo2021enabling, singh2021fpga, singh2021accelerating, oliveira2021pimbench, boroumand2021google, boroumand2021google_arxiv,denzler2021casper}}, but limitations in memory technology prevented {commercial} hardware from {successfully} materializing.
More recently, difficulties in DRAM scaling (i.e., {challenges in} increasing density {and performance} while {maintaining reliability,} latency and energy consumption){~\cite{kang.memoryforum14,liu.isca13,mutlu.imw13,kim-isca2014,mutlu2017rowhammer,ghose2018vampire,mutlu.superfri15,kim2020revisiting,mutlu2020retrospective,frigo2020trr,kim2018solar,raidr,mutlu2015main, mandelman.ibmjrd02, lee-isca2009,cojocar2020susceptible,yauglikcci2021blockhammer,patel2017reaper,khan.sigmetrics14,khan.dsn16,khan.micro17,lee.hpca15,lee.sigmetrics17,chang.sigmetrics17,chang.sigmetrics16,chang.hpca14,meza.dsn15,david2011memdvfs,deng2011memscale,hong2010memory,kanev.isca15,qureshi.dsn15, orosa2021deeper, hassan2021uncovering, patel2021harp}} have motivated innovations such as 3D-stacked memory{~\cite{hmc.spec.2.0, jedec.hbm.spec,lee.taco16,ghose2019demystifying,ramulator,ahn.tesseract.isca15,gokhale2015hmc}} and nonvolatile memory{~\cite{lee-isca2009, kultursay.ispass13, strukov.nature08, wong.procieee12,lee.cacm10,qureshi.isca09,zhou.isca09,lee.ieeemicro10,wong.procieee10,yoon-taco2014,yoon2012row,girard2020survey}} which present {new} opportunities to redesign the memory subsystem while integrating processing capabilities.
3D-stacked memory integrates DRAM layers with a logic layer, which can embed processing elements.
Several works explore this approach, called \emph{processing-near-memory} {(\emph{PNM})}, to implement different types of processing components in the logic layer, such as general-purpose cores{~\cite{deoliveira2021,boroumand.arxiv17,boroumand2016pim, boroumand.asplos18,boroumand2019conda,ahn.tesseract.isca15,syncron,singh2019napel, oliveira2021pimbench}}, application-specific accelerators{~\cite{zhu2013accelerating, DBLP:conf/isca/AkinFH15, DBLP:conf/sigmod/BabarinsaI15, kim.arxiv17, kim.bmc18, liu-spaa17, kim.isca16, DBLP:conf/IEEEpact/LeeSK15,cali2020genasm,alser2020accelerating,fernandez2020natsa,impica,singh2020nero,lloyd2015memory,landgraf2021combining,rodrigues2016scattergather,lloyd2017keyvalue,gokhale2015rearr, asgarifafnir, herruzo2021enabling, singh2021fpga, singh2021accelerating, boroumand2021google, boroumand2021google_arxiv,denzler2021casper}}, simple functional units{~\cite{ahn.pei.isca15,nai2017graphpim,hadidi2017cairo, lee2022isscc, kwon202125, lee2021hardware}}, GPU cores{~\cite{zhang.hpdc14, pattnaik.pact16, hsieh.isca16,kim.sc17}}, or reconfigurable logic~\cite{DBLP:conf/hpca/GaoK16, guo2014wondp, asghari-moghaddam.micro16, ke2021near}. 
However, 3D-stacked memory suffers from high cost and limited capacity, and the logic layer has {hardware} area and thermal dissipation constraints, which limit the capabilities of the embedded processing components.
On the other hand, \emph{processing-using-memory} {(\emph{PUM})} takes advantage of the analog {operational} properties of memory cells in SRAM{~\cite{aga.hpca17,eckert2018neural,fujiki2019duality,kang.icassp14}}, DRAM{~\cite{seshadri2020indram,seshadri.bookchapter17.arxiv,seshadri.bookchapter17,Seshadri:2015:ANDOR,seshadri.micro17,seshadri.arxiv16,seshadri2018rowclone,seshadri2013rowclone,angizi2019graphide,kim.hpca18,kim.hpca19,gao2020computedram,chang.hpca16,xin2020elp2im,li.micro17,deng.dac2018,hajinazarsimdram,rezaei2020nom,wang2020figaro,ali2019memory,ferreira2021pluto,olgun2021quactrng, olgun2021pidram}}, or nonvolatile memory{~\cite{li.dac16,angizi2018pima,angizi2018cmp,angizi2019dna,levy.microelec14,kvatinsky.tcasii14,shafiee2016isaac,kvatinsky.iccd11,kvatinsky.tvlsi14,gaillardon2016plim,bhattacharjee2017revamp,hamdioui2015memristor,xie2015fast,hamdioui2017myth,yu2018memristive,puma-asplos2019, ankit2020panther,chi2016prime,ambrosi2018hardware,bruel2017generalize,huang2021mixed,zheng2016tcam,xi2020memory, yavits2021giraf}} to perform specific types of operations efficiently.
However, processing-using-memory is either limited to simple bitwise operations (e.g., majority, AND, OR)~\cite{aga.hpca17, seshadri.micro17,seshadri.arxiv16,Seshadri:2015:ANDOR,seshadri2020indram}, requires high area overheads to perform more complex operations~\cite{li.micro17, deng.dac2018,ferreira2021pluto}, {or requires significant changes to data organization, manipulation, and handling mechanisms to enable bit-serial computation, while still having limitations on certain operations~\cite{hajinazarsimdram,ali2019memory,angizi2019graphide}.}\footnote{{PUM approaches performing bit-serial computation~\cite{hajinazarsimdram,ali2019memory,angizi2019graphide} need to layout {data} elements vertically (i.e., all bits of an element in the same bitline), which (1) does \emph{not} allow certain data manipulation operations (e.g., shuffling {of data elements in an array}) and (2) requires {paying} the overhead of bit transposition, {when the format of data needs to change~\cite{hajinazarsimdram}, i.e., prior to performing bit-serial computation}}.} 
Moreover, processing-using-memory {approaches are usually} efficient {mainly} for regular computations, since they naturally operate on a large number of memory cells (e.g., entire rows {across many subarrays~\cite{salp,seshadri.micro17,seshadri.arxiv16,seshadri2013rowclone,seshadri2018rowclone,hajinazarsimdram,seshadri2020indram,seshadri.bookchapter17.arxiv,seshadri.bookchapter17,Seshadri:2015:ANDOR}}) simultaneously.
For these reasons, complete PIM {systems} based on 3D-stacked memory or processing-using-memory have not {yet} been {commercialized} in real hardware.

The UPMEM PIM architecture~\cite{upmem2018, devaux2019} is the first PIM {system} to be {commercialized} in real hardware.
To avoid the aforementioned limitations, it uses conventional 2D DRAM arrays and combines them with general-purpose processing cores, called \emph{DRAM Processing Units} (\emph{DPUs}), on the same chip.
Combining memory and processing components on the same chip imposes serious design challenges.
For example, DRAM designs use only {three metal layers~\cite{weber2005current,peng2015design}}, while conventional processor designs typically use more than {ten}~\cite{devaux2019,yuffe2011,christy2020,singh2017}.
While these challenges prevent the fabrication of fast {logic} transistors, UPMEM overcomes these challenges via DPU cores that are {relatively} deeply pipelined and fine-grained multithreaded{~\cite{ddca.spring2020.fgmt,henessy.patterson.2012.fgmt,burtonsmith1978,smith1982architecture,thornton1970}} to run at several hundred megahertz.
The UPMEM PIM architecture provides several key advantages with respect to other PIM proposals.
First, it relies on mature 2D DRAM design and fabrication process, avoiding the drawbacks of {emerging} 3D-stacked memory technology.
Second, the {general-purpose} DPUs support a {wide} variety of computations and {data} types, {similar to simple modern general-purpose processors}.
Third, the architecture is suitable for irregular computations because the threads in a DPU can execute independently of each other (i.e., they are not bound by {lockstep execution as in} SIMD\footnote{{\emph{Single Instruction Multiple Data} (\emph{SIMD})~\cite{ddca.spring2020.simd,henessy.patterson.2012.simd,flynn1966very} refers to an execution paradigm where multiple processing elements execute the \emph{same} operation on \emph{multiple} data elements simultaneously.}}).
{Fourth, UPMEM provides} a complete software stack that enables DPU programs to be written in the {commonly-used} C language~\cite{upmem-guide}.

{Rigorously understanding the UPMEM PIM architecture, the first publicly-available PIM architecture, and its suitability to various workloads can provide valuable insights to programmers, users and architects of this architecture as well as of future PIM systems. 
To this end, our work provides the first comprehensive {experimental characterization and} analysis of the first publicly-available real-world PIM architecture. {To enable our experimental studies and} analyses, we develop new microbenchmarks and a new benchmark suite, {which we openly and freely make available}~\cite{gomezluna2021repo}.}

We develop a set of microbenchmarks to {evaluate, characterize, and understand} the limits of the UPMEM-based PIM system, {yielding new insights}.
First, we obtain the compute throughput of a DPU for different arithmetic operations and data types.
Second, we measure the bandwidth of two {different} memory spaces that a DPU can {directly access using load/store instructions}: (1) a DRAM bank called \emph{Main RAM} (\emph{MRAM}), and (2) an SRAM-based scratchpad {memory} called \emph{Working RAM} (\emph{WRAM}). 
We employ streaming (i.e., unit-stride), strided, and random memory access patterns to measure the {sustained} bandwidth of both {types of} memories.
Third, we measure the {sustained} bandwidth between the standard main memory and the MRAM banks for different types and sizes of transfers, {which is important for the communication of the DPU with the host CPU and other DPUs}.

We present \emph{PrIM} (\emph{\underline{Pr}ocessing-\underline{I}n-\underline{M}emory benchmarks}), the first benchmark suite for a real PIM architecture. 
PrIM includes 16 workloads from different application domains {(e.g., dense/sparse linear algebra, databases, data analytics, graph processing, neural networks, bioinformatics, image processing)}, which we identify as memory-bound using the roofline model for a conventional CPU~\cite{roofline}. 
We perform strong scaling\footnote{{\emph{Strong scaling} refers to how the execution time of a program solving a particular problem varies with the number of processors for a fixed problem size~\cite{hager2010introduction.scaling,amdahl1976validity}.}} and weak scaling\footnote{{\emph{Weak scaling} refers to how the execution time of a program solving a particular problem varies with the number of processors for a fixed problem size per processor~\cite{hager2010introduction.scaling,gustafson1988reevaluating}.}} experiments with the 16 benchmarks on a system with 
2,556 DPUs, and compare {their performance and energy consumption to their modern CPU and GPU counterparts}.
{Our extensive evaluation 
provides new insights about suitability of different workloads to the PIM system, programming recommendations for software designers, and suggestions and hints for hardware and architecture designers of future PIM systems.}
All our microbenchmarks and PrIM benchmarks 
are publicly {and freely} available~\cite{gomezluna2021repo} to serve as programming samples for {real} PIM architectures, {evaluate and compare {current and} future PIM systems,} and help further advance PIM architecture, {programming, and software} research.\footnote{{We refer the reader to a recent overview paper~\cite{mutlu2020modern} on the state-of-the-art challenges in PIM research.}}

The main contributions of this work are as follows: 
\begin{itemize}[noitemsep,topsep=0pt,leftmargin=8pt]
    \item
        We perform the first comprehensive {characterization and} analysis of the first publicly-available real-world PIM architecture. {We} analyze the {new} architecture's {potential,} limitations and bottlenecks. We analyze (1) memory bandwidth at different levels of the {DPU} memory hierarchy for different memory access patterns, (2) {DPU} compute throughput of different arithmetic operations for different data types, and (3) strong and weak {scaling characteristics} for different {computation} patterns. We find that (1) the UPMEM PIM architecture is fundamentally compute bound, since workloads with more complex operations than integer addition fully utilize the instruction pipeline before they can potentially saturate the memory bandwidth, and (2) workloads that require inter-DPU communication do \emph{not} scale well, since there is no direct communication channel among DPUs, {and} therefore, all inter-DPU communication takes place via the host CPU, i.e., through the narrow memory bus.
    \item
        We present {and open-source} PrIM, the first benchmark suite for a real PIM architecture, composed of 16 {real-world} workloads that are memory-bound on conventional processor-centric systems. The workloads have different characteristics, {exhibiting heterogeneity in their} memory access patterns, operations and data types, {and} communication patterns. The PrIM benchmark suite provides a common set of workloads to evaluate the UPMEM PIM architecture with and can be useful for programming, architecture and systems researchers all alike to improve multiple aspects of {future} PIM hardware and software.{$^5$}
    \item
        We compare the performance {and energy consumption} of {PrIM} benchmarks on two UPMEM-based PIM systems {with 2,556 DPUs and 640 DPUs} to {modern} conventional processor-centric systems, {i.e.,} CPUs and GPUs. {Our analysis {reveals} several {new and interesting} findings. 
        {We highlight three major findings.} 
        First, both UPMEM-based PIM systems outperform a modern CPU (by \jieee{93.0$\times$ and 27.9$\times$}, on average, respectively) for 13 of the PrIM benchmarks, which do \emph{not} require intensive inter-DPU {synchronization} {or floating point operations}.\footnote{{Two of the other three PrIM benchmarks, Breadth-first Search (BFS) and Needleman-Wunsch (NW), pay the huge overhead of inter-DPU synchronization via the host CPU. The third one, Sparse Matrix-Vector Multiply (SpMV), makes intensive use of floating point multiplication and addition.}}
        {Section~\ref{sec:comparison} provides a detailed analysis of our comparison of PIM systems to modern CPU and GPU.}
        Second, the 2,556-DPU PIM system is faster than a modern GPU (by \jieee{2.54$\times$}, on average) for 10 PrIM benchmarks with (1) streaming memory accesses, (2) little {or no} inter-DPU synchronization, 
        {and (3)} little {or no} use of complex arithmetic operations (i.e., integer multiplication/division, floating point operations).\footnote{We also evaluate the 640-DPU PIM system and find that it is \emph{slower} than the GPU for most PrIM benchmarks, but the performance gap between {the two systems (640-DPU PIM and GPU)} is significantly smaller for the 10 PrIM benchmarks that do \emph{not} need (1) heavy inter-DPU communication or (2) intensive use of multiplication operations. The 640-DPU PIM system is faster than the GPU for two benchmarks, which are not well-suited for the GPU. {Section~\ref{sec:comparison} provides a detailed analysis of our comparison.}} 
        {Third}, energy consumption {comparison of the PIM, CPU, and GPU systems} follows the same trends as the performance comparison}: {the PIM system yields large energy savings over the CPU and the CPU, for workloads where it largely outperforms the CPU and the GPU.}
        We are comparing the first ever commercial PIM system to CPU and GPU systems that have been heavily optimized for decades in terms of architecture, software, and manufacturing. Even then, we see significant advantages of PIM over CPU and GPU in most PrIM benchmarks (Section~\ref{sec:comparison}). We believe the architecture, software, and manufacturing of PIM systems will continue to improve (e.g., we suggest optimizations and areas for future improvement in Section
       ~\ref{sec:discussion}). As such, more fair comparisons to CPU and GPU systems would be possible and can reveal higher benefits for PIM systems in the future. 
\end{itemize}

\section{UPMEM PIM Architecture}\label{sec:upmem_pim}

{We describe} the organization of a UPMEM PIM-enabled system (Section~\ref{sec:sys-org}), the architecture of a DPU core (Section~\ref{sec:dpu-architecture}), and important aspects of programming DPUs (Section~\ref{sec:dpu-programming}).

\subsection{System Organization}
\label{sec:sys-org}

\begin{figure*}[h]
    \centering
    \includegraphics[width=\linewidth]{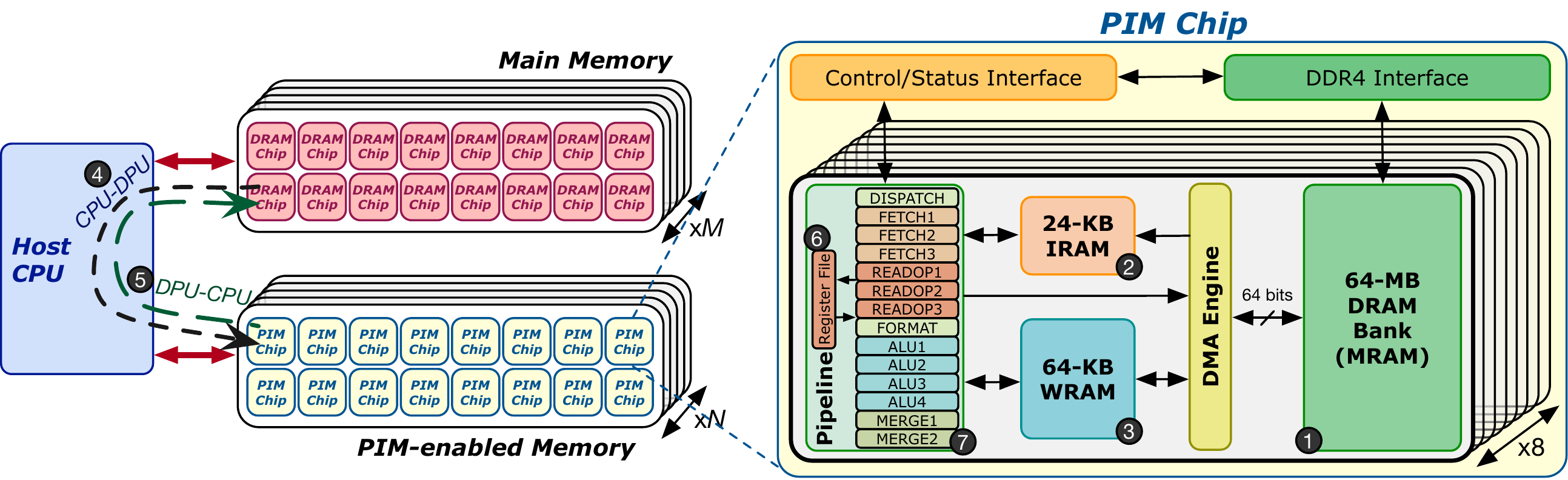}
    \vspace{-4mm}
    \caption{UPMEM-based PIM system with a host CPU, standard main memory, and PIM-enabled memory (left), and internal components of a UPMEM PIM chip (right)~\cite{upmem2018, devaux2019}.}
    \label{fig:scheme}
\end{figure*}

\begin{figure}[h]
    \centering
    \includegraphics[width=\linewidth]{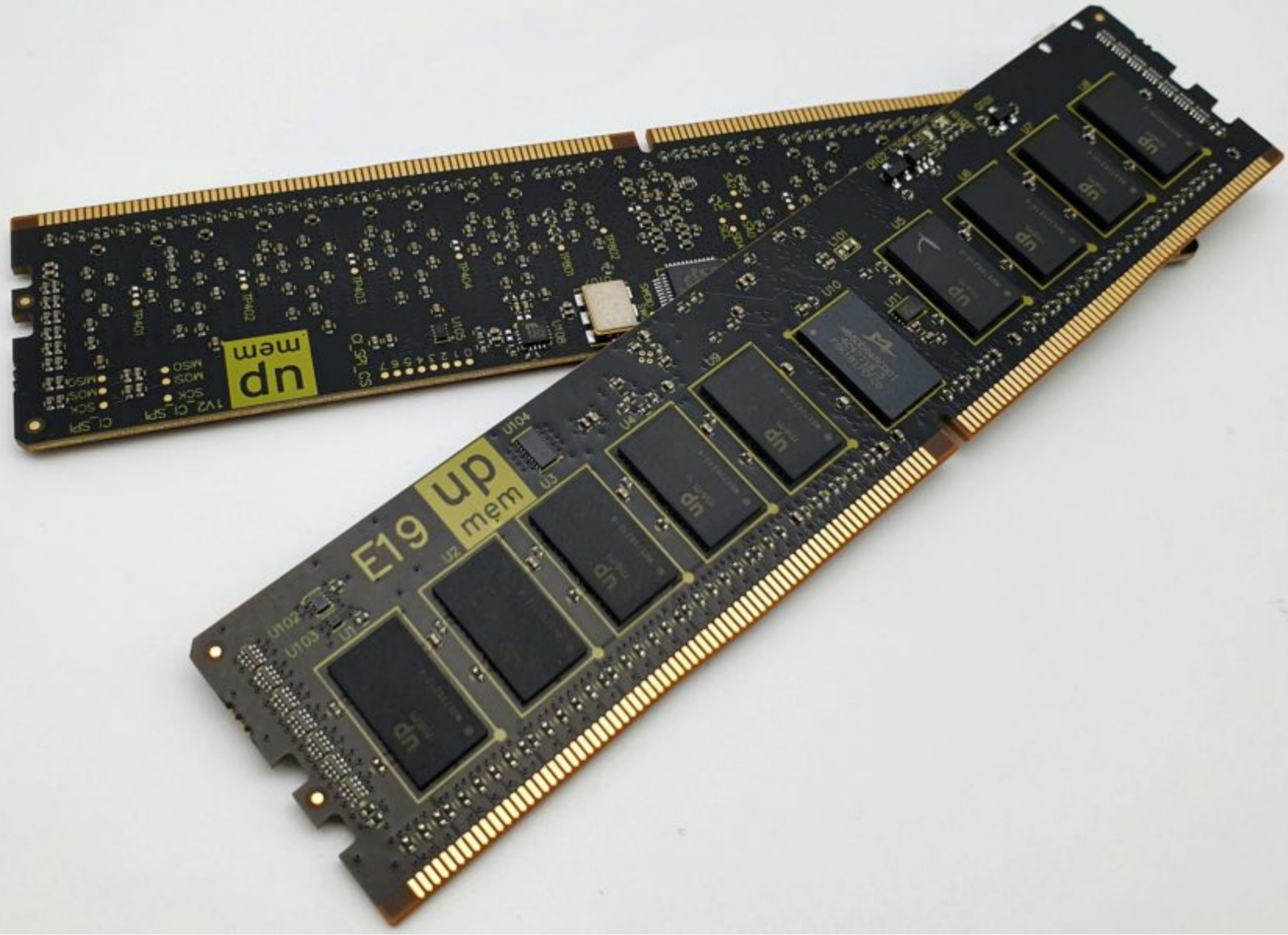}
    \caption{UPMEM-based PIM modules (downloaded from~\cite{upmem}).}
    \label{fig:dimm}
\end{figure}

\sloppy
Figure~\ref{fig:scheme} (left) depicts a UPMEM-based PIM system with (1) a \emph{host} CPU {(e.g., an x86~\cite{saini1993design}, ARM64~\cite{jaggar1997arm}, or 64-bit RISC-V~\cite{waterman2016design} {multi-core system})}, (2) standard main memory (DRAM memory modules~\cite{kim2014memory,ca.fall2020.refresh,ca.fall2020.challenges,ca.fall2020.solution}), and (3) PIM-enabled memory (UPMEM modules)~\cite{upmem2018, devaux2019}. {PIM-enabled memory can reside on one or more memory channels. 
A UPMEM module is a standard {DDR4-2400 DIMM (module)}~\cite{jedec2012ddr4} with several PIM chips. 
Figure~\ref{fig:dimm} shows two UPMEM modules. 
All DPUs in the UPMEM {modules} operate together as a parallel coprocessor {to the host CPU}.}

Inside each UPMEM PIM chip (Figure~\ref{fig:scheme} (right)), there are 8 DPUs. Each DPU has exclusive access to (1) a 64-MB DRAM bank, called \emph{Main RAM} (\emph{MRAM}) \boldone, (2) a 24-KB instruction memory, called \emph{Instruction RAM} (\emph{IRAM}) \boldtwo, and (3) a 64-KB scratchpad memory, called \emph{Working RAM} (\emph{WRAM}) \boldthree. 
MRAM {is} accessible by the host CPU (Figure~\ref{fig:scheme} (left)) for {\emph{copying}} input data (from main memory to MRAM) \boldfour~ and \emph{retrieving} results (from MRAM to main memory) \boldfive. 
These {CPU-DPU and DPU-CPU} data transfers can be performed in parallel (i.e., concurrently across multiple MRAM banks), if the buffers transferred from/to all MRAM banks {are of the same size}. Otherwise, the data transfers happen serially {(i.e., {a} transfer from/to another MRAM bank starts after the transfer from/to an MRAM bank completes)}. 
There is no support for direct communication between DPUs. All inter-DPU communication takes place through the host CPU by \emph{retrieving} results {from the DPU to the CPU} and {\emph{copying}} data {from the CPU to the DPU}.
\juanfinal{In current UPMEM-based PIM systems, concurrent host CPU and DPU accesses to the same MRAM bank are \emph{not} possible.}

{The programming interface for serial transfers~\cite{upmem-guide}} provides functions for copying a buffer to (\texttt{dpu\_copy\_to}) and from (\texttt{dpu\_copy\_from}) a specific MRAM bank.
{The programming interface for parallel transfers~\cite{upmem-guide}} provides functions for assigning buffers to specific MRAM banks (\texttt{dpu\_prepare\_xfer}) and then initiating the actual {CPU-DPU or DPU-CPU} transfers to execute in parallel (\texttt{dpu\_push\_xfer}).
{Parallel transfers require} that the transfer sizes {to/from all MRAM banks be} the same.
If the buffer to copy to all MRAM banks is the same, we can execute a broadcast {CPU-DPU} memory transfer (\texttt{dpu\_broadcast\_to}).

Main memory and PIM-enabled memory require different data layouts. While main memory uses the conventional horizontal DRAM mapping~\cite{GS-DRAM,devaux2019}, which maps consecutive 8-bit words onto consecutive DRAM chips, PIM-enabled memory needs entire 64-bit words mapped onto the same MRAM bank (in one PIM chip)~\cite{devaux2019}. 
The reason for this special data layout in PIM-enabled memory is that {each DPU has access to only a single MRAM bank, but it} can operate on data types of up to 64 bits. 
The UPMEM SDK includes a transposition library~\cite{devaux2019} to perform the necessary data shuffling when transferring data between main memory and MRAM banks. 
These data layout transformations are transparent to programmers. 
The {UPMEM SDK-provided} functions for serial/parallel/broadcast {CPU-DPU} and serial/parallel {DPU-CPU} transfers call the transposition library internally, and {the library} ultimately performs data layout conversion, as needed.

\juanfinal{The host CPU can allocate the desired number of DPUs, i.e., a \emph{DPU set}, to execute a DPU function or \emph{kernel}. 
Then, the host CPU launches the DPU kernel \emph{synchronously} or \emph{asynchronously}~\cite{upmem-guide}. Synchronous execution suspends the host CPU thread until the DPU set completes the kernel execution. Asynchronous execution returns immediately the control to the host CPU thread, which can later check the completion status~\cite{upmem-guide}.}\footnote{\juanfinal{In this work, we only use synchronous execution, since our benchmarks (Section~\ref{sec:benchmarks})) do \emph{not} require the host CPU to compute while a DPU kernel is running. We believe exploring the use of asynchronous execution is a promising topic for future work.}}

In current {UPMEM-based PIM system} configurations~\cite{upmem}, the maximum number of UPMEM DIMMs is 20. 
{A UPMEM-based PIM system with 20 UPMEM modules} can contain up to 2,560 DPUs which amounts to 160 GB of PIM-capable memory.

{Table~\ref{tab:pim-setups} presents the two {real} UPMEM-based PIM systems that we use in this work.} 

\begin{table*}[h]
\scriptsize
        \captionof{table}{UPMEM-based PIM Systems.}
        \label{tab:pim-setups}
        \begin{minipage}{\textwidth}
        \begin{center}
        \vspace{-1mm}
        \subcaption{Memory Parameters.}
        \vspace{-1mm}
        \resizebox{0.81\linewidth}{!}{
        \begin{tabular}{|l||ccccccc|cc|}
    \hline
    \multirow{3}{*}{\textbf{System}} & \multicolumn{7}{c|}{\textbf{PIM-enabled Memory}} & \multicolumn{2}{c|}{\textbf{DRAM Memory}} \\
    \cline{2-10}
     & \textbf{DIMM} & \textbf{Number of} & \textbf{Ranks/} & \textbf{DPUs/} & \textbf{Total} & \textbf{DPU} & \textbf{Total} & \textbf{Number of} & \textbf{Total} \\
     & \textbf{Codename} & \textbf{DIMMs} & \textbf{DIMM} & \textbf{DIMM} & \textbf{DPUs} & \textbf{Frequency} & \textbf{Memory} & \textbf{DIMMs} & \textbf{Memory} \\
    \hline
    \hline
\textbf{2,556-DPU System} & P21 & 20 & 2 & 128 & 2,556$^9$ & 350 MHz & 159.75 GB & 4 & 256 GB \\ \hline
 \textbf{640-DPU System} & E19 & 10 & 1 & 64 & 640 & 267 MHz & 40 GB & 2 & 64 GB \\ \hline
\end{tabular}

        }
        \end{center}
        \vspace{1mm}
        \end{minipage}
        \begin{minipage}{\textwidth}
        \begin{center}
        \subcaption{CPU Parameters.}
        \vspace{-1mm}
        \resizebox{0.71\linewidth}{!}{
        \begin{tabular}{|l||ccccc|}
    \hline
    \multirow{3}{*}{\textbf{System}} & \multicolumn{5}{c|}{\textbf{CPU}} \\
    \cline{2-6}
     & \textbf{CPU} & \textbf{CPU} & \multirow{2}{*}{\textbf{Sockets}} & \textbf{Mem. Controllers/} & \textbf{Channels/}\\
     & \textbf{Processor} & \textbf{Frequency} &  & \textbf{Socket} & \textbf{Mem. Controller}\\
    \hline
    \hline
\textbf{2,556-DPU System} & Intel Xeon Silver 4215~\cite{xeon-4215} & 2.50 GHz & 2 & 2 & 3\\ \hline
 \textbf{640-DPU System} & Intel Xeon Silver 4110~\cite{xeon-4110} & 2.10 GHz & 1 & 2 & 3\\ \hline
\end{tabular}

        }
        \end{center}
        \end{minipage}
\end{table*}

\begin{figure}[h]
    \centering
    \includegraphics[width=\linewidth]{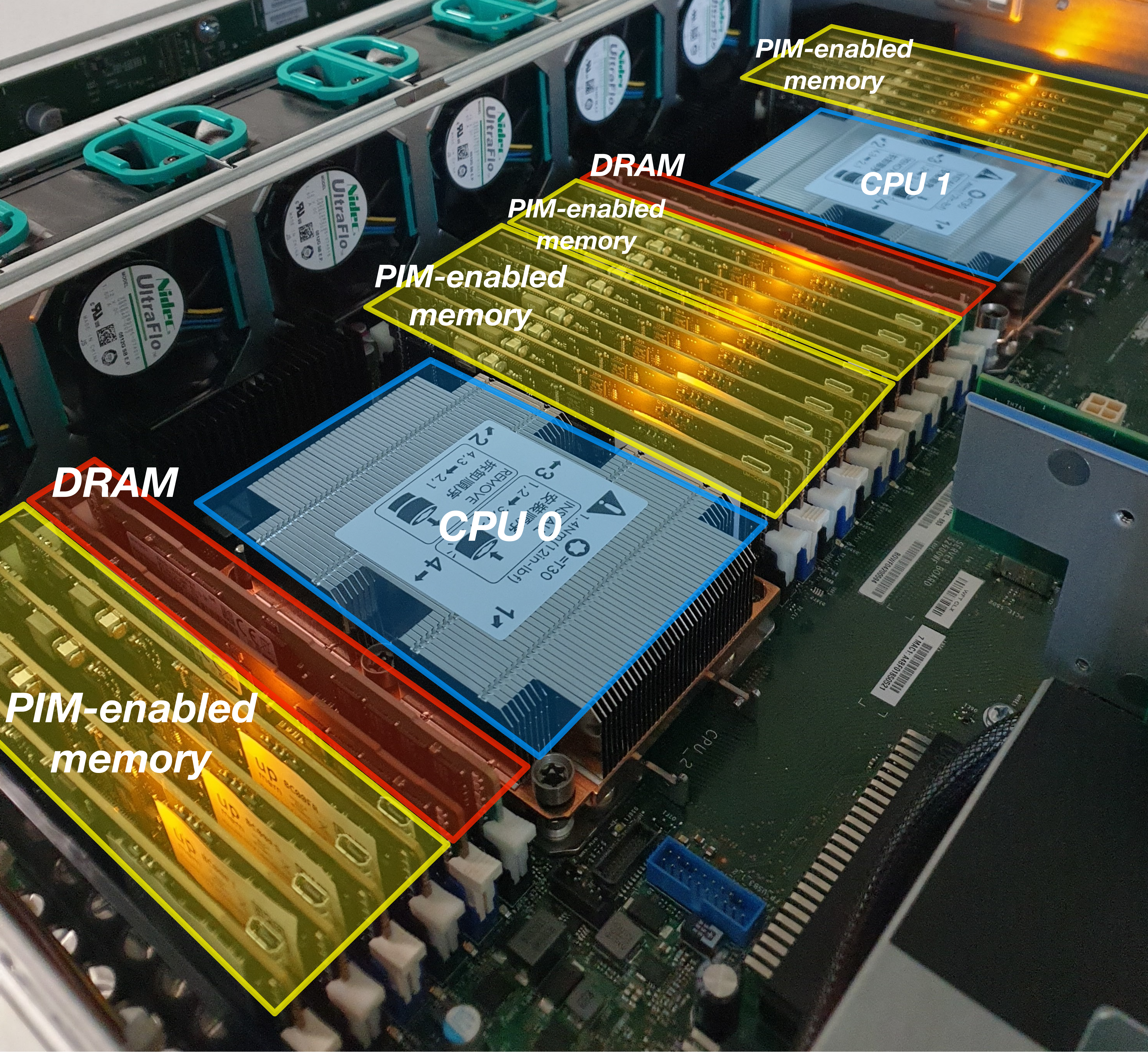}
    \caption{UPMEM-based PIM system with 2,560 DPUs. See Table~\ref{tab:pim-setups} for the specifications.}
    \label{fig:dual_socket}
    \vspace{-4mm}
\end{figure}

We use a real UPMEM-based PIM system that contains 2,556 DPUs, {and a total of {159.75 GB} MRAM}. The DPUs are organized into 20 double-rank DIMMs, with 128 DPUs {per DIMM}.\footnote{There are four faulty DPUs in the system where we run our experiments. They cannot be used and do not affect system functionality or the correctness of our results, but take away from the system's full computational power of 2,560 DPUs.}
{Each DPU runs} at 350 MHz.
The 20 UPMEM DIMMs are in a dual x86 socket with 2 memory controllers per socket. Each memory controller has 3 memory channels~\cite{xeon-4215}.
In each socket, two DIMMs of conventional DRAM (employed as main memory of the host CPU) are on {one} channel of one of the memory controllers. 
Figure~\ref{fig:dual_socket} shows a UPMEM-based PIM system with 20 UPMEM DIMMs.

We also use {an older real} system with 640 DPUs.  {The DPUs are organized into 10 single-rank DIMMs, with 64 DPUs {per DIMM}}.
The total amount of MRAM is thus 40 GB.
{Each DPU in this system runs} at 267 MHz.
The 10 UPMEM DIMMs are in an x86 socket with 2 memory controllers. Each memory controller has 3 memory channels~\cite{xeon-4110}.
Two DIMMs of conventional DRAM are on one channel {of one of the memory controllers}. 


\subsection{DRAM Processing Unit (DPU) Architecture}\label{sec:dpu-architecture}

A DPU {(Figure~\ref{fig:scheme} (right))} is a multithreaded in-order 32-bit RISC core with a {specific} Instruction Set Architecture (ISA)~\cite{upmem-guide}. 
The DPU {has} 24 hardware threads, each with 24 32-bit general-purpose registers {(\boldsix~ in Figure~\ref{fig:scheme} (right))}. 
These hardware threads share an instruction memory {(IRAM)} \boldtwo~ and a scratchpad memory {(WRAM)} \boldthree~ to store operands.
The DPU has a pipeline depth of 14 stages \boldseven, however, only the last three stages of the pipeline (i.e., ALU4, MERGE1, and MERGE2 in Figure~\ref{fig:scheme} (right)) can execute in parallel with the DISPATCH and FETCH stages of the next instruction in the same thread.
Therefore, instructions from the same thread must be dispatched 11 cycles apart, requiring at least 11 threads to fully utilize the pipeline~\cite{comm-upmem}.

The 24 KB IRAM can hold up to 4,096 48-bit encoded instructions.
The WRAM has a capacity of 64 KB. 
The DPU can access the WRAM through 8-, 16-, 32-, and 64-bit load/store instructions. 
{The ISA provides DMA instructions~\cite{upmem-guide} to move instructions from the MRAM bank to the IRAM, and data between the MRAM bank and the WRAM.} 

The frequency of a DPU {can potentially} reach 
\juanfinal{more than 400 MHz}~\cite{upmem}.
At 400 MHz, the {maximum possible} MRAM-WRAM bandwidth per DPU can achieve 
around 800 MB/s. 
Thus, the maximum aggregated {MRAM} bandwidth for a configuration with 2,560 DPUs can potentially be 2 TB/s.
{However, the DPUs run at 350 MHz in our 2,556-DPU setup and at 267 MHz in the 640-DPU {system}.} 
For this reason, the {maximum possible} MRAM-WRAM bandwidth per DPU in our setup is 700 MB/s (534 MB/s in the 640-DPU setup), and the maximum aggregated bandwidth for the 2,556 DPUs is 1.7 TB/s (333.75 GB/s {in the 640-DPU system}).

\subsection{DPU Programming}\label{sec:dpu-programming}

{UPMEM-based PIM systems use {the} \emph{Single Program Multiple Data} (\emph{SPMD})~\cite{ddca.spring2020.gpu} programming model, where software threads, called \emph{tasklets}, (1) execute the same code but operate on different pieces of data, and (2) can execute different control-flow {paths} at runtime.}

{Up to 24 tasklets can run on a DPU, since the number of hardware threads is 24. 
Programmers determine the number of tasklets per DPU at compile time, {and tasklets are statically assigned to each DPU}.} 

Tasklets inside the same DPU can share data {among each other} in MRAM and in WRAM, and can synchronize via \emph{mutexes}, \emph{barriers}, \emph{handshakes}, and \emph{semaphores}~\cite{rauber2013parallel}.

{Tasklets in different DPUs do \emph{not} share memory or any direct communication channel. As a result, they cannot directly communicate or synchronize}.
{As mentioned in Section~\ref{sec:sys-org}}, the host CPU handles communication of intermediate data between DPUs, and merges partial results into final ones.

\subsubsection{{Programming Language and Runtime Library}}
\label{sec:language-library}

DPU programs are written in the C language with some library calls~\cite{upmem2018, upmem-guide}.\footnote{{In this work, we use UPMEM SDK 2021.1.1~\cite{upmem-sdk}.}} 
{The UPMEM SDK~\cite{upmem-sdk} supports common data types supported in the C language and the LLVM compilation framework~\cite{lattner2004llvm}.
For the complete list of supported instructions, we refer the reader to the UPMEM user manual~\cite{upmem-guide}.}

The UPMEM runtime library~\cite{upmem-guide} provides library calls to move (1) instructions from the MRAM bank to the IRAM, and (2) data between the MRAM bank and the WRAM (namely, \texttt{mram\_read()} for MRAM-WRAM transfers, and \texttt{mram\_write()} for WRAM-MRAM transfers).

\sloppy
The UPMEM runtime library also provides functions to (1) lock and unlock mutexes (\texttt{mutex\_lock()}, \texttt{mutex\_unlock()}), which create critical sections, (2) access barriers (\texttt{barrier\_wait()}), which suspend tasklet execution until all tasklets in the DPU reach the same point in the program, (3) wait for and notify a handshake (\texttt{handshake\_wait\_for()}, \texttt{handshake\_notify()}), which enables one-to-one tasklet synchronization, and (4) increment and decrement semaphore counters (\texttt{sem\_give()}, \texttt{sem\_take()}).

{Even though} using {the C language to program the DPUs} ensures a low learning curve, programmers need to deal with several challenges. 
First, programming thousands of DPUs running up to 24 tasklets requires careful workload partitioning and orchestration. 
Each tasklet has a tasklet ID that programmers can use for that purpose. 
Second, programmers have to explicitly move data between the standard main memory and the MRAM banks, and {ensuring data coherence between the CPU and DPUs} (i.e., ensuring that CPU and DPUs use up-to-date and correct copies of data) is their responsibility.
Third, DPUs do \emph{not} employ cache memories.
The data movement between the MRAM banks and the WRAM is explicitly managed by the programmer.

\subsubsection{General Programming Recommendations}
\label{sec:general-recommendations}

General programming recommendations of the UPMEM-based PIM system that we find in the UPMEM programming guide~\cite{upmem-guide}, presentations~\cite{devaux2019}, and white papers~\cite{upmem2018} are as follows.

The first recommendation is to \textbf{execute on the DPUs portions of parallel code that are as long as possible}, avoiding frequent interactions with the host CPU. 
This recommendation minimizes CPU-DPU and DPU-CPU transfers, which happen through the narrow memory bus (Section~\ref{sec:sys-org}), and {thus cause a} data movement bottleneck~\cite{mutlu2019enabling,mutlu2020modern,ghoseibm2019,ghose2019arxiv}, which the PIM paradigm promises to alleviate.

The second recommendation is to \textbf{split the workload into independent data blocks}, which the DPUs operate on independently {(and concurrently)}. This recommendation maximizes parallelism and minimizes the need for inter-DPU communication and synchronization, which incurs high overhead, as it happens via the host CPU using CPU-DPU and DPU-CPU transfers.

The third recommendation is to \textbf{use as many working DPUs in the system as possible}, as long as the workload is sufficiently large to keep {the DPUs} busy performing actual work. This recommendation maximizes parallelism and increases utilization of the compute resources. 

The fourth recommendation is to \textbf{launch at least 11 tasklets in each DPU}, in order to fully utilize the fine-grained multithreaded pipeline, as mentioned in Section~\ref{sec:dpu-architecture}. 


\gpboxbegin{\gptask{gpr}}
\begin{enumerate}[1., wide, labelsep=0.5em]
\item \textbf{Execute on the {\emph{DRAM Processing Units} (\emph{DPUs})} portions of parallel code that are as long as possible.}
\item \textbf{Split the workload into independent data blocks, which the DPUs operate on independently.}
\item \textbf{Use as many working DPUs in the system as possible.}
\item \textbf{Launch at least 11 \emph{tasklets} {(i.e., software threads)} per DPU.}
\end{enumerate}
\gpboxend

In this work, we perform the first comprehensive characterization and analysis of the UPMEM PIM architecture, which allows us to (1) validate these programming recommendations and identify for which workload characteristics they hold, as well as (2) propose additional programming recommendations and suggestions for future PIM software designs, {and (3) propose suggestions and hints for future PIM hardware designs, which can enable easier programming as well as broad applicability of the hardware to more workloads.}

\section{Performance Characterization of a UPMEM DPU}
\label{sec:microbench}

This section presents {the first} performance characterization of a UPMEM DPU using microbenchmarks to assess various {architectural} limits {and bottlenecks}.
Section~\ref{sec:arith-throughput-wram} evaluates the throughput of arithmetic operations and WRAM bandwidth of a DPU using a streaming {microbenchmark}.
Section~\ref{sec:mram-bandwidth} evaluates the {sustained} bandwidth between MRAM and WRAM.
Section~\ref{sec:throughput-oi} evaluates the impact of the operational intensity of a workload on the {arithmetic} throughput of the DPU.
Finally, Section~\ref{sec:cpu-dpu} evaluates the bandwidth between the main memory of the host and {the MRAM banks}. 
Unless otherwise stated, {we report experimental results} on the {larger,} 2,556-DPU system presented in Section~\ref{sec:sys-org}.
{All observations and trends identified in this section also apply to the {older 640-DPU system (we verified this experimentally)}.} 
All microbenchmarks used in this section are publicly {and freely} available~\cite{gomezluna2021repo}.

\subsection{Arithmetic Throughput and WRAM Bandwidth}\label{sec:arith-throughput-wram}

The DPU pipeline is capable of performing one integer {addition/subtraction} operation every cycle and up to one 8-byte {WRAM load/store} every cycle when the pipeline is full~\cite{devaux2019}.
Therefore, at 350 MHz, the theoretical peak arithmetic throughput is {350 Millions of OPerations per Second (MOPS), assuming only integer addition operations are issued into the pipeline,} and the theoretical peak WRAM bandwidth is 2,800~MB/s.
In this section, we evaluate the arithmetic throughput and {sustained} WRAM bandwidth {that can be achieved} by a streaming {microbenchmark (i.e., a benchmark with unit-stride access to memory locations)} and how {the arithmetic throughput and WRAM bandwidth} vary with the number of tasklets deployed.

\subsubsection{\textbf{Microbenchmark Description}}\label{sec:wram-benchmark-desc}

To evaluate arithmetic throughput and WRAM bandwidth {in streaming workloads}, we implement {a set of microbenchmarks~\cite{gomezluna2021repo}} where every tasklet loops over elements of an array in WRAM and performs read-modify-write operations.
We measure the time it takes {to perform} WRAM loads, arithmetic operations, WRAM stores, and loop control.
We do \emph{not} measure the time it takes {to perform} MRAM-WRAM DMA transfers {(we will study them separately in Section~\ref{sec:mram-bandwidth})}.

\noindent\paragraph{\textbf{Arithmetic Throughput.}} For arithmetic throughput, we examine the addition, subtraction, multiplication, and division operations for 32-bit integers, 64-bit integers, floats, and doubles.
Note that the throughput for unsigned integers is the same as that for signed integers. 
As we indicate {at} the beginning of Section~\ref{sec:arith-throughput-wram}, the DPU pipeline is capable of performing one integer addition/subtraction operation every cycle, assuming that the pipeline is full~\cite{devaux2019}. 
However, real-world workloads do not execute {\emph{only}} integer addition/subtraction operations. Thus, the theoretical peak arithmetic throughput of 350 MOPS is not realistic {for full execution of real workloads}. 
Since the DPUs store operands in WRAM (Section~\ref{sec:dpu-architecture}), a realistic evaluation of arithmetic throughput should consider the accesses to WRAM to read source operands and write destination operands. One access to WRAM involves one WRAM address calculation and one load/store operation.

Listing~\ref{lst:ubench_arith} shows an example microbenchmark for {the} throughput evaluation of 32-bit integer addition. Listing~\ref{sublst:codea} shows our microbenchmark written in C. The operands are stored in \texttt{bufferA}, which we allocate in WRAM using \texttt{mem\_alloc}~\cite{upmem-guide} (line 2). 
The \texttt{for} loop in line 3 goes through {each element of} \texttt{bufferA} and {adds} a scalar value \texttt{scalar} to each element. 
In each iteration of the loop, we load one element of \texttt{bufferA} {into} a temporal variable \texttt{temp} (line 4), add \texttt{scalar} {to it} (line 5), and store the result {back} into the same position of \texttt{bufferA} (line 6). 
Listing~\ref{sublst:codeb} shows the compiled code, which we can inspect using UPMEM's Compiler Explorer~\cite{upmem-explorer}. 
The loop contains 6 instructions: WRAM address calculation (\texttt{lsl\_add}, line 3), WRAM load (\texttt{lw}, line 4), addition (\texttt{add}, line 5), WRAM store (\texttt{sw}, line 6), loop index update (\texttt{add}, line 7), and conditional branch (\texttt{jneq}, line 8). 
For a 32-bit integer subtraction (\texttt{sub}), the number of instructions in the streaming loop is also 6, but for other operations and data types the number of instructions can be different {(as we show below)}.

\begin{figure}[h]
    \setcaptiontype{lstlisting}
    \begin{minipage}{\linewidth}
	    \begin{lstlisting}[style=myC]
#define SIZE 256
int* bufferA = mem_alloc(SIZE*sizeof(int));
%\HilightYellow%for(int i = 0; i < SIZE; i++){
%\HilightGreen%    int temp = bufferA[i];
%\HilightPink%    temp += scalar;
%\HilightBlue%    bufferA[i] = temp;
%\HilightYellow%}
        \end{lstlisting}
        \vspace{-8mm}
        \subcaption{C-based code.}
        \label{sublst:codea}
    \end{minipage}
   \vspace{2mm}
    \begin{minipage}{\linewidth}
	    \begin{lstlisting}[style=myC]
%\HilightYellow%  move r2, 0
.LBB0_1: // Loop header
%\HilightGreen%  lsl_add r3, r0, r2, 2
%\HilightGreen%  lw r4, r3, 0
%\HilightPink%  add r4, r4, r1
%\HilightBlue%  sw r3, 0, r4
%\HilightYellow%  add r2, r2, 1
%\HilightYellow%  jneq r2, 256, .LBB0_1
       \end{lstlisting}
       \vspace{-8mm}
       \subcaption{Compiled code in UPMEM {DPU} ISA.}
       \label{sublst:codeb}
   \end{minipage}
    \caption{Microbenchmark for throughput evaluation of 32-bit integer addition~\cite{gomezluna2021repo}.}
    \label{lst:ubench_arith}
\end{figure}

{Given the instructions in the loop of the streaming microbenchmark (Listing~\ref{sublst:codeb})}, we can obtain the expected throughput of arithmetic operations. 
{Only one out of the six instructions is an arithmetic operation (\texttt{add} in line 5 in Listing~\ref{sublst:codeb}).} 
Assuming that the pipeline is full, the DPU issues (and retires) one instruction every cycle~\cite{devaux2019}. As a result, we need as many cycles as instructions in the streaming loop to perform one arithmetic operation. If the number of instructions {in the loop} is $n$ and the DPU frequency is $f$, we calculate the arithmetic throughput in operations per second (OPS) as expressed in Equation~\ref{eq:throughput}.

\begin{equation}
Arithmetic\ Throughput\ (in\ OPS) = \frac{f}{n} 
\label{eq:throughput}
\end{equation}

For a 32-bit integer addition (Listing~\ref{lst:ubench_arith}), the expected arithmetic throughput on a DPU running at 350 MHz {is} 58.33 {millions of operations per second (MOPS). We verify this on real hardware in Section~\ref{sec:arith-throughput}.}


\noindent\paragraph{\textbf{WRAM Bandwidth.}}
{To evaluate sustained} WRAM bandwidth, we examine the four versions of the STREAM benchmark~\cite{mccalpin1995}, which are {COPY, ADD, SCALE, and TRIAD}, for 64-bit integers. 
These {microbenchmarks} access two ({COPY, SCALE}) or three ({ADD, TRIAD}) arrays {in a streaming manner (i.e., with unit-stride or sequentially)}.
The operations performed by {ADD, SCALE, and TRIAD} are addition, multiplication, and addition+multiplication, respectively.

{In our experiments, we measure the \emph{sustained bandwidth} of WRAM, which is the average bandwidth that we measure over a relatively long period of time (i.e., while streaming through an entire array in WRAM).}

{We can obtain the maximum theoretical WRAM bandwidth of our STREAM microbenchmarks, which depends on the number of instructions needed to execute the operations in each version of STREAM. Assuming that the DPU pipeline is full, we calculate the maximum theoretical WRAM bandwidth in bytes per second (B/s) with Equation~\ref{eq:bandwidth}, where $b$ is the total number of bytes read and written, $n$ is the number of instructions in a version of STREAM to read, modify, and write the $b$ bytes, and $f$ is the DPU frequency.}

\begin{equation}
WRAM\ Bandwidth\ (in\ B/s) = \frac{b \times f}{n} 
\label{eq:bandwidth}
\end{equation}

For example, COPY executes one WRAM load (\texttt{ld}) and one WRAM store (\texttt{sd}) per 64-bit element. 
These two instructions require 22 cycles to execute for a single tasklet. 
When the pipeline is full (i.e., {with} 11 tasklets or more), $11 \times 16 = 176$ bytes are read and written in 22 cycles. 
As a result, {$b = 176$ and $n = 22$, and thus,} the maximum theoretical WRAM bandwidth for COPY, {at $f$=350 MHz}, is 
2,800 MB/s. {We verify this on real hardware in Section~\ref{sec:wram-bandwidth}.}

\subsubsection{\textbf{Arithmetic Throughput}}\label{sec:arith-throughput}

\begin{figure}[h]
    \centering
    \includegraphics[width=\linewidth]{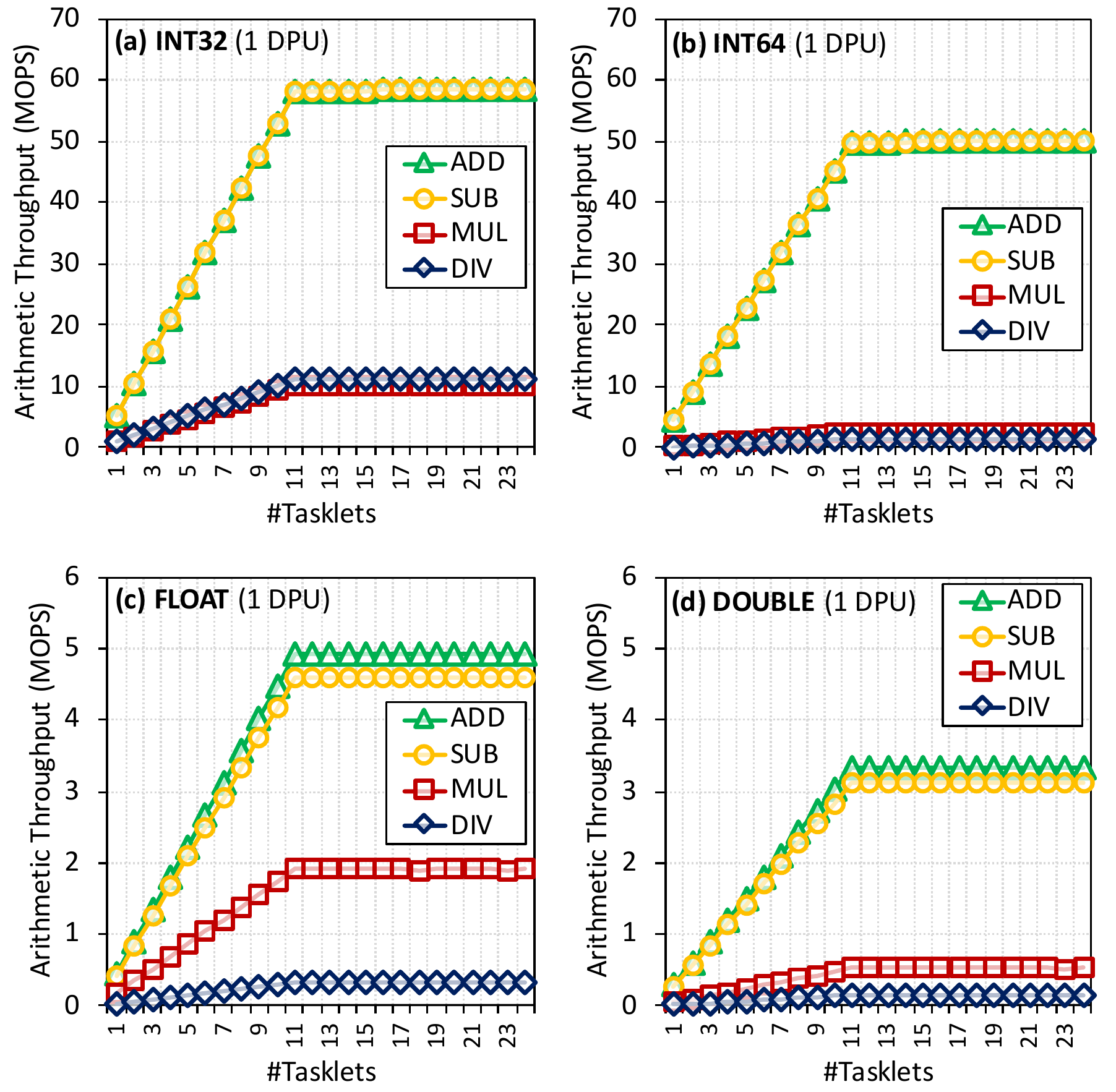}
    \caption{Throughput of arithmetic operations (ADD, SUB, MUL, DIV) on one DPU for four different data types: {(a) INT32, (b) INT64, (c) FLOAT, (d) DOUBLE.}}
    \label{fig:throughput-dpu}
\end{figure}

Figure~\ref{fig:throughput-dpu} shows how the {measured} arithmetic throughput on one DPU {(in MOPS)} varies with the number of tasklets. We use 1 to 24 tasklets, which is the maximum number of hardware threads.

We make four key observations from Figure~\ref{fig:throughput-dpu}. 

First, the throughput of all arithmetic operations and data types saturates after 11 tasklets.
This observation is consistent with the description of the pipeline in Section~\ref{sec:dpu-architecture}.
Recall that {the DPU uses} {fine-grained multithreading across tasklets} to fully utilize {its} pipeline.
Since instructions in the same tasklet are dispatched 11 cycles apart, 11 tasklets is the minimum number of tasklets needed to fully utilize the pipeline.

\gboxbegin{\rtask{ko}}
\textbf{The arithmetic throughput of a DRAM Processing Unit saturates at 11 or more tasklets}. 
{This observation is consistent for} different data types (INT32, INT64, UINT32, UINT64, FLOAT, DOUBLE) and operations (ADD, SUB, MUL, DIV).
\gboxend

Second, the throughput of addition/subtraction is 
58.56 {MOPS} for 32-bit integer values {(Figure~\ref{fig:throughput-dpu}a)}, and 
50.16 {MOPS} for 64-bit integer values {(Figure~\ref{fig:throughput-dpu}b)}. 
{The number of instructions inside the streaming loop for 32-bit integer additions/subtractions is 6 (Listing~\ref{lst:ubench_arith}).} 
Hence, the expected throughput at 
350 MHz is 58.33 {MOPS (obtained with Equation~\ref{eq:throughput})}, which is close to {what we measure (58.56 {MOPS})}.
A loop {with} 64-bit integer additions/subtractions contains 7 instructions: the same 6 instructions as the 32-bit version plus an addition/subtraction with carry-in bit (\texttt{addc/subc}) for the 
{upper 32} bits of the {64-bit} operands. 
Hence, the expected throughput at 
350 MHz is 50 {MOPS} which is also close to {what we measure (50.16 {MOPS})}.

Third, the throughput of integer multiplication and division is significantly lower than that of integer addition and subtraction {(note the large difference in y-axis scale between Figure~\ref{fig:throughput-dpu}a,b and Figure~\ref{fig:throughput-dpu}c,d)}.
A major reason is that the DPU pipeline does not include a complete $32\times32$-bit multiplier {due to} hardware cost concerns and limited number of available metal layers~\cite{devaux2019}.
Multiplications and divisions of 32-bit operands are implemented {using two instructions} (\texttt{mul\_step}, \texttt{div\_step})~\cite{upmem-guide}, which are based on bit shifting and addition.
With these {instructions}, multiplication and division can take up to 32 {cycles (32 \texttt{mul\_step} or \texttt{div\_step} instructions)} to perform, depending on the values of the operands. 
{In case multiplication and division take 32 cycles}, the expected throughput {({Equation~\ref{eq:throughput}})} is 
10.94 {MOPS}, which is similar {to what we measure (10.27 MOPS for 32-bit multiplication and 11.27 MOPS for 32-bit division, {as shown in} Figure~\ref{fig:throughput-dpu}a)}. 
For multiplication and division of 64-bit integer operands, programs call two {UPMEM runtime} library functions (\texttt{\_\_muldi3}, \texttt{\_\_divdi3})~\cite{upmem-guide, llvm-builtin} with 123 and 191 instructions, respectively. The expected throughput for these {64-bit operations} is significantly lower than for 32-bit operands, as our measurements confirm (2.56 MOPS for 64-bit multiplication and 1.40 MOPS for 64-bit division, {as shown in} Figure~\ref{fig:throughput-dpu}b).

Fourth, the throughput of floating point operations {(as shown in Figures~\ref{fig:throughput-dpu}c and ~\ref{fig:throughput-dpu}d)} is more than an order of magnitude lower than that of integer operations.
A major reason is that the DPU pipeline does \emph{not} feature native floating point ALUs.
The UPMEM runtime library emulates these operations {in} software~\cite{upmem-guide, llvm-builtin}.
As a result, for each 32-bit or 64-bit floating point operation, the number of instructions executed in the pipeline is between several tens (32-bit floating point addition) and more than 2000 (64-bit floating point division). 
This explains the low throughput. {We measure 4.91/4.59/1.91/0.34 MOPS for FLOAT add/sub/multiply/divide (Figure~\ref{fig:throughput-dpu}c) and 3.32/3.11/0.53/0.16 MOPS for DOUBLE add/sub/multiply/divide (Figure~\ref{fig:throughput-dpu}d).}

\gboxbegin{\rtask{ko}}
\begin{itemize}[wide, labelsep=0.5em]
\item \textbf{{DRAM Processing Units (DPUs) provide native hardware} support for 32- and 64-bit integer addition and subtraction}, {leading to high throughput for these operations}. 
\item \textbf{DPUs do \emph{not} natively support 32- and 64-bit multiplication and division, and floating point operations}. These operations are emulated by the UPMEM runtime library, leading to much lower throughput.
\end{itemize}
\gboxend

\subsubsection{\textbf{Sustained WRAM Bandwidth}}
\label{sec:wram-bandwidth}

\begin{figure}[h]
        \centering
        \includegraphics[width=\linewidth]{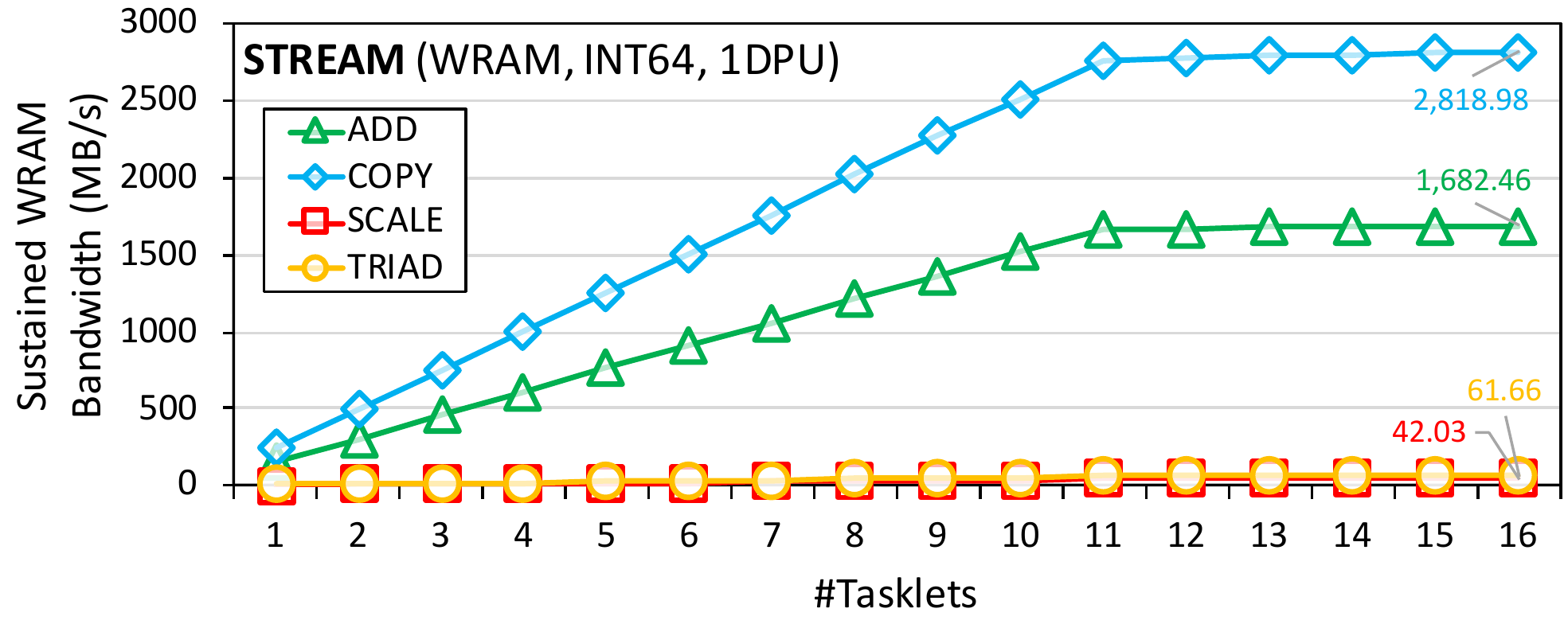}
        \captionof{figure}{{Sustained} WRAM bandwidth for streaming access patterns.
        }
        \label{fig:wram-stream}
\end{figure}

Figure~\ref{fig:wram-stream} shows how the {sustained} WRAM bandwidth varies with the number of tasklets (from 1 to 16 tasklets).
In these experiments, we unroll the loop of the STREAM {microbenchmarks}, in order to {exclude} loop control {instructions}, and achieve the highest possible {sustained WRAM bandwidth. We make three major observations.}

{First,} similar to arithmetic throughput, we observe that WRAM bandwidth saturates after 11 tasklets which is the number of tasklets needed to fully utilize the DPU pipeline.

{Second,} the maximum {measured} sustained WRAM bandwidth depends on the number of instructions needed to execute the operation. 
{For COPY, we measure 
2,818.98 MB/s, which is similar to the maximum theoretical WRAM bandwidth of 2,800 MB/s, which we obtain with Equation~\ref{eq:bandwidth}} {(see Section~\ref{sec:wram-benchmark-desc})}.
ADD executes 5 instructions per 64-bit element: two WRAM loads (\texttt{ld}), one addition (\texttt{add}), one addition with carry-in bit (\texttt{addc}), and one WRAM store (\texttt{sd}). 
In this case, $11 \times 24 = 264$ bytes are accessed in 55 cycles when the pipeline is full. 
Therefore, the maximum {theoretical WRAM} bandwidth for ADD is 
1,680 MB/s, which is similar to {what we measure} 
(1,682.46 MB/s).
The {maximum sustained WRAM} bandwidth {for} SCALE and TRIAD is significantly smaller {(42.03 and 61.66 MB/s, respectively)}, since {these microbenchmarks} use the costly multiplication operation, {which is a library function with 123 instructions (Section~\ref{sec:arith-throughput})}. 

{Third, and importantly (but not shown in Figure~\ref{fig:wram-stream}),} \textbf{WRAM bandwidth is independent of the access pattern} (streaming, strided, random),\footnote{We have verified this observation using a microbenchmark (which we {also provide as part of our open source release~\cite{gomezluna2021repo}}), but do not show the {detailed} results here for brevity. This microbenchmark uses three arrays in WRAM, $a$, $b$, and $c$. Array $a$ is a list of addresses to copy from $b$ to $c$ (i.e., $c[a[i]] = b[a[i]]$). This list of addresses can be (1) unit-stride (i.e., $a[i] = a[i - 1] + 1$), (2) strided (i.e., $a[i] = a[i - 1] + stride$), or (3) random (i.e., $a[i] = rand()$). For a given number of tasklets and size of the arrays, we measure the \emph{same} execution time for \emph{any} access pattern (i.e., unit-stride, strided, or random), which verifies that WRAM bandwidth is independent of the access pattern.} since all 8-byte WRAM {loads and stores} take one cycle {when the DPU pipeline is full, same as any other native instruction executed in the pipeline~\cite{devaux2019}}.

\gboxbegin{\rtask{ko}}
\textbf{The sustained bandwidth provided by the DRAM Processing Unit's internal Working memory (WRAM) is independent of the {memory} access pattern} (either streaming, strided, or random access pattern).

\textbf{All 8-byte WRAM {loads and stores} take one cycle}, when the DRAM Processing Unit's pipeline is full (i.e., with 11 or more tasklets).
\gboxend

\subsection{MRAM Bandwidth and Latency}\label{sec:mram-bandwidth}

{Recall that a DPU, so as to be able to access data from WRAM via load/store instructions, should first transfer the data from its associated MRAM bank to its WRAM via a DMA engine.}
This section evaluates the bandwidth that can be sustained from MRAM, including read and write bandwidth (Section~\ref{sec:mram-read-write}), streaming access bandwidth (Section~\ref{sec:mram-streaming}), and strided/random access bandwidth (Section~\ref{sec:mram-strided-random}).

\subsubsection{\textbf{Read and Write Latency and Bandwidth}}\label{sec:mram-read-write}

In this experiment, we measure the latency of a single DMA transfer of different sizes for a single tasklet, and compute the corresponding {MRAM} bandwidth.
These DMA transfers are performed via the \texttt{mram\_read(mram\_source,  wram\_destination, SIZE)} and \texttt{mram\_write(wram\_source, mram\_destination, SIZE)} functions, where \texttt{SIZE} is the transfer size in bytes and must be a multiple of 8 between 8 and 2,048 according to 
{UPMEM SDK 2021.1.1}~\cite{upmem-guide}.

\noindent\paragraph{\textbf{Analytical Modeling.}}
We can {analytically} model the MRAM access latency {(in cycles) using the linear expression in Equation~\ref{eq:mramlatency}},
where $\alpha$ is the fixed cost of a {DMA} transfer, $\beta$ {represents the variable cost (i.e., cost per byte)}, and $size$ is the transfer size {in bytes}.

\vspace{-1mm}
\begin{equation}
MRAM\ Latency\ (in\ cycles) = \alpha + \beta \times size 
\label{eq:mramlatency}
\end{equation}

After modeling the MRAM access latency {using Equation~\ref{eq:mramlatency}}, we can {analytically} model the MRAM bandwidth (in B/s) using Equation~\ref{eq:mrambandwidth}, where $f$ is the DPU frequency.

\vspace{-1mm}
\begin{multline}
MRAM\ Bandwidth\ (in\ B/s) = \frac{size \times f}{MRAM\ Latency} = \\ = \frac{size \times f}{\alpha + \beta \times size}
\end{multline}
\label{eq:mrambandwidth}

\begin{figure}[h]
        \centering
        \includegraphics[width=0.7\linewidth]{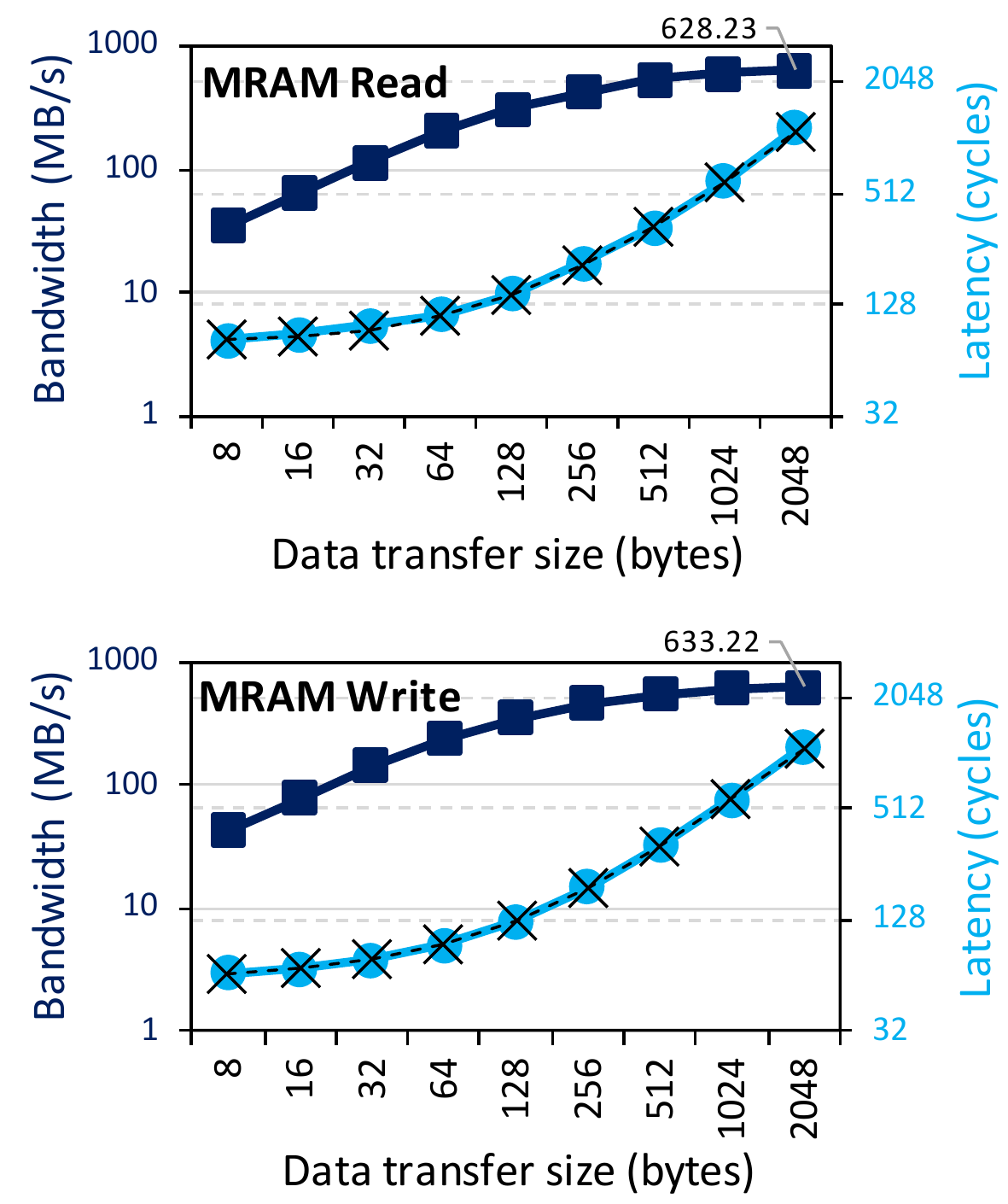}
        \caption{MRAM read and write latency {(log scale)} and bandwidth (log scale) for {data} transfer sizes between 8 and 2,048 bytes. 
        The black dashed line represents latency estimates with a linear model {(Equation~\ref{eq:mramlatency})}.}
        \label{fig:mram-bandwidth}
\end{figure}

\vspace{-4mm}
\noindent\paragraph{\textbf{Measurements.}}
Figure~\ref{fig:mram-bandwidth} shows how the measured MRAM read and write latency and bandwidth vary with transfer size {and how well the measured latency follows the analytical model we develop above.}

\vspace{-4mm}
\vspace{-4mm}

In our measurements, we find that $\alpha$ is $\sim$$77$ cycles for \texttt{mram\_read} and $\sim$$61$ cycles for \texttt{mram\_write}. For both types of transfers, the value $\beta$ is 0.5~cycles/B. {The inverse of $\beta$ is the maximum theoretical MRAM bandwidth (assuming the fixed cost $\alpha = 0$), which results in 2 B/cycle.}
The latency values estimated with {our analytical model in Equation~\ref{eq:mrambandwidth} (as shown by the black dashed lines} in Figure~\ref{fig:mram-bandwidth}) accurately match the latency measurements (light blue {lines} in Figure~\ref{fig:mram-bandwidth}). 

\gboxbegin{\rtask{ko}}
\vspace{-2mm}
\begin{itemize}[wide, labelsep=0.5em]
\item \textbf{The {DRAM Processing Unit's Main memory (MRAM) bank} access latency increases linearly with the transfer size. 
\item The maximum theoretical MRAM bandwidth is 2 bytes per cycle.}
\end{itemize}
\vspace{-3mm}
\gboxend
\vspace{-2mm}

{We make four observations from Figure~\ref{fig:mram-bandwidth}.} 

First, we observe that read and write accesses to MRAM are symmetric. 
The latency and bandwidth of read and write transfers are very similar for a given {data} transfer size.

Second, we observe 
that the sustained MRAM bandwidth (both read and write) increases with {data} transfer size.
{The maximum sustained MRAM bandwidth we measure is 628.23 MB/s for read and 633.22 MB/s for write transfers (both for 2,048-byte transfers). 
Based on this observation, a general recommendation to maximize {MRAM} bandwidth utilization is to \textbf{use large DMA transfer sizes when all the accessed data is going to be used}.
According to Equation~\ref{eq:mrambandwidth}, the theoretical maximum MRAM bandwidth is 
700 MB/s at {a} DPU frequency of 350 MHz (assuming no fixed transfer cost, i.e., $\alpha = 0$). 
Our measurements are within 
12\% of this theoretical maximum.}

\pboxbegin{\ptask{pr}}
\vspace{-2mm}
For data movement between the {DRAM Processing Unit's Main memory (MRAM) bank and the internal Working memory (WRAM)}, \textbf{use large DMA transfer sizes when all the accessed data is going to be used}.
\vspace{-2mm}
\pboxend

{Third, we observe that MRAM} latency changes 
{slowly} between 8-byte and 128-byte transfers. 
According to Equation~\ref{eq:mramlatency}, the read latency for 128 bytes is 141 cycles {and the read latency for 8 bytes is 81 cycles. In other words, latency increases by only 74\% while transfer size increases by 16$\times$}. 
The reason is that, for small {data transfer sizes}, the fixed cost ($\alpha$) of the transfer latency dominates the variable cost ($\beta \times size$). 
{For large data transfer sizes, the fixed cost ($\alpha$) does \emph{not} dominate the variable cost ($\beta \times size$), and in fact the opposite starts becoming true.} 
{We observe that, for} read transfers, $\alpha$ (77 cycles) represents 95\% of the latency for 8-byte reads and 55\% of the latency for 128-byte reads. 
{Based on this observation, one recommendation for programmers} is to \textbf{fetch more bytes than necessary within a {128-byte} limit when using {small data transfer sizes}}.
{Doing so increases} the probability of finding data in WRAM {for} later accesses, {eliminating future MRAM accesses}.
The program can simply check if the desired data has been fetched in a previous MRAM-WRAM transfer, before issuing a new {small data} transfer.

\vspace{-1mm}
\pboxbegin{\ptask{pr}}
\vspace{-2mm}
For small transfers between the {DRAM Processing Unit's Main memory (MRAM) bank and the internal Working memory (WRAM)}, \textbf{fetch more bytes than necessary within a 128-byte limit}. {Doing so increases the {likelihood} of finding data in WRAM for} later accesses {(i.e., the program can check {whether the desired data is} in WRAM before issuing a new MRAM access)}.
\vspace{-2mm}
\pboxend

{Fourth, MRAM bandwidth} scales almost linearly between 8 and 128 bytes {due to the slow MRAM latency increase.}
{After 128 bytes, MRAM bandwidth begins to saturate.} 
The reason the {MRAM} bandwidth saturatesat large data transfer sizes {is related} to the inverse relationship of bandwidth and latency (Equation~\ref{eq:mrambandwidth}). 
The fixed cost ($\alpha$) of the transfer latency {becomes} negligible with respect to the variable cost ($\beta \times size$) as the {data} transfer size increases. 
For example, $\alpha$ for read transfers (77 cycles) represents {only} 23\%, 13\%, and 7\% of the MRAM latency for 512-, 1,024-, and 2,048-byte read transfers, respectively. 
As a result, the MRAM read bandwidth increases by only 13\% and 17\% for 1,024- and 2,048-byte transfers over 512-byte transfers. 
Based on this observation, \textbf{the recommended data transfer size, when all the accessed data is going to be used, depends on a program's WRAM usage}, since WRAM has a limited size (only 64 KB). For example, if each tasklet of a DPU program needs to allocate 3 temporary WRAM buffers for data from 3 different arrays stored in MRAM, using 2,048-byte data transfers {requires that the size of each WRAM buffer is 2,048 bytes. 
This limits the number of tasklets to 10, which is less than the recommended minimum of 11 tasklets (Sections~\ref{sec:general-recommendations} and~\ref{sec:arith-throughput}), since {$\frac{64 KB}{3 \times 2,048} < 11$}.} 
In such a case, using 1,024-byte data transfers is preferred, since the bandwidth of 2,048-byte transfers is only 4\% higher than that of 1,024-byte transfers, according to our measurements ({shown in Figure~\ref{fig:mram-bandwidth}}).

\vspace{-1mm}
\pboxbegin{\ptask{pr}}
\vspace{-2.5mm}
\textbf{Choose the data transfer size {between the DRAM Processing Unit's Main memory (MRAM) bank and the internal Working memory (WRAM)} based on the program's WRAM usage}, as it imposes a tradeoff between the sustained MRAM bandwidth and the number of tasklets that can run in the {DRAM Processing Unit (which is dictated by the limited WRAM capacity)}.
\vspace{-2.5mm}
\pboxend


\vspace{-2mm}
\subsubsection{{Sustained} Streaming Access Bandwidth}\label{sec:mram-streaming}

In this experiment, we use the same four versions of the STREAM benchmark~\cite{mccalpin1995} described in Section~\ref{sec:wram-benchmark-desc}, but include the MRAM-WRAM DMA transfer time in our measurements.
We also add another version of {the copy benchmark}, COPY-DMA, which copies data from MRAM to WRAM and back without performing any WRAM loads/stores in the {DPU} core.
We use 1024-byte DMA transfers.
We scale the number of tasklets from 1 to 16.
The tasklets collectively stream 2M 8-byte elements \izzat{(total of 16 MB)}, which are divided evenly across the tasklets.

\begin{figure}[h]
\vspace{-2mm}
        \centering
        \includegraphics[width=\linewidth]{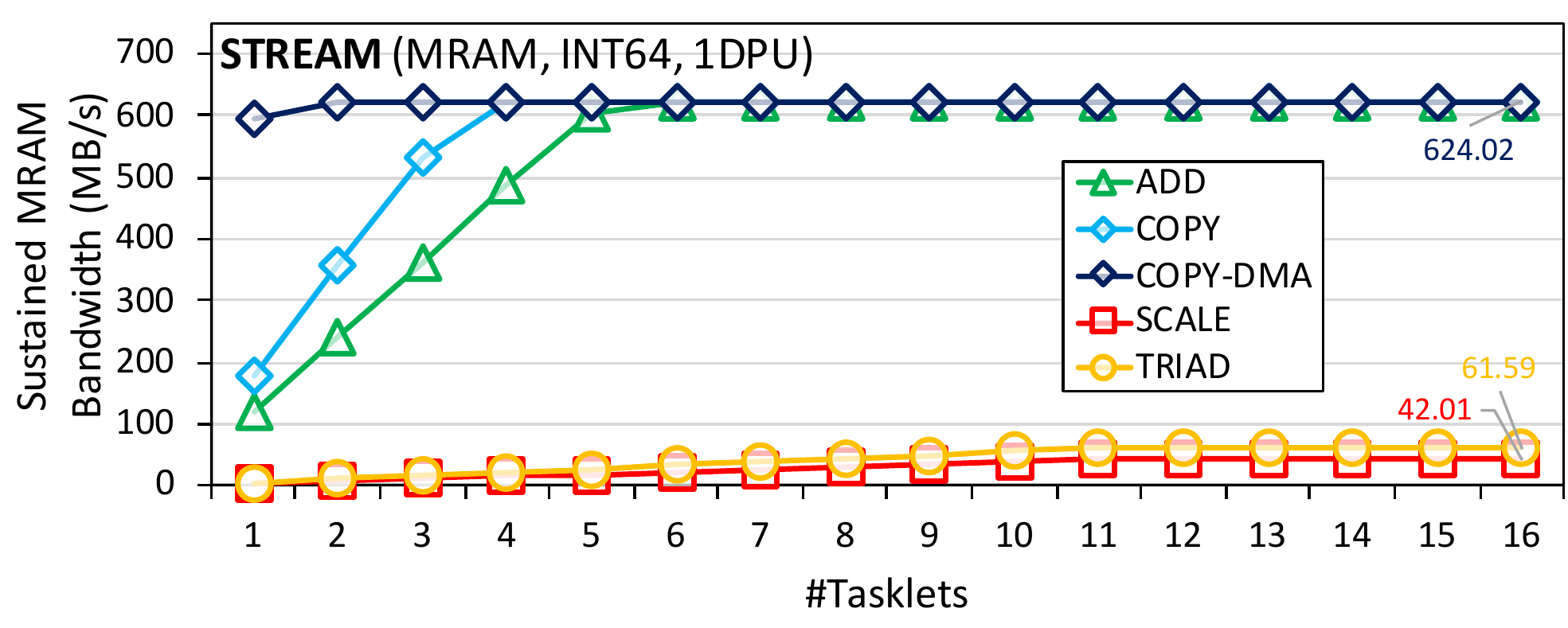}
        \vspace{-5mm}
        \captionof{figure}{{Sustained} MRAM bandwidth for streaming access patterns.}
        \label{fig:mram-stream}
\vspace{-4mm}
\end{figure}

Figure~\ref{fig:mram-stream} shows how the MRAM streaming access bandwidth varies with the number of tasklets.

We make four key observations. 

{First,} the {sustained MRAM} bandwidth of COPY-DMA is 
624.02 MB/s, which is close to the {theoretical maximum} bandwidth 
(700 MB/s derived in Section~\ref{sec:dpu-architecture}). 
The measured aggregate {sustained} bandwidth for 2,556 DPUs is 1.6~TB/s. 
In the 640-DPU system, we measure {the sustained MRAM bandwidth to be} 470.50~MB/s {per DPU} {(theoretical maximum = 534 MB/s), resulting in aggregate sustained MRAM bandwidth of 301~GB/s} for 640 DPUs.

{Second, the MRAM} bandwidth of COPY-DMA saturates with two tasklets. 
{Even though} the DMA engine can perform {only one data} transfer at a time~\cite{comm-upmem}, using two or more tasklets in COPY-DMA guarantees that there is always a DMA request enqueued to keep the DMA engine busy when a previous DMA request completes, {thereby} achieving {the highest MRAM} bandwidth.

{Third, the MRAM} bandwidth for COPY and ADD saturates at 4 and 6 tasklets, respectively, i.e., {earlier than} the 11 tasklets needed to fully utilize the pipeline.
This observation indicates that {these microbenchmarks are limited by access to MRAM (and not the instruction pipeline)}. 
When the COPY benchmark uses fewer than 4 tasklets, the latency of pipeline instructions (i.e., WRAM loads/stores) is longer than the latency of MRAM accesses (i.e., MRAM-WRAM and WRAM-MRAM DMA transfers). 
After 4 tasklets, this trend flips, and the latency of MRAM accesses becomes longer. The reason is that the MRAM accesses are serialized, such that the MRAM access latency increases linearly with the number of tasklets. 
Thus, after 4 tasklets, the overall latency is {dominated by} the MRAM access latency, which hides the pipeline latency. 
{As a result, the sustained MRAM bandwidth of COPY saturates with 4 tasklets at the highest MRAM bandwidth, same as COPY-DMA.}
Similar observations apply to the ADD benchmark with 6 tasklets.

{Fourth,} the {sustained MRAM} bandwidth of SCALE and TRIAD is {approximately one order of} magnitude smaller than that of COPY-DMA, COPY, and ADD.
{In addition}, {SCALE and TRIAD's MRAM} bandwidth saturates at 11 tasklets, i.e., the number of tasklets needed to fully utilize the pipeline. 
This observation indicates that {SCALE and TRIAD performance} {is} limited by pipeline throughput, not MRAM {access}.
Recall that SCALE and TRIAD use costly multiplications, which are based on the \texttt{mul\_step} {instruction}, as explained in Section~\ref{sec:arith-throughput}.
As a result, instruction execution in the pipeline has {much} higher latency than MRAM access.
Hence, it makes sense that {SCALE and TRIAD} are bound by pipeline throughput, and thus the {maximum sustained WRAM} bandwidth of SCALE and TRIAD (Figure~\ref{fig:wram-stream}) is the same as {the maximum sustained MRAM} bandwidth (Figure~\ref{fig:mram-stream}).

\vspace{-2mm}
\gboxbegin{\rtask{ko}}
\vspace{-1mm}
\begin{itemize}[wide, labelsep=0.5em]
\item \textbf{When the access latency {to a DRAM Processing Unit's Main memory (MRAM) bank for a streaming} benchmark (COPY-DMA, COPY, ADD) {is larger than} the pipeline latency} (i.e., execution {latency} of arithmetic operations and WRAM accesses), \textbf{the performance of the {DRAM Processing Unit (DPU)} saturates at a number of tasklets (i.e., software threads) {smaller} than 11. This is a memory-bound workload.} 
\item \textbf{When the pipeline latency for a {streaming} benchmark (SCALE, TRIAD) {is larger than the MRAM access latency}, the performance of a DPU saturates at 11 tasklets. This is a compute-bound workload.}
\end{itemize}
\vspace{-1mm}
\gboxend

\subsubsection{Sustained Strided and Random Access Bandwidth}\label{sec:mram-strided-random}
We evaluate the sustained MRAM bandwidth of strided and random access patterns.

{To evaluate strided access bandwidth in MRAM, we devise an experiment in which we write a new microbenchmark that accesses MRAM in a strided manner. 
The microbenchmark accesses an array at a constant stride (i.e., constant distance between consecutive memory accesses), copying elements from the array into another array using the same stride.
We implement two versions of the microbenchmark, \emph{coarse-grained DMA} and \emph{fine-grained DMA}, to test both coarse-grained and fine-grained MRAM access.}
{In coarse-grained DMA, the microbenchmark accesses via DMA} a large contiguous segment (1024~B) of the array in MRAM, and the strided access happens in WRAM. 
{The coarse-grained DMA} approach resembles what {modern} CPU hardware does (i.e., reads large cache lines {from main memory} and strides through them in the cache).
{In fine-grained DMA, the microbenchmark transfers via DMA} only the data {that will be used by the microbenchmark from} MRAM.
{The fine-grained DMA} approach results in more DMA requests, but less total {amount of} data transferred {between MRAM and WRAM}.

{To evaluate random access bandwidth in MRAM}, we implement the GUPS benchmark~\cite{luszczek_hpcc2006}, which performs read-modify-write operations on random positions of an array.
{We use only} fine-grained DMA for random access, since {random} memory accesses in GUPS do not benefit from fetching large chunks of data, {because they are \emph{not} spatially correlated}. 

In {our} experiments, we scale the number of tasklets from 1 to 16.
The tasklets collectively access arrays in MRAM with (1) coarse-grained strided access, (2) fine-grained strided access, or (3) fine-grained random access. 
{Each array contains} 2M 8-byte elements (total of 16MB), which are divided evenly across the tasklets. 

\begin{figure*}[h]
    \centering
    \includegraphics[width=1.0\linewidth]{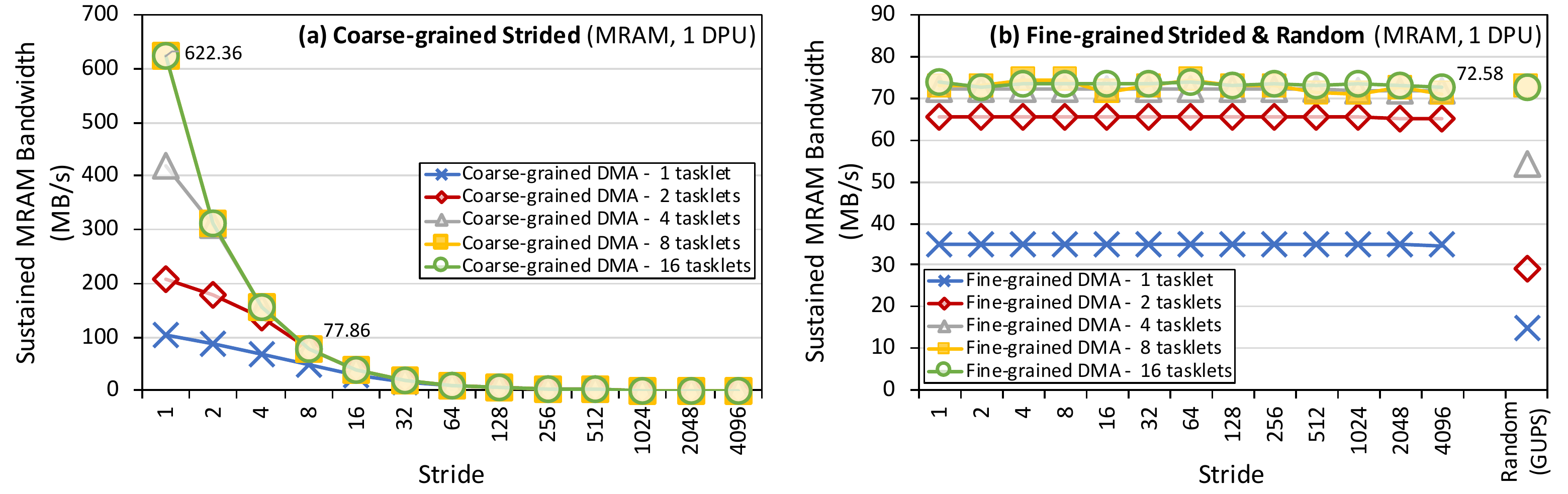}
    \caption{{Sustained} MRAM bandwidth for (a) coarse-grained strided and (b) fine-grained strided and random access patterns.}
    \label{fig:mram-patterns}
\end{figure*}

Figure~\ref{fig:mram-patterns} shows how the {sustained MRAM bandwidth varies with access pattern (strided and random access)} as well as with the number of tasklets.

{We make four key observations.}

{First, we measure maximum sustained MRAM bandwidth to be 
622.36 MB/s for coarse-grained DMA (with 16 tasklets and a stride of 1, Figure~\ref{fig:mram-patterns}a), and 
{72.58 MB/s} for fine-grained DMA (with 16 tasklets, Figure~\ref{fig:mram-patterns}b).} 
{This difference in the sustained MRAM bandwidth values of coarse-grained DMA and fine-grained DMA is related to the difference in MRAM bandwidth for different transfer sizes ({as we analyze in} Section~\ref{sec:mram-read-write}). 
While coarse-grained DMA uses 1,024-byte transfers, fine-grained DMA uses 8-byte transfers.}

{Second, we observe that the sustained MRAM bandwidth of coarse-grained DMA (Figure~\ref{fig:mram-patterns}a)} decreases as the stride {becomes larger}. 
{This is due to the effective utilization of the transferred data, which decreases for larger strides (e.g., a stride of 4 means that only one fourth of the transferred data is effectively used).}

Third, the coarse-grained DMA approach has higher {sustained MRAM} bandwidth for {smaller strides} while the fine-grained DMA approach has higher {sustained MRAM} bandwidth for larger strides.
The larger the stride in coarse-grained DMA, {the larger the amount of} fetched data {that} remains unused, causing fine-grained DMA to become more efficient with larger strides. 
In these experiments, 
the coarse-grained DMA approach achieves higher sustained MRAM bandwidth {than the fine-grained DMA approach} for strides between 1 and 8. For a stride of 16 or larger, the fine-grained DMA approach achieves higher sustained MRAM bandwidth.
This 
is because with larger strides, the fraction of {transferred} data that is actually used by the microbenchmark becomes smaller (i.e., effectively-used MRAM bandwidth becomes smaller).
With a stride of 16 {and coarse-grained DMA, the microbenchmark} uses only one sixteenth of the fetched data. 
As a result, we measure {the sustained MRAM bandwidth to be} 38.95 MB/s for coarse-grained DMA, which is only one sixteenth of the maximum sustained MRAM bandwidth of 622.36 MB/s, and is lower than the sustained MRAM {bandwidth of fine-grained DMA 
(72.58 MB/s)}.

Fourth, the maximum sustained MRAM bandwidth for random access is 72.58 MB/s (with 16 tasklets, {as shown in} Figure~\ref{fig:mram-patterns}b). This bandwidth value is {very} similar to the maximum MRAM bandwidth of the fine-grained DMA approach for strided access (e.g., 72.58 MB/s with 16 tasklets and stride 4,096, {as shown in} Figure~\ref{fig:mram-patterns}b), since {our microbenchmark uses} fine-grained DMA for random access.

Based on these observations, we recommend that programmers \textbf{use the coarse-grained DMA approach for workloads with small {strides} and the fine-grained DMA approach for workload with large {strides or} random access patterns}.

\pboxbegin{\ptask{pr}}
\begin{itemize}[wide, labelsep=0.5em]
\item For strided access patterns with a \textbf{stride smaller than 16 8-byte elements, fetch a large contiguous chunk} (e.g., 1,024 bytes) {from a DRAM Processing Unit's Main memory (MRAM) bank}. 
\item For {strided access patterns with} \textbf{larger strides and random access} patterns, fetch \textbf{only the data elements that are needed} {from an MRAM bank}.
\end{itemize}
\pboxend

\subsection{\textbf{Arithmetic Throughput versus Operational Intensity}}\label{sec:throughput-oi}
Due to its fine-grained multithreaded architecture~\cite{ddca.spring2020.fgmt,henessy.patterson.2012.fgmt,burtonsmith1978,smith1982architecture,thornton1970}, a DPU overlaps instruction execution {latency} in the pipeline and MRAM access {latency}~\cite{upmem-guide,devaux2019}. 
As a result, the overall DPU performance is determined by the dominant {latency} (either instruction execution {latency} or MRAM access {latency}). 
We observe this behavior in our experimental results in Section~\ref{sec:mram-streaming}, where the dominant latency (pipeline latency or MRAM access latency) determines the
sustained MRAM bandwidth for {different} versions of {the STREAM benchmark~\cite{mccalpin1995}}.

To further understand the DPU architecture, we design a new microbenchmark where we vary the {number} of pipeline instructions with respect to the {number} of MRAM accesses, and measure performance in terms of arithmetic throughput (in {MOPS}, as defined in Section~\ref{sec:wram-benchmark-desc}). 
{By varying the number of pipeline instructions per MRAM access, we move from microbenchmark configurations where the MRAM access latency dominates (i.e., \emph{memory-bound regions}) to microbenchmark configurations where the pipeline latency dominates (i.e., \emph{compute-bound regions}).}

Our microbenchmark includes MRAM-WRAM DMA transfers, WRAM load/store accesses, and a variable number of arithmetic operations. 
The number of MRAM-WRAM DMA transfers in the microbenchmark is constant, and thus the total MRAM latency is {also} constant. However, the latency of instructions executed in the pipeline varies with the variable number of arithmetic operations.

{Our} experiments aim to show how arithmetic throughput varies with operational intensity. 
We define \emph{operational intensity} as the number of arithmetic operations performed per byte accessed from MRAM {(OP/B)}. 
As explained in Section~\ref{sec:arith-throughput}, an arithmetic operation in the UPMEM PIM architecture takes multiple instructions to execute. 
The experiment is inspired by the roofline model~\cite{roofline}, a performance analysis methodology that shows the performance of a program (arithmetic instructions executed per second) as a function of the \emph{arithmetic intensity} (arithmetic instructions executed per byte accessed from memory) of the program, {as compared to} the peak performance of the machine (determined by the compute throughput and the L3 and DRAM memory bandwidth).

\begin{figure*}[h]
        \centering
        \includegraphics[width=1.0\linewidth]{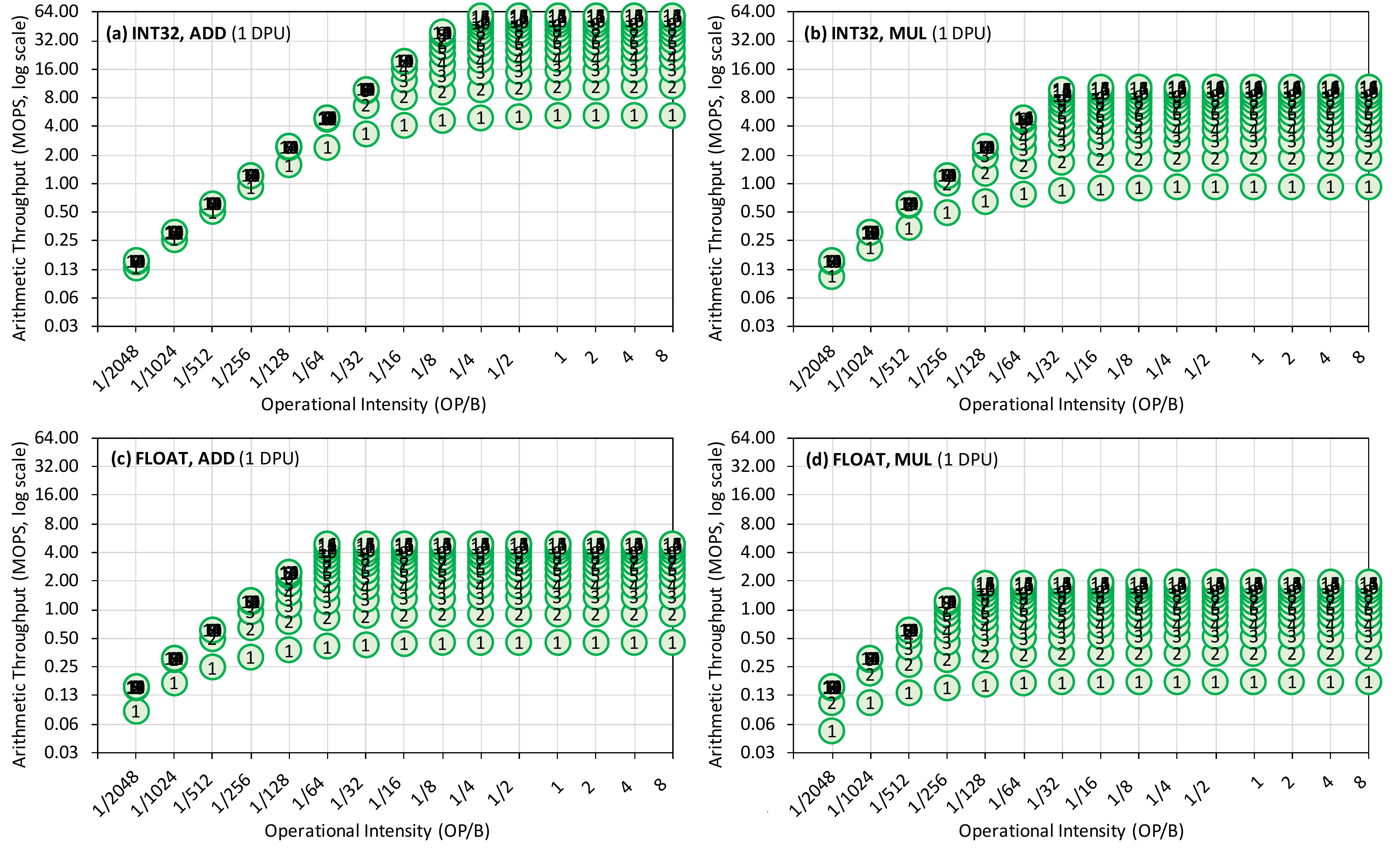}
        \captionof{figure}{{Arithmetic} throughput versus operational intensity for (a) 32-bit integer addition, (b) 32-bit integer multiplication, (c) 32-bit floating point addition, and (d) 32-bit floating point multiplication. The number inside each dot indicates the number of tasklets. 
        {Both x- and y-axes are log scale.}}
        \label{fig:ai-dpu}
\end{figure*}

Figure~\ref{fig:ai-dpu} shows results of arithmetic throughput versus operational intensity for representative data types and operations: (a) 32-bit integer addition, (b) 32-bit {integer} multiplication, (c) 32-bit floating point addition, and (d) 32-bit floating point multiplication. 
Results for other data types (64-bit integer and 64-bit floating point) and arithmetic operations (subtraction and division) follow similar trends.
We change the operational intensity from very low values 
($\frac{1}{2048}$ operations/byte, i.e., one operation {per} every 512 32-bit elements {fetched}) to high values (8 operations/byte, i.e., 32 operations per {every} 32-bit element {fetched}), {and} measure the resulting throughput for different numbers of tasklets (from 1 to 16).

We make four key observations from Figure~\ref{fig:ai-dpu}.

First, the four plots in Figure~\ref{fig:ai-dpu} show (1) {the} memory-bound region (where arithmetic throughput increases with operational intensity) and (2) {the} compute-bound region (where arithmetic throughput is flat at its maximum value) for each number of tasklets. For a {given} number of tasklets, the transition between the memory-bound {region} and the compute-bound region occurs when the latency of instruction execution in the pipeline surpasses the MRAM latency. We refer to the operational intensity value where the transition between {the} memory-bound region and {the} compute-bound region happens as the \emph{throughput saturation point}.

Second, arithmetic throughput saturates at low (e.g., 
$\frac{1}{4}$~OP/B for integer addition, i.e., 1 {integer} addition per {every} 32-bit element {fetched}) or very low (e.g., 
$\frac{1}{128}$~OP/B for floating point multiplication, i.e., 1 multiplication {per} every 32 32-bit elements {fetched}) operational intensity. 
This result demonstrates that \textbf{the DPU is fundamentally a compute-bound processor} designed for workloads with low data reuse.

\gboxbegin{\rtask{ko}}
\textbf{The arithmetic throughput of a DRAM Processing Unit (DPU) saturates at low or very low operational intensity} (e.g., 1 integer addition per 32-bit element). 
{Thus, \textbf{the DPU is fundamentally a compute-bound processor}.} 

We expect \textbf{most real-world workloads be compute-bound in the UPMEM PIM architecture}.
\gboxend

Third, the throughput saturation point is lower for data types and operations that require more instructions per operation. For example, the throughput for 32-bit multiplication (Figure~\ref{fig:ai-dpu}b), which requires up to 32 \texttt{mul\_step} instructions (Section~\ref{sec:arith-throughput}), saturates at $\frac{1}{32}$~OP/B, while the throughput for 32-bit addition (Figure~\ref{fig:ai-dpu}a), which is natively supported (it requires a single \texttt{add} instruction), saturates at $\frac{1}{4}$~OP/B. 
Floating point operations saturate earlier than integer operations, since they require from several tens to hundreds of instructions: 32-bit floating point addition (Figure~\ref{fig:ai-dpu}c) and multiplication (Figure~\ref{fig:ai-dpu}d) saturate at $\frac{1}{64}$ and $\frac{1}{128}$~OP/B, respectively.

Fourth, we observe that in the compute-bound regions (i.e., after the saturation points), {arithmetic} throughput saturates with 11 tasklets, which is the number of tasklets needed to fully utilize the pipeline.
On the other hand, in the memory-bound region, throughput saturates with fewer tasklets because the memory bandwidth limit is reached before the pipeline is fully utilized.
For example, at very low operational intensity values ($\leq$ 
$\frac{1}{64}$~OP/B), throughput of 32-bit integer addition saturates with just two tasklets which is consistent with the observation in Section~\ref{sec:mram-streaming} where COPY-DMA bandwidth saturates with two tasklets.
However, an operational intensity of 
$\frac{1}{64}$~OP/B is extremely low, as it entails only one addition for every 64~B accessed (16 32-bit integers).
We expect higher operational intensity (e.g., greater than $\frac{1}{4}$~OP/B) in most real-world workloads~\cite{roofline,deoliveira2021} and, thus, {arithmetic throughput to saturate with 11 tasklets in real-world workloads}.

In the Appendix (Section~\ref{app:throughput-oi}), we present a different view of these results, where we show how arithmetic throughput varies with the number of tasklets at different operational intensities.

\subsection{\textbf{CPU-DPU Communication}}\label{sec:cpu-dpu}
The host CPU and the DPUs in PIM-enabled memory communicate via the memory bus. 
The host CPU can access MRAM banks to (1) transfer input data from main memory to MRAM {(i.e., CPU-DPU)}, and (2) transfer results {back} from MRAM to main memory {(i.e., DPU-CPU)}, as Figure~\ref{fig:scheme} shows. We call these data transfers {CPU-DPU} and {DPU-CPU} transfers, respectively. 
As explained in Section~\ref{sec:sys-org}, these data transfers can be \emph{serial} (i.e., {performed} sequentially across multiple MRAM banks) or \emph{parallel} (i.e., {performed} concurrently across multiple MRAM banks). 
The UPMEM SDK~\cite{upmem-guide} provides functions for serial and parallel transfers. 
For serial transfers, \texttt{dpu\_copy\_to} copies a buffer from {the host} main memory to a specific MRAM bank {(i.e., CPU-DPU)}, and \texttt{dpu\_copy\_from} copies a buffer {from} one MRAM bank to {the host} main memory {(i.e., DPU-CPU)}.
For parallel transfers, a program needs to use two functions. 
First, \texttt{dpu\_prepare\_xfer} prepares the parallel transfer by assigning different buffers to specific MRAM banks. 
Second, \texttt{dpu\_push\_xfer} launches the actual transfers to execute in parallel. One argument of \texttt{dpu\_push\_xfer} defines whether the parallel data transfer happens from {the host} main memory to the MRAM banks (i.e., {CPU-DPU}) or from the MRAM banks to {the host} main memory (i.e., {DPU-CPU}). 
Parallel transfers have the limitation (in UPMEM SDK 2021.1.1~\cite{upmem-guide}) that the transfer sizes to all MRAM banks involved in the same parallel transfer need to be the same.
A special case of parallel {CPU-DPU} transfer (\texttt{dpu\_broadcast\_to}) broadcasts the same buffer from main memory to all MRAM banks.

In this section, we measure the sustained bandwidth of all types of {CPU-DPU} and {DPU-CPU} transfers between the {host} main memory and MRAM banks.
We perform two different experiments. 
The first experiment transfers {a} buffer of {varying} size to/from a single MRAM bank. Thus, we obtain the sustained bandwidth of {CPU-DPU} and {DPU-CPU} transfers of different {sizes} for one MRAM bank.
In this experiment, we use \texttt{dpu\_copy\_to} and \texttt{dpu\_copy\_from} and vary the transfer size from 8 bytes to 32 MB.
The second experiment transfers buffers of size 32 MB per MRAM bank from/to a set of 1 to 64 MRAM banks within the same rank.
We experiment with both serial and parallel transfers (\texttt{dpu\_push\_xfer}), including broadcast {CPU-DPU transfers} {(\texttt{dpu\_broadcast\_to})}. Thus, we obtain the sustained bandwidth of serial/parallel/broadcast {CPU-DPU} transfers and serial/parallel {DPU-CPU} transfers for a number of MRAM banks in the same rank between 1 and 64. 

\begin{figure*}[h]
    \centering
    \includegraphics[width=\linewidth]{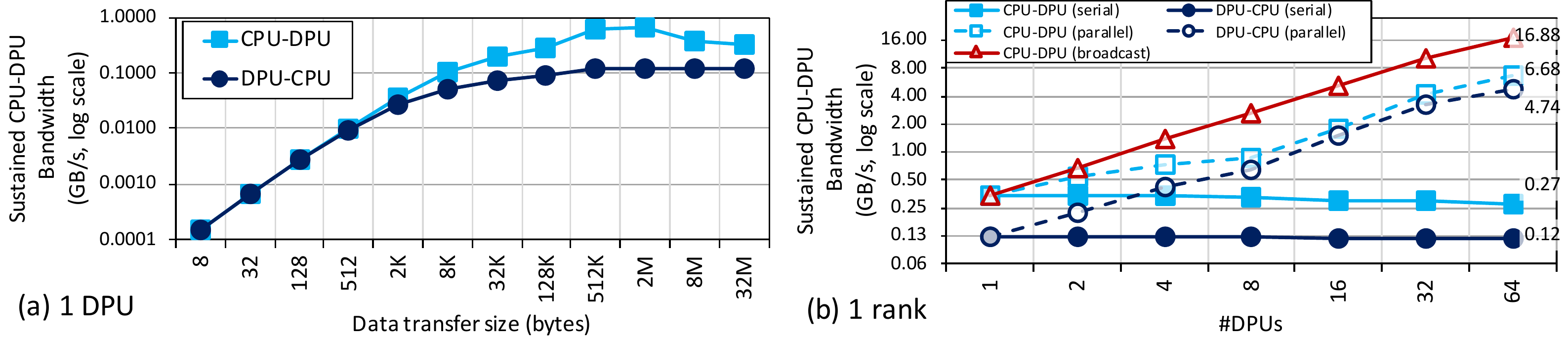}
    \vspace{1mm}
    \caption{Sustained bandwidth (log scale {x- and y-axes}) of (a) {CPU-DPU} ({host} main memory to one MRAM bank) and {DPU-CPU} (one MRAM bank to {host} main memory) transfers of different sizes for one DPU, and (b) serial/parallel/broadcast {CPU-DPU} ({host} main memory to several MRAM banks) and serial/parallel {DPU-CPU} (several MRAM banks to {host} main memory) transfers of 32 MB for a set of 1-64 DPUs within one rank.}
    \label{fig:cpudpu}
    \vspace{2mm}
\end{figure*}

Figure~\ref{fig:cpudpu} presents the sustained bandwidth results of both experiments.

We make seven key observations.\footnote{Note that our measurements of and observations about CPU-DPU and DPU-CPU transfers are {both} platform-dependent (i.e., measurements and observations may change for a different host CPU) and UPMEM SDK-dependent (i.e., the implementation of CPU-DPU/DPU-CPU transfers may change in future releases of the UPMEM SDK). For example, our bandwidth measurements on the 640-DPU system {(not shown)} differ from {those on} the 2,556-DPU system ({but we find the trends we observe to be similar on both systems}).} 

First, sustained bandwidths of {CPU-DPU} and {DPU-CPU} transfers for a single DPU (Figure~\ref{fig:cpudpu}a) are similar for transfer sizes between 8 and 512 bytes. 
For transfer sizes greater than 512 bytes, sustained bandwidth of {CPU-DPU} transfers is higher than that of {DPU-CPU} transfers. 
For the largest transfer size we {evaluate} (32 MB), {CPU-DPU} and {DPU-CPU} bandwidths are 
0.33 GB/s and 0.12 GB/s, respectively. 

Second, the sustained bandwidths of {CPU-DPU} and {DPU-CPU} transfers for a single DPU (Figure~\ref{fig:cpudpu}a) increase linearly between 8 bytes and 2 KB, and tend to saturate for larger {transfer} sizes. 

\vspace{2mm}
\gboxbegin{\rtask{ko}}
\textbf{Larger {CPU-DPU} and {DPU-CPU} transfers between the host main memory and the DRAM Processing Unit's Main memory (MRAM) banks result in higher sustained bandwidth.}
\gboxend
\vspace{2mm}

Third, for one rank (Figure~\ref{fig:cpudpu}b) the sustained bandwidths of serial {CPU-DPU} and {DPU-CPU} transfers remain flat for different numbers of DPUs.
Since these transfers are executed serially, latency increases proportionally with the number of DPUs (hence, the total amount of data transferred). As a result, the sustained bandwidth does not increase.

Fourth, the sustained bandwidth of the parallel transfers increases with the number of DPUs, reaching the highest sustained bandwidth {values at} 64 DPUs.
The maximum sustained {CPU-DPU} bandwidth that we measure is 
6.68 GB/s, while the maximum sustained {DPU-CPU} bandwidth is 
4.74 GB/s.
However, we observe that the increase in sustained bandwidth {with DPU count} is sublinear. The sustained {CPU-DPU} bandwidth for 64 DPUs is 20.13$\times$ higher than that for one DPU. For {DPU-CPU} transfers, the sustained bandwidth increase of 64 DPUs to one DPU is 38.76$\times$.

\gboxbegin{\rtask{ko}}
\textbf{The sustained bandwidth of parallel {CPU-DPU} and {DPU-CPU} transfers between the host main memory and the DRAM Processing Unit's Main memory (MRAM) banks increases with the number of DRAM Processing Units inside a rank.}
\gboxend

Fifth, we observe large differences between sustained bandwidths of {CPU-DPU} and {DPU-CPU} transfers for both serial and parallel transfers. 
These differences are due to different implementations of {CPU-DPU} and {DPU-CPU} transfers in UPMEM SDK 2021.1.1~\cite{comm-upmem}. 
While {CPU-DPU} transfers use x86 AVX write instructions~\cite{avx2011intel}, which are asynchronous, {DPU-CPU} transfers use AVX read instructions~\cite{avx2011intel}, which are synchronous. 
As a result, {DPU-CPU} transfers cannot sustain {as many memory accesses as {CPU-DPU} transfers, which results} in lower sustained bandwidths of both serial and parallel {DPU-CPU} transfers than the {CPU-DPU} transfer counterparts.

Sixth, sustained bandwidth of broadcast {CPU-DPU} transfers reaches up to 16.88 GB/s. 
One reason why this maximum sustained bandwidth is significantly higher {than} that of parallel {CPU-DPU} transfers is better locality in the cache hierarchy of the host CPU~\cite{comm-upmem}. 
While a broadcast transfer copies the \emph{same} buffer to \emph{all} MRAM banks, which increases temporal locality in the CPU cache hierarchy, a parallel {CPU-DPU} transfer copies \emph{different} buffers to \emph{different} MRAM banks. {These buffers are more likely to miss in the CPU cache hierarchy and need to be fetched from main memory into CPU caches before being copied to MRAM banks.} 

Seventh, in all our experiments across an entire rank, the sustained bandwidth is lower than the theoretical maximum bandwidth of DDR4-2400 DIMMs (19.2 GB/s)~\cite{jedec2012ddr4}. We attribute this bandwidth loss to the transposition library~\cite{devaux2019} that the UPMEM SDK uses to map entire 64-bit words onto the same MRAM bank (Section~\ref{sec:sys-org}).

\gboxbegin{\rtask{ko}}
\vspace{-2mm}
\textbf{The sustained bandwidth of parallel {CPU-DPU} transfers between the host main memory and the DRAM Processing Unit's Main memory (MRAM) banks is higher than the sustained bandwidth of parallel {DPU-CPU} transfers between the MRAM banks and the host main memory} due to different implementations of {CPU-DPU} and {DPU-CPU} transfers in the UPMEM runtime library.

\textbf{The sustained bandwidth of broadcast {CPU-DPU} transfers (i.e., the same buffer is copied to multiple MRAM banks) is higher than that of parallel {CPU-DPU} transfers (i.e., different buffers are copied to different MRAM banks)} due to higher temporal locality in the CPU cache hierarchy.
\vspace{-2mm}
\gboxend

\vspace{-4mm}
\section{PrIM Benchmarks}
\label{sec:benchmarks}
\vspace{-1mm}

We present the benchmarks included in 
our open-source \emph{PrIM} (\emph{\underline{Pr}ocessing-\underline{I}n-\underline{M}emory}) {\emph{benchmark suite}}, the first benchmark suite for a real PIM architecture. 
PrIM benchmarks are publicly {and freely} available~\cite{gomezluna2021repo}.

For each benchmark, we include {in this section} a description of its implementation on a UPMEM-based PIM system with multiple DPUs. 
Table~\ref{tab:benchmarks} shows a summary of the benchmarks. 
{We group benchmarks by the application domain they belong to. 
Within each application domain, we sort benchmarks by (1) incremental complexity of the PIM implementation (e.g., we explain VA before GEMV) and (2) alphabetical order. 
We use the order of the benchmarks in Table~\ref{tab:benchmarks} consistently throughout the rest of the paper.}
For each benchmark, the table includes (1) the benchmark's short name, which we use in the remainder of the paper, (2) memory access patterns of the benchmark (sequential, strided, random), (3) computation pattern (operations and data types), and (4) communication/synchronization type of the PIM implementation (intra-DPU, inter-DPU). For intra-DPU communication, the table specifies the synchronization primitives, such as barriers, handshakes, and mutexes, that the benchmark uses (Section~\ref{sec:language-library}).

\vspace{-2mm}
\begin{table*}[h]
        \begin{center}
        \captionof{table}{{PrIM} benchmarks.}
        \label{tab:benchmarks}
        \resizebox{1.0\linewidth}{!}{
        \begin{tabular}{|l|l|c||c|c|c|c|c|c|c|}
    \hline
    \multirow{2}{*}{\textbf{Domain}} & \multirow{2}{*}{\textbf{Benchmark}} & \multirow{2}{*}{\textbf{Short name}} & \multicolumn{3}{c|}{\textbf{Memory access pattern}} & \multicolumn{2}{c|}{\textbf{Computation pattern}} & \multicolumn{2}{c|}{\textbf{Communication/synchronization}}  \\
    \cline{4-10}
     & & & \textbf{Sequential} & \juancrr{\textbf{Strided}} & \textbf{Random} & \textbf{Operations} & \textbf{Datatype} & \textbf{Intra-DPU} & \textbf{Inter-DPU}  \\
    \hline
    \hline
    \multirow{2}{*}{Dense linear algebra}
      & Vector Addition & VA & Yes & & & add & int32\_t &  &  \\
    \cline{2-10}
      & Matrix-Vector Multiply & GEMV & Yes & & & add, mul & uint32\_t &  &  \\
    \hline
    \multirow{1}{*}{Sparse linear algebra}
      & Sparse Matrix-Vector Multiply & SpMV & Yes & & Yes & add, mul & float &  &  \\
    \hline
    \multirow{2}{*}{Databases}
      & Select & SEL & Yes & & & add, compare & int64\_t & handshake, barrier & Yes \\
    \cline{2-10}
      & Unique & UNI & Yes & & & add, compare & int64\_t & handshake, barrier & Yes \\
    \hline
    \multirow{2}{*}{Data analytics}
      & Binary Search & BS & \juanc{Yes} & & Yes & compare & int\ivan{64}\_t &  &  \\
    \cline{2-10}
      & Time Series Analysis & TS & Yes & & & add, sub, mul, div & int32\_t &  &  \\
    \hline
    \multirow{1}{*}{Graph processing}
      & Breadth-First Search & BFS & Yes & & Yes & bitwise logic & uint64\_t & barrier, mutex & Yes \\
    \hline
    \multirow{1}{*}{Neural networks}
      & Multilayer Perceptron & MLP & Yes & & & add, mul, compare & int32\_t &  &  \\
    \hline
    \multirow{1}{*}{Bioinformatics}
      & Needleman-Wunsch & NW & Yes & Yes & & add, sub, compare & int32\_t & barrier & Yes \\
    \hline
    \multirow{2}{*}{Image processing}
      & Image histogram (short) & \juancrr{HST-S} & Yes & & Yes & add & \jieee{uint32\_t} & barrier & Yes \\
    \cline{2-10}
      & Image histogram (long) & \juancrr{HST-L} & Yes & & Yes & add & \jieee{uint32\_t} & barrier, mutex & Yes \\
    \hline
    \multirow{4}{*}{Parallel primitives}
      & Reduction & RED & Yes & Yes & & add & int64\_t & barrier & Yes \\
    \cline{2-10}
      & Prefix sum (scan-scan-add) & SCAN-SSA & Yes & & & add & int64\_t & handshake, barrier & Yes \\
    \cline{2-10}
      & Prefix sum (reduce-scan-scan) & SCAN-RSS & Yes & & & add & int64\_t & handshake, barrier & Yes \\
    \cline{2-10}
      & Matrix transposition & TRNS & Yes & & Yes & add, sub, mul & int64\_t & mutex & \\
    \hline
\end{tabular}
        }
        \end{center}
\end{table*}

\vspace{2mm}
All implementations of PrIM benchmarks follow the general programming recommendations presented in Section~\ref{sec:general-recommendations}. 
Note that our goal is not {to provide} extremely optimized implementations, but implementations that follow the general programming recommendations and make good use of the resources in PIM-enabled memory with {reasonable programmer} effort. 
For several benchmarks, where we can design more than one implementation that is suitable to the UPMEM-based PIM system, we develop all alternative implementations and compare them. 
As a result, we provide two versions of two of the benchmarks, Image histogram (HST) and Prefix sum (SCAN). In the Appendix (Section~\ref{sec:appendix-results}), we compare these versions and find the cases (i.e., dataset characteristics) where each version of each of these benchmarks results in higher performance. 
We also design and develop three versions of Reduction (RED). However, we do not provide them as separate benchmarks, since one of the three versions {always} provides higher performance than {(or at least equal to)} the other two ({see Appendix,} Section~\ref{sec:appendix-results}).\footnote{We provide the three versions of RED as part of the same benchmark. Users can select the version they want to test via compiler flags.}

\begin{figure}[h]
    \centering
    \includegraphics[width=\linewidth]{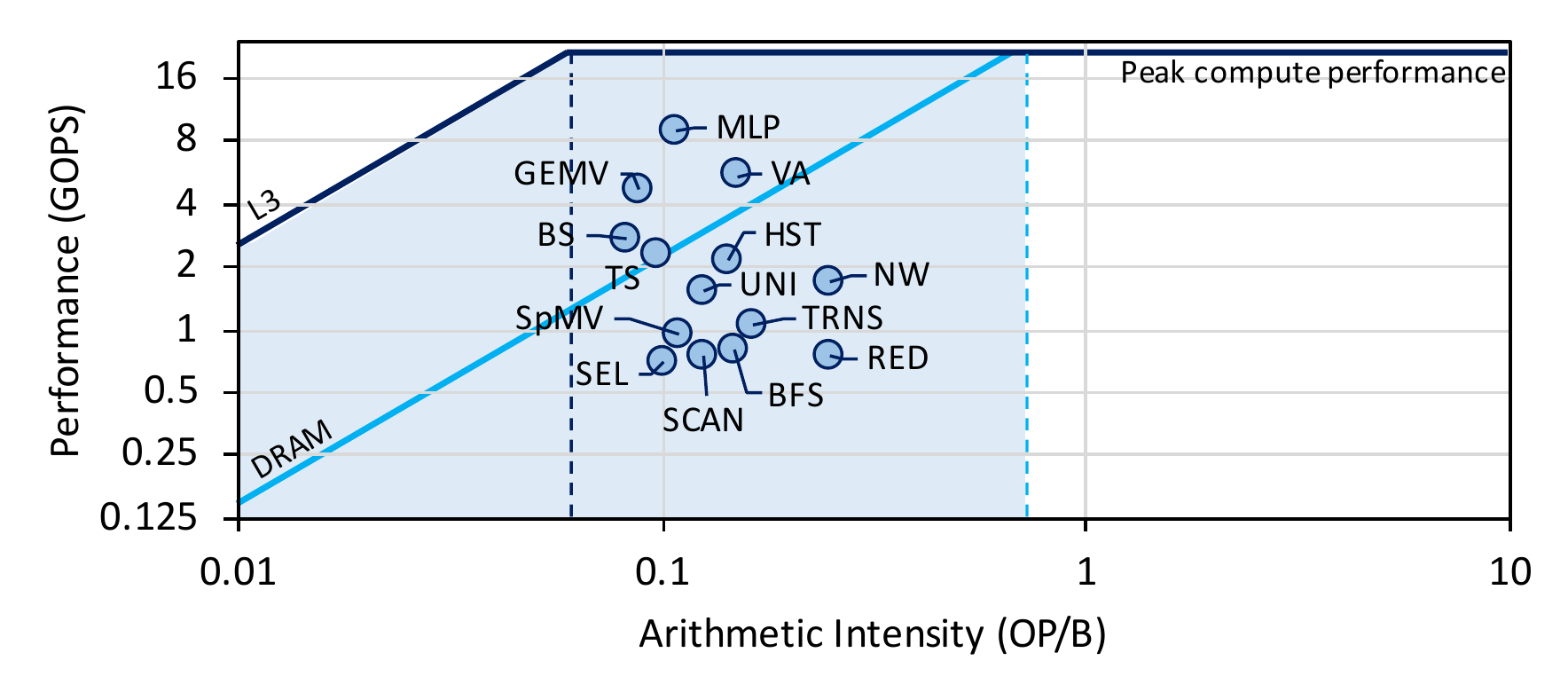}
    \vspace{-4mm}
    \caption{Roofline model for the CPU versions of the {14} PrIM workloads on an Intel Xeon E3-1225 v6 CPU.}
    \label{fig:roofline}
    \vspace{-3mm}
\end{figure}

Our benchmark selection is based on several criteria: (1) suitability for PIM, (2) domain diversity, and (3) diversity of memory access, computation, and communication/synchronization patterns, as shown in Table~\ref{tab:benchmarks}. 
We identify the suitability of these workloads for PIM by studying their memory boundedness. 
We employ the roofline model~\cite{roofline}, {as described} in Section~\ref{sec:throughput-oi}, to quantify the memory boundedness of the {CPU versions of the} workloads. Figure~\ref{fig:roofline} shows the roofline model on an Intel Xeon E3-1225 v6 CPU~\cite{xeon-e3-1225} with Intel Advisor~\cite{advisor}. 
{In} these experiments, we use the first dataset for each workload in {Table~\ref{tab:datasets} (see Section~\ref{sec:evaluation})}.

{We observe} from Figure~\ref{fig:roofline} that all {of the CPU versions of the PrIM} workloads are in the {memory-bounded} area of the roofline model (i.e., {the shaded region on the left side of the intersection between the DRAM bandwidth line and the peak compute performance line}). 
Hence, these workloads are all limited by memory. 
We conclude that {all 14 CPU versions of PrIM} workloads are potentially suitable for PIM~\cite{deoliveira2021}. 
{We briefly describe each PrIM benchmark and its PIM implementation next.}

\subsection{Vector Addition}
Vector Addition (VA)~\cite{blackford2002updated} takes two vectors $a$ and $b$ as inputs and performs their element-wise addition. 

Our PIM implementation divides the input vectors {$a$ and $b$} into as many equally-sized chunks as {the number of} DPUs in the system, {and makes a linear assignment (i.e., chunk $i$ assigned to DPU $i$).}
{The host CPU loads one chunk of both vectors $a$ and $b$ to the MRAM bank of each DPU.} 
Inside each DPU, {we assign blocks of elements from $a$ and $b$ to tasklets in a cyclic manner (i.e., block $j$ assigned to tasklet $j \% T$ for a total number $T$ of tasklets per DPU).}
{Each tasklet (1) moves the blocks of elements from $a$ and $b$ to the WRAM, (2) performs the element-wise addition, and (3) moves the results to the MRAM bank}. 
Tasklets iterate as many times as needed until the whole chunk assigned to a DPU is processed. 
{At the end of the execution on the DPUs, the CPU retrieves the output vector chunks from the MRAM banks to the host main memory and constructs the complete output vector.}

\subsection{Matrix-Vector Multiply}
\label{sec:gemv}
Matrix-Vector multiply (GEMV)~\cite{blackford2002updated} {is a dense linear algebra routine that} takes a matrix of size $m \times n$ and a vector of size $n \times 1$ as inputs and performs the multiplication between them, producing a new $m \times 1$ vector as a result.

Our PIM implementation of GEMV partitions the matrix across the DPUs available in the system, assigning a fixed {number of consecutive} rows to each {DPU}, while the input vector is replicated {across} all DPUs. 
{The host CPU assigns each set of consecutive rows to a DPU using linear assignment (i.e., set of rows $i$ assigned to DPU $i$).}
Inside each DPU, tasklets are in charge of computing {on the set} of the rows assigned to that DPU. 
{We assign a subset of consecutive rows from the set {of rows assigned to a DPU} to each tasklet (i.e., subset of rows $j$ assigned to tasklet $j$).}
First, each tasklet reads a block of elements, both from {one row of the input matrix and from the vector, and places these elements} in the WRAM. 
Second, {each tasklet performs} multiply and accumulation of those elements, and {it jumps} to the first step until {it reaches} the end of the row. 
Third, {each tasklet stores} the sums of multiplications in MRAM.
{Fourth, each tasklet repeats these three steps as many times as there are rows in its subset.} 
{Fifth, each DPU produces a contiguous chunk of elements of the output vector. The CPU retrieves the output vector chunks and builds the complete output vector.}

\vspace{-1mm}
\subsection{Sparse Matrix-Vector Multiply}
Sparse Matrix-Vector multiply (SpMV)~\cite{saad2003iterative.parallel} is a {sparse} linear algebra routine where a sparse matrix is multiplied by a dense vector.

Our {PIM} implementation of SpMV uses the Compressed Sparse Row (CSR) storage format~\cite{saad2003iterative.sparsematrices,liu2015csr5,kanellopoulos2019smash} to represent the matrix.
{First, the host CPU distributes the rows of the matrix evenly across DPUs, using linear assignment (i.e., set of rows $i$ assigned to DPU $i$) as in GEMV (Section~\ref{sec:gemv})}. 
{Within} each DPU, {the rows of the matrix are distributed evenly across tasklets (i.e., subset of rows $j$ assigned to tasklet $j$, same as in GEMV).}
The input vector is replicated across DPUs.
Each tasklet multiplies its {subset of} rows with the input vector and produces a contiguous chunk of the output vector.
{At the end of the execution on the DPUs,} the CPU copies back the output vector chunks {from the MRAM banks to the host main memory}, in order to construct the entire output vector.

\vspace{-1mm}
\subsection{Select}
Select (SEL)~\cite{ceri1985translating} is a database operator that, given an input array, filters the array elements according to a {given input} predicate. 
Our version of SEL removes the elements that satisfy the predicate, and keeps the elements that do not. 

Our PIM implementation of SEL partitions the array across the DPUs available in the system. 
The tasklets inside a DPU coordinate using handshakes {(Section~\ref{sec:language-library})}. 
First, each tasklet moves a block of elements to WRAM. 
Second, {each tasklet filters the elements and counts} the number of filtered elements. 
Third, each tasklet passes its number of filtered elements to the next tasklet using handshake-based communication, which inherently performs a prefix-sum operation~\cite{blelloch1989, gomezluna2015ds, yan2013streamscan} to determine where in MRAM to store the filtered elements. 
The tasklet {then} moves its filtered elements to MRAM. 
Fourth, the host CPU performs the final merge of the filtered arrays returned by each DPU {via serial {DPU-CPU} transfers, since parallel {DPU-CPU} transfers are not feasible because each DPU may return a different number of filtered elements}.

\vspace{-1mm}
\subsection{Unique}
Unique (UNI)~\cite{ceri1985translating} is a database operator that, for each group of consecutive array elements with the same value, removes all but the first of these elements. 

Our PIM implementation {of} UNI follows a similar approach to SEL. 
The main difference lies {in} the more complex handshake-based communication that UNI needs. 
Besides the number of unique elements, each tasklet has to pass the value of its last unique element to the next tasklet. 
This way, the next tasklet can check whether its first element is unique or not in the context of the entire array.

\vspace{-1mm}
\subsection{Binary Search}
Binary Search (BS)~\cite{knuth1971optimum} takes {a sorted array} as input and finds the position of some {query} values within {the sorted array}.

Our PIM implementation of binary search distributes the sorted array across the DPUs. Inside each DPU, tasklets are in charge of a subpartition of the assigned {query} values. First, each tasklet checks the assigned set of {query} values to find, moving them from {the MRAM bank} to WRAM and iterating over them using a for loop. Second, each tasklet {performs} the binary search algorithm, moving from left to right or vice-versa, depending on the current value to find. 
{Third, the tasklet stops the algorithm when it finds one query value. 
Fourth, at the end of the execution on the DPUs, the host CPU retrieves the positions of the found query values.}

\vspace{-1mm}
\subsection{Time Series Analysis}
Time Series analysis (TS)~\cite{yeh2016matrix} {aims} to find anomalies and similarities between subsequences of a given time series. 
Our version of time series analysis is based on Matrix Profile~\cite{zhu2018matrix}, an algorithm that works in a streaming-like manner, where subsequences (or query sequences) coming from a source of data are compared to a well-known time series that {has} the expected behavior.

Our PIM implementation of time series analysis divides the time series across the DPUs, adding the necessary overlapping between them, and replicating the query sequence {across the tasklets to compare to the time series}. Different slices of the time series are assigned to different tasklets. 
First, {each tasklet performs the dot product of its} slice of the time series and the query sequence.
Second, {each tasklet calculates} the similarity between the slice of the time series and the query sequence by computing the z-normalized Euclidean distance~\cite{zhu2018matrix}. 
Third, {each tasklet compares} the calculated similarity to the minimum similarity (or maximum, depending on the application) found so far, and {updates} it if the calculated similarity is a new minimum (or maximum). 
{Fourth, at the end of the execution on the DPUs, the host CPU retrieves the minimum (or maximum) similarity values and their positions from all DPUs, and finds the overall minimum (or maximum) and its position.}

\vspace{-2mm}
\subsection{Breadth-First Search}
Breadth-First Search (BFS)~\cite{bundy1984breadth} is a graph algorithm that labels each node in the graph with its distance from a given source node.
In our version, all edges have the same weight, therefore the distance represents the number of edges.

Our PIM implementation of BFS uses a {Compressed Sparse Row (CSR)}~\cite{saad2003iterative.sparsematrices,liu2015csr5,kanellopoulos2019smash} representation of the \emph{adjacency matrix}, {which represents the graph. Each element $(i, j)$ of the adjacency matrix indicates whether vertices $i$ and $j$ are connected by an edge.} 
Vertices are distributed evenly across DPUs, with each DPU receiving the \emph{neighbor lists} for the vertices that it owns. {The neighbor list of vertex $i$ contains the vertex IDs of the vertices that are connected to vertex $i$ by an edge.}
Each DPU maintains its own local copy of the list of visited vertices in the graph, which is represented as a bit-vector. 
{At the end of each iteration of the BFS algorithm, the host CPU merges all local per-DPU copies of the list of visited vertices.}
The whole list of visited vertices is called the \emph{frontier}.

{At the beginning of each iteration, the host CPU broadcasts} the complete current frontier to all the DPUs.
Each DPU uses the current frontier to update its local copy of the visited list. The DPU keeps the vertices of the current frontier that correspond to the vertices that it owns and discards the rest. 
The tasklets in the DPU (1) go through these vertices concurrently, (2) visit their neighbors, and (3) add the neighbors to the next frontier if they have not previously been visited. {This approach to BFS is called \emph{top-down} approach~\cite{hwukirk2016.bfs,luo2010effective}.} 
{Tasklets use critical sections (implemented via mutexes)} to update the next frontier concurrently without {data conflicts}.
{At the end of each iteration, the CPU retrieves} the next frontier produced by each DPU, and computes their union to {construct} the complete next frontier.
The iterations continue until the next frontier is empty {at the end of an iteration}.

\vspace{-1mm}
\subsection{Multilayer Perceptron}
Multilayer perceptron (MLP)~\cite{hinton1987learning} is a class of feedforward artificial neural network with at least three layers: input, hidden and output. 

Our PIM implementation of MLP performs MLP inference. 
In each layer, the weights are arranged as a matrix and the input is a vector. 
The computation in each layer is a matrix-vector multiplication. The implementation of each layer is based on our implementation of GEMV (Section~\ref{sec:gemv}). Thus, in each layer of MLP, the distribution of the workload among DPUs and tasklets is the same as in GEMV. 
ReLU is the activation function at the end of each layer.
\jieee{When a layer terminates, the host CPU (1) retrieves the output vector chunks from the MRAM banks, (2) constructs the complete vector, (3) feeds this vector to the DPUs as the input of the next layer, and (4) transfers the weights matrix of the next layer to the DPUs.
At the end of the output layer, the host CPU retrieves the output vector chunks, and constructs the final output vector.}

\vspace{-1mm}
\subsection{Needleman-Wunsch}
\label{sec:nw}
Needleman-Wunsch (NW)~\cite{Needleman1970Ageneral} is a bioinformatics algorithm that performs global sequence alignment, i.e., it compares two biological sequences over their \emph{entire length} to find out the optimal alignment of these sequences. NW is a dynamic programming algorithm that consists of three steps: (i) initialize a 2D score matrix $m \times n$, where $m$, $n$ are the lengths of the sequences (i.e., the number of \emph{base pairs}, \emph{bps}, in each sequence); (ii) fill the score matrix by calculating the score for each cell {in the matrix}, which is the maximum of the scores of the neighboring cells (left, top, or top-left cells) plus a penalty in case of a mismatch; and (iii) trace back the optimal alignment by marking a path from the cell on the bottom right back to the cell on the top left of the score matrix. Note that there may be {more than one} possible optimal alignments between two sequences.

Our PIM implementation first fills the upper triangle (top-left part) of the 2D score matrix, and then the lower triangle (bottom-right part) of it. The matrix is partitioned into large 2D {blocks}, and the algorithm iterates over the diagonals at a large {block} granularity (from the top-left diagonal to the bottom-right diagonal). In each iteration, all large {blocks} that belong to the same diagonal of the 2D score matrix are calculated in parallel by evenly distributing them across DPUs. Inside the DPU, each large 2D {block} is further partitioned into small 2D sub-blocks. The tasklets of each DPU work on the diagonals at a small sub-block granularity, i.e., in each iteration {the tasklets of a DPU concurrently} calculate different small sub-blocks that belong to the same large block of one diagonal.

For each diagonal of the 2D score matrix, the host CPU retrieves the large {blocks} produced by all DPUs. Then, it uses the filled cells of {the} last row and the last column of each large {block} as input to the next iteration (i.e., the next diagonal), since only these cells are neighboring cells with the next diagonal {blocks}. The iterations continue until all diagonal large {blocks} of the whole 2D score matrix are calculated. The host CPU finally uses the resulting 2D score matrix to trace back the optimal alignment.

In the Appendix (Section~\ref{app:nw}), we show additional experimental results for NW to {demonstrate that the computation of the complete problem and the computation of the longest diagonal scale differently across one rank of DPUs.}

\vspace{-1mm}
\subsection{Image Histogram}
\label{sec:histogram}
Image histogram (HST)~\cite{gomez2013optimized} calculates the histogram of an image, i.e., it counts the number of occurrences of each possible pixel value in an input image and stores the {aggregated counts of occurrences} into a set of bins.

We develop two PIM implementations of image histogram: {short (HST-S) and long (HST-L).}

HST-S distributes chunks of the input image across tasklets running on a DPU. Each tasklet creates a local histogram in WRAM. When the local histograms are created, the tasklets synchronize with a barrier, and the local histograms are merged in a parallel manner. 
Since each tasklet features a local histogram in WRAM, the maximum histogram size is relatively small (e.g., 256 32-bit bins, {when} running 16 tasklets).\footnote{{256 32-bit bins is the maximum histogram size for 16 tasklets (1) assuming power-of-two size of the histogram and (2) taking into account that each tasklet allocates a WRAM buffer for its chunk of the input image.}} 

HST-L can generate larger histograms, the size of which is limited {only} by the total amount of WRAM, since only one local histogram per DPU is stored in WRAM. 
{Same as HST-S, HST-L distributes chunks of the input image across tasklets, which update the histogram in WRAM by using a mutex, in order to ensure that only a single tasklet updates the histogram at a time.}

{Both HST-S and HST-L merge all per-DPU histograms into a single final histogram in the host CPU.}

We compare HST-S and HST-L for different histogram sizes in the Appendix (Section~\ref{app:histogram}), in order to find out which HST version is preferred {on the UPMEM-based PIM system} for each histogram size.

\vspace{-1mm}
\subsection{Reduction}
\label{sec:reduction}
Reduction (RED)~\cite{rabenseifner2004optimization} computes the sum of the elements in an input array.

Our PIM implementation of reduction has two steps. In the first step, each tasklet inside a DPU is assigned a chunk of the array. The tasklet accumulates all values of the chunk and produces a local reduction result. 
In the second step, 
after a barrier, a single tasklet reduces the partial results of all tasklets from the first step. 
{At the end of the second step, the host CPU retrieves the reduction result.}

{Alternatively, we can implement the second step as a parallel tree reduction~\cite{harris2007optimizing,degonzalo2019automatic}. We implement two versions of this parallel tree reduction, which use different intra-DPU synchronization primitives. 
One of the versions uses handshakes for communication between tasklets from one level of the tree to the next one. 
The other version uses barriers between levels of the tree.
In the Appendix (Section~\ref{app:reduction}), we compare the single-tasklet implementation to the two versions of parallel tree reduction.}

\vspace{-1mm}
\subsection{Prefix Sum (Scan)}
\label{sec:scan}
Prefix sum or scan (SCAN)~\cite{blelloch1989} is a parallel primitive that computes the prefix sum of the values in an array. We implement an exclusive scan: the $i$-th element of the output contains the sum of all elements of the input array from the first element to the ($i$-1)-th element.

We implement two PIM versions of scan: Scan-Scan-Add (SCAN-SSA)~\cite{yan2013streamscan, hwukirk2016.scan, sengupta2008efficient} and Reduce-Scan-Scan (SCAN-RSS)~\cite{yan2013streamscan, hwukirk2016.scan, dotsenko2008fast}.
Both versions assign a large chunk of the input array to each DPU. 

SCAN-SSA has three steps. First, it computes the scan operation locally inside each DPU. Second, {it copies} the last element of the local scan to the host CPU, and {places it} in a vector in the position corresponding to the DPU order. The host {CPU} scans this vector and {moves} each result value to the corresponding DPU. Third, {it adds} the value computed in the host {CPU} to all elements of the local scan output in each DPU. 
{Fourth, the host CPU retrieves the complete scanned array from the MRAM banks.}

SCAN-RSS also has three steps. First, it computes the reduction operation in each DPU. Second, {it copies} the reduction results to the host CPU, where {the host CPU scans them}. Third, {it moves} the result values of the scan operation in the host {CPU} to the corresponding DPUs, where {the tasklets perform} a local scan (including the value coming from the host {CPU}).
{Fourth, the host CPU retrieves the complete scanned array from the MRAM banks.}

The advantage of SCAN-RSS over SCAN-SSA is that {SCAN-RSS} requires fewer accesses to MRAM.
For an array of size $N$, SCAN-RSS needs $3 \times N$ + 1 accesses: 
$N$ reads and 1 write for Reduce, and $N$ reads and $N$ writes for Scan.
SCAN-SSA needs $4 \times N$ accesses: $N$ reads and $N$ writes for Scan, and $N$ reads and $N$ writes for Add.
The advantage of SCAN-SSA over SCAN-RSS is that it requires less synchronization.
The reduction operation in SCAN-RSS requires 
a barrier, but the addition operation in SCAN-SSA does not require any synchronization.
We expect SCAN-RSS {to perform} better for large arrays where access to MRAM dominates {the execution time}, and SCAN-SSA to perform better for smaller arrays where the 
reduction that requires synchronization constitutes a larger fraction of the {entire} computation.
We compare both implementations of SCAN for arrays of different sizes in Appendix Section~\ref{app:scan}.

\vspace{-1mm}
\subsection{Matrix Transposition}
\label{sec:trns}
\vspace{-1mm}
Matrix transposition (TRNS)~\cite{cayley1858ii} converts an $M \times N$ array into an $N \times M$ array. 
We focus on in-place transposition, where the transposed array occupies the same physical storage locations as the original array. 
In-place transposition is a permutation, which can be factored into disjoint cycles~\cite{hungerford1997abstract}. 
A straightforward parallel implementation can assign {entire} cycles to threads. However, {in such a straightforward implementation}, (1) the length of cycles is \emph{not} uniform in rectangular matrices, causing load imbalance, and (2) the memory accesses are random {as operations are done on} single matrix elements ({without} exploiting spatial locality). Thus, efficient parallelization is challenging.

Our PIM implementation follows an efficient 3-step tiled approach~\cite{sung2014matrix, gomez2016matrix} that (1) exploits spatial locality by {operating on} tiles of matrix elements, as opposed to single elements, and (2) balances the workload by partitioning the cycles across tasklets. 
To perform the three steps, we first factorize the dimensions of the $M \times N$ array as an $M' \times m \times N' \times n$ array, where $M = M' \times m$ and $N = N'\times n$.

The first step {operates on} tiles of size $n$. This step performs the transposition of an $M \times N'$ array, where each element is a tile of size $n$. The resulting array has dimensions $N' \times M \times n = N' \times M' \times m \times n$.
In the UPMEM-based PIM system, we perform this step using $n$-sized {CPU-DPU} transfers that copy the input array from the main memory of the host CPU to the corresponding MRAM banks. 

The second step performs $N' \times M'$ transpositions of $m \times n$ tiles. 
In each DPU, one tasklet transposes an $m \times n$ tile in WRAM. The resulting array has dimensions $N' \times M' \times n \times m$. 

The third step {operates on} tiles of size $m$. This step performs transpositions of $N'$ arrays of dimensions $M' \times n$, where each element is a tile of size $m$. The resulting array has dimensions $N' \times n \times M' \times m$. 
In each DPU, the available tasklets collaborate on the transposition of an $M' \times n$ array (with $m$-sized elements) using the algorithm presented in~\cite{sung2012dl}. {Differently from the algorithm in~\cite{sung2012dl}, which uses atomic instructions for synchronization}~\cite{gomez2013atomics}, our PIM implementation uses mutexes for synchronization of tasklets via an array of flags that keeps track of the moved tiles ({we choose this implementation because} the UPMEM ISA~\cite{upmem-guide} does \emph{not} include atomic instructions).

{After the three steps, the host CPU retrieves the transposed matrix from the MRAM banks.}

\vspace{-3mm}
\section{Evaluation of PrIM Benchmarks}
\label{sec:evaluation}

In this section, we evaluate the 16 PrIM benchmarks on 
the 2,556-DPU system (Section~\ref{sec:sys-org}), unless otherwise stated.
{Our evaluation uses} the datasets presented in Table~\ref{tab:datasets}, which are publicly {and freely} available~\cite{gomezluna2021repo}. 
Since these datasets are large and do not fit in WRAM, we need to use MRAM-WRAM DMA transfers repeatedly.
The results we present are for the best performing transfer sizes, which we include in Table~\ref{tab:datasets} {to facilitate the} reproducibility of our results. 
{We provide the command lines we use to execute each benchmark along with all parameters in~\cite{gomezluna2021repo}.}

\begin{table*}[h]
     \astretch{1.2}
            \begin{center}
                \captionof{table}{{Evaluated} Datasets.}
                \label{tab:datasets}
                \centering
                \resizebox{1.0\linewidth}{!}{
                    \begin{tabular}{|l||l|l|l|}
    \hline
     \multirow{2}{*}{\textbf{Benchmark}} & \multirow{2}{*}{\textbf{Strong Scaling Dataset}} & \multirow{2}{*}{\textbf{Weak Scaling Dataset}} & \multirow{2}{*}{\shortstack{\textbf{MRAM-WRAM} \\ \textbf{Transfer Sizes}}} \\
                       &                                 &                               &      \\
    \hline
    \hline
    VA & 1 DPU-1 rank: \jieee{2.5M elem. (10 MB)} | 32 ranks: 160M elem. (640 MB) & \jieee{2.5M elem./DPU (10 MB)} & 1024 bytes \\ \hline
    GEMV & 1 DPU-1 rank: $8192\times 1024$ elem. (32 MB) | 32 ranks: $163840\times 4096$ elem. (2.56 GB) & $1024\times 2048$ elem./DPU (8 MB) & 1024 bytes \\ \hline    
    SpMV & \emph{bcsstk30}~\cite{matrixmarket} (12 MB) & \emph{bcsstk30}~\cite{matrixmarket} & 64 bytes \\ \hline
    SEL & 1 DPU-1 rank: 3.8M elem. (30 MB) | 32 ranks: 240M elem. (1.9 GB) & 3.8M elem./DPU (30 MB) & 1024 bytes \\ \hline
    UNI & 1 DPU-1 rank: 3.8M elem. (30 MB) | 32 ranks: 240M elem. (1.9 GB) & 3.8M elem./DPU (30 MB) & 1024 bytes \\ \hline
    BS & 2M elem. (16 MB). 1 DPU-1 rank: 256K queries. (2 MB) | 32 ranks: 16M queries. (128 MB) & 2M elem. (16 MB). 256K queries./DPU (2 MB). & 8 bytes \\ \hline
    TS & 256 elem. query. 1 DPU-1 rank: 512K elem. (2 MB) | 32 ranks: 32M elem. (128 MB) & 512K elem./DPU (2 MB) & 256 bytes \\ \hline
    BFS & \emph{loc-gowalla}~\cite{graphgowalla} (22 MB) & \emph{rMat}~\cite{rmat} ($\approx$100K vertices and $1.2M$ edges per DPU) & 8 bytes \\ \hline
    MLP & \jieee{3} fully-connected layers. 1 DPU-1 rank: 2K neurons (32 MB) | 32 ranks: $\approx$160K neur. (2.56 GB) & \jieee{3} fully-connected layers. 1K neur./DPU (\jieee{4} MB) & 1024 bytes \\ \hline
    NW & 1 DPU-1 rank: 2560 bps (50 MB), large/small sub-block=$\frac{2560}{\#DPUs}$/2 | 32 ranks: 64K bps (32 GB), l./s.=32/2 & 512 bps/DPU (2MB), l./s.=512/2 & 8, 16, 32, 40 bytes \\ \hline
    \juancrr{HST-S} & 1 DPU-1 rank: $1536\times1024$ input image~\cite{vanhateren1998} (6 MB) | 32 ranks: 64 $\times$ input image & $1536\times1024$ input image~\cite{vanhateren1998}/DPU (6 MB) & 1024 bytes \\ \hline
    \juancrr{HST-L} & 1 DPU-1 rank: $1536\times1024$ input image~\cite{vanhateren1998} (6 MB) | 32 ranks: 64 $\times$ input image & $1536\times1024$ input image~\cite{vanhateren1998}/DPU (6 MB) & 1024 bytes \\ \hline
    RED & 1 DPU-1 rank: 6.3M elem. (50 MB) | 32 ranks: 400M elem. (3.1 GB) & 6.3M elem./DPU (50 MB) & 1024 bytes \\ \hline
    SCAN-SSA & 1 DPU-1 rank: 3.8M elem. (30 MB) | 32 ranks: 240M elem. (1.9 GB) & 3.8M elem./DPU (30 MB) & 1024 bytes \\ \hline
    SCAN-RSS & 1 DPU-1 rank: 3.8M elem. (30 MB) | 32 ranks: 240M elem. (1.9 GB) & 3.8M elem./DPU (30 MB) & 1024 bytes \\ \hline
    TRNS & 1 DPU-1 rank: $12288 \times 16 \times \jieee{64} \times 8$ (\jieee{768} MB) | 32 ranks: $12288 \times 16 \times \jieee{2048} \times 8$ (\jieee{24 GB}) & $12288 \times 16 \times 1 \times 8$/DPU (\jieee{12} MB) & 128, 1024 bytes \\    
    \hline
\end{tabular}

                }
            \end{center}

\end{table*}

First, we present performance and scaling results.  
We evaluate strong scaling$^3$ for the 16 {PrIM} benchmarks {(Section~\ref{sec:strong})} on the 2,556-DPU system by running the experiments on (1) 1 DPU, (2) 1 rank (from 1 to 64 DPUs), and 
(3) 32 ranks (from 256 to 2,048 DPUs). 
The {goal} of these experiments is to evaluate how the performance of the UPMEM-based PIM system scales with the number of DPUs for a fixed problem size. 
The ideal strong scaling is linear scaling, i.e., the {ideal speedup for strong scaling with $N$ DPUs} over the execution on a single DPU {should be $N$}.

We also evaluate weak scaling$^4$ for the 16 {PrIM} benchmarks {(Section~\ref{sec:weak})} on 1 rank (from 1 to 64 DPUs). 
In this experiment, we evaluate how the performance of the UPMEM-based PIM system scales with the number of DPUs for a fixed problem size per DPU. 
In an ideal weak scaling scenario, the execution time remains constant for any number of DPUs.

Second, we compare the performance and energy consumption of two full-blown UPMEM-based PIM 
systems (Table~\ref{tab:pim-setups}) with 2,556 DPUs (newer system) and with 640 DPUs (older system) to those of a {modern} Intel Xeon E3-1240 CPU~\cite{xeon-e3-1225} and a {modern} NVIDIA Titan V GPU~\cite{titanv} {(Section~\ref{sec:comparison})}. 

{In Section~\ref{sec:discussion},} we provide new insights about suitability of different workloads to the PIM system, programming recommendations for software designers, and suggestions and hints for hardware and architecture designers of future PIM systems.

\vspace{-3mm}
\subsection{Performance and Scaling Results}
\label{sec:performance}


We evaluate the performance of all the benchmarks with strong and weak scaling experiments using the datasets in Table~\ref{tab:datasets}.
Section~\ref{sec:strong} presents strong scaling results for {a single DPU, a single rank (from 1 to 64 DPUs), and for sets of 4 to 32 ranks (from 256 to 2,048 DPUs)}. We also evaluate the cost of inter-DPU synchronization.
In Section~\ref{sec:weak}, we analyze weak scaling on an entire rank for 1 to 64 DPUs. We include in the analysis the cost of inter-DPU synchronization {via the host CPU, {as well as} {CPU-DPU} and {DPU-CPU} latencies.}


\subsubsection{\textbf{Strong Scaling Results}}
\label{sec:strong}
We evaluate strong scaling with three different configurations: (1) 1-16 tasklets inside one DPU, (2) 1-64 DPUs inside one rank, and (3) 4-32 ranks. For the experiments inside one rank and multiple ranks, we use the best-performing number of tasklets {from the experiment on one DPU.}

\begin{figure*}[h]
\includegraphics[width=1.0\linewidth]{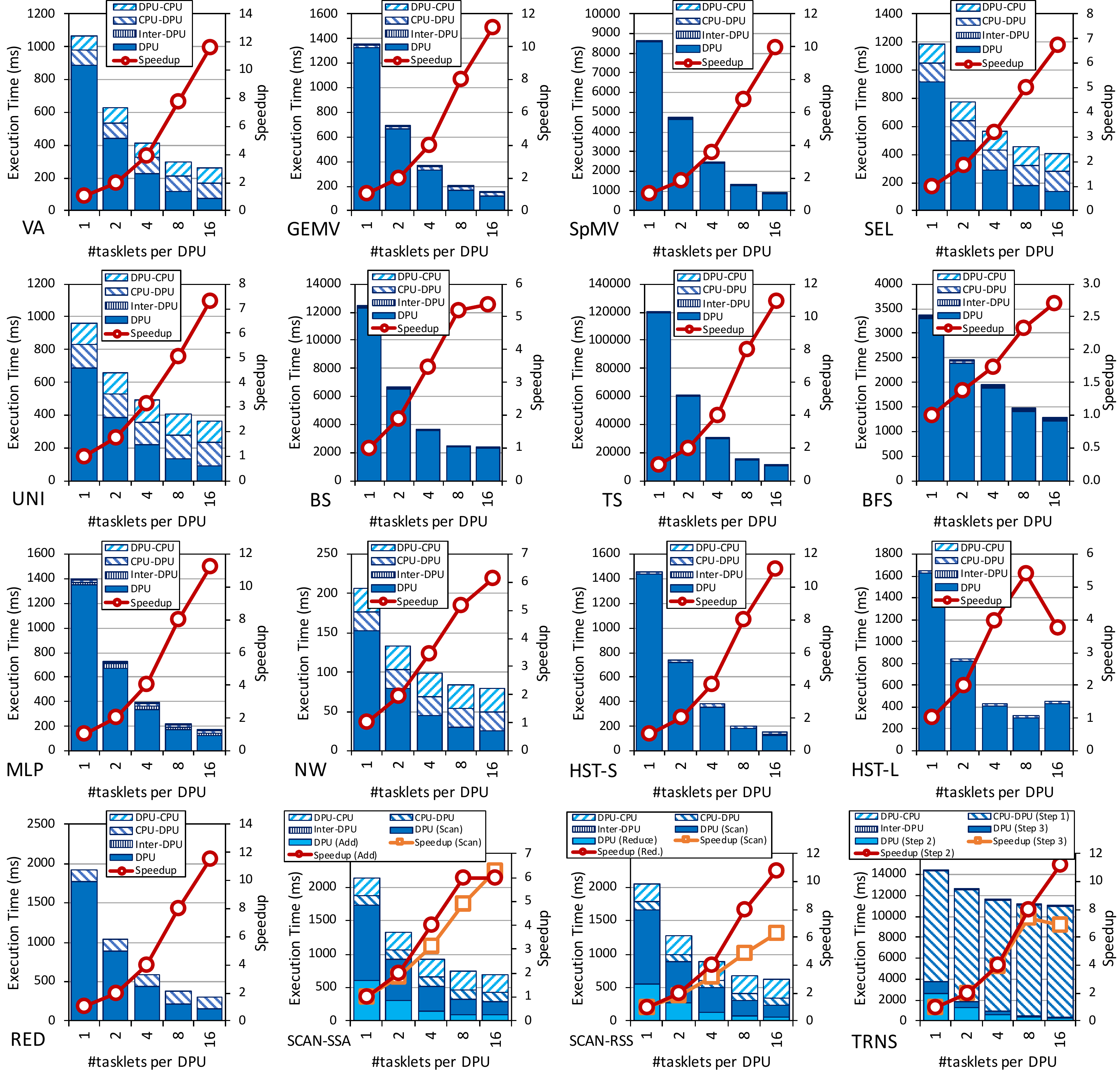}
\vspace{-3mm}
\caption{Execution time (ms) of 16 benchmarks on 1, 2, 4, 8, and 16 tasklets in one DPU (left y-axis), and speedup ({considering only the portion of execution time spent on the DPU}) provided by more tasklets normalized to the performance of 1 tasklet (right y-axis).} \label{fig:1dpu_strong}
\vspace{-3mm}
\end{figure*}

\noindent\paragraph{\textbf{One DPU}}
\label{sec:1dpu}
Figure~\ref{fig:1dpu_strong} presents execution time and speedup scaling (versus tasklet count) results for 16 benchmarks on a single DPU. 
The speedup results (right y-axis of each plot) correspond to only the execution time {portion spent} on the DPU ({i.e., "DPU" portion of each bar} in Figure~\ref{fig:1dpu_strong}).
In these experiments, we set the number of tasklets to 1, 2, 4, 8, and 16.
The benchmarks distribute computation among the available tasklets in a data-parallel manner.
The datasets and their sizes are in Table~\ref{tab:datasets}. 
The times shown in Figure~\ref{fig:1dpu_strong} are broken down into {the} execution time on the DPU ("DPU"), the time for inter-DPU communication via the host CPU ({"Inter-DPU"}), the time for CPU to DPU transfer of input data ({"CPU-DPU"}), and the time for DPU to CPU transfer of final results ({"DPU-CPU"}). 

We make the following {five} observations from Figure~\ref{fig:1dpu_strong}. 

{First,} in 
VA, GEMV, SpMV, SEL, UNI, 
TS, MLP, NW, HST-S, RED, SCAN-SSA (Scan kernel), SCAN-RSS (both kernels), and TRNS (Step 2 kernel), the best performing number of tasklets is 16. 
This is in line with our observations in Section~\ref{sec:arith-throughput}: a number of tasklets greater than 11 is usually a good choice to achieve the best performance from the pipeline. 
These benchmarks show good scaling from 1 to 8 tasklets with speedups between 
$1.5\times$ and $2.0\times$ {as we double} the number of tasklets until 8. From 8 to 16 tasklets, the speedups are between 
$1.2\times$ and \jieee{$1.5\times$} due to the pipeline throughput saturating at 11 tasklets. 
\jieee{For BS and BFS, 16 tasklets provide the highest performance too. However, scaling in BS and BFS is more limited than in the kernels listed in the beginning of this paragraph, as we discuss later in this section.}

\gboxbegin{\rtask{ko}}
\textbf{A number of tasklets greater than 11 is a good choice for most real-world workloads} we tested (16 kernels out of 19 kernels from 16 benchmarks), {as it fully utilizes the DRAM Processing Unit's pipeline}.
\gboxend

{Second,} some of these benchmarks 
(VA, GEMV, SpMV, BS, TS, MLP, HST-S, TRNS (Step 2)) do \emph{not} use synchronization primitives, while in others 
(SEL, UNI, NW, RED, SCAN-SSA (Scan kernel), SCAN-RSS (both kernels)), synchronization across tasklets (via handshakes and/or barriers) is lightweight.


{Third,} BFS, HST-L, and TRNS (Step 3) show limited scaling when increasing the number of tasklets because they use mutexes, which cause contention when accessing shared data structures (i.e., output frontier in BFS, local per-DPU histogram in HST-L, array of flags in TRNS (Step 3)). While in BFS using 16 tasklets {provides the highest performance since it can compensate for the large synchronization cost, in HST-L and TRNS (Step 3) the best performing number of tasklets is 8 due to the high synchronization overheads beyond 8 tasklets.}

\gboxbegin{\rtask{ko}}
Intensive use of \textbf{intra-DPU synchronization across tasklets (e.g., mutexes, {barriers, handshakes}) may limit scalability}, {sometimes} causing the best performing number of tasklets {to} be lower than 11.
\gboxend

Fourth, SCAN-SSA (Add kernel) {experiences} speedups between $1.5\times$ and $2.0\times$ when {we double} the number of tasklets until 8. However, there is no speedup from 8 to 16 tasklets, even though this {step of the SCAN-SSA benchmark} does \emph{not} use any synchronization primitives. 
We observe the same behavior for the STREAM ADD {microbenchmark} in Figure~\ref{fig:mram-stream}, i.e., performance saturation happens before the 11 tasklets {required} to fully utilize the pipeline. As explained in Section~\ref{sec:mram-streaming}, the reason is that both STREAM {ADD} and SCAN-SSA (Add kernel) are \emph{not} compute-intensive kernels, since they perform only {one} integer addition per input element accessed from MRAM. As a result, the overall latency is dominated by the MRAM access latency, which hides the pipeline latency {(and thus performance saturates at fewer than 11 tasklets required to fully utilize the pipeline).} 
\jieee{The same reason explains that BS obtains almost no speedup (only 3\%) from 8 to 16 tasklets, since BS performs only one comparison per input element.}

\gboxbegin{\rtask{ko}}
\textbf{Most real-world workloads are in the compute-bound region} of the {DRAM Processing Unit} (all kernels {except} SCAN-SSA (Add kernel)), i.e., the pipeline latency dominates the MRAM access latency.
\gboxend

Fifth, while the amount of time spent on {CPU-DPU transfer}s and {DPU-CPU transfers} is relatively low compared to the time spent on DPU {execution} for most benchmarks, we observe that {CPU-DPU transfer} time is very high in TRNS. 
The {CPU-DPU transfer} of TRNS performs step 1 of the matrix transposition algorithm~\cite{sung2014matrix,gomez2016matrix} by issuing $M' \times m$ data transfers of $n$ elements, {as explained in Section~\ref{sec:trns}.} 
Since we use a small $n$ value in the experiment ({$n = 8$}, as indicated in Table~\ref{tab:datasets}), the {sustained {CPU-DPU} bandwidth is far from the maximum {CPU-DPU} bandwidth} ({see sustained CPU-DPU bandwidth for different transfer sizes in Figure~\ref{fig:cpudpu}a}). 

\gboxbegin{\rtask{ko}}
\textbf{{{Transferring} large data chunks from/to the host CPU is preferred}} for input data and output results due to higher sustained {CPU-DPU}/{DPU-CPU} bandwidths.
\gboxend

\begin{figure*}[h]
\includegraphics[width=1.0\linewidth]{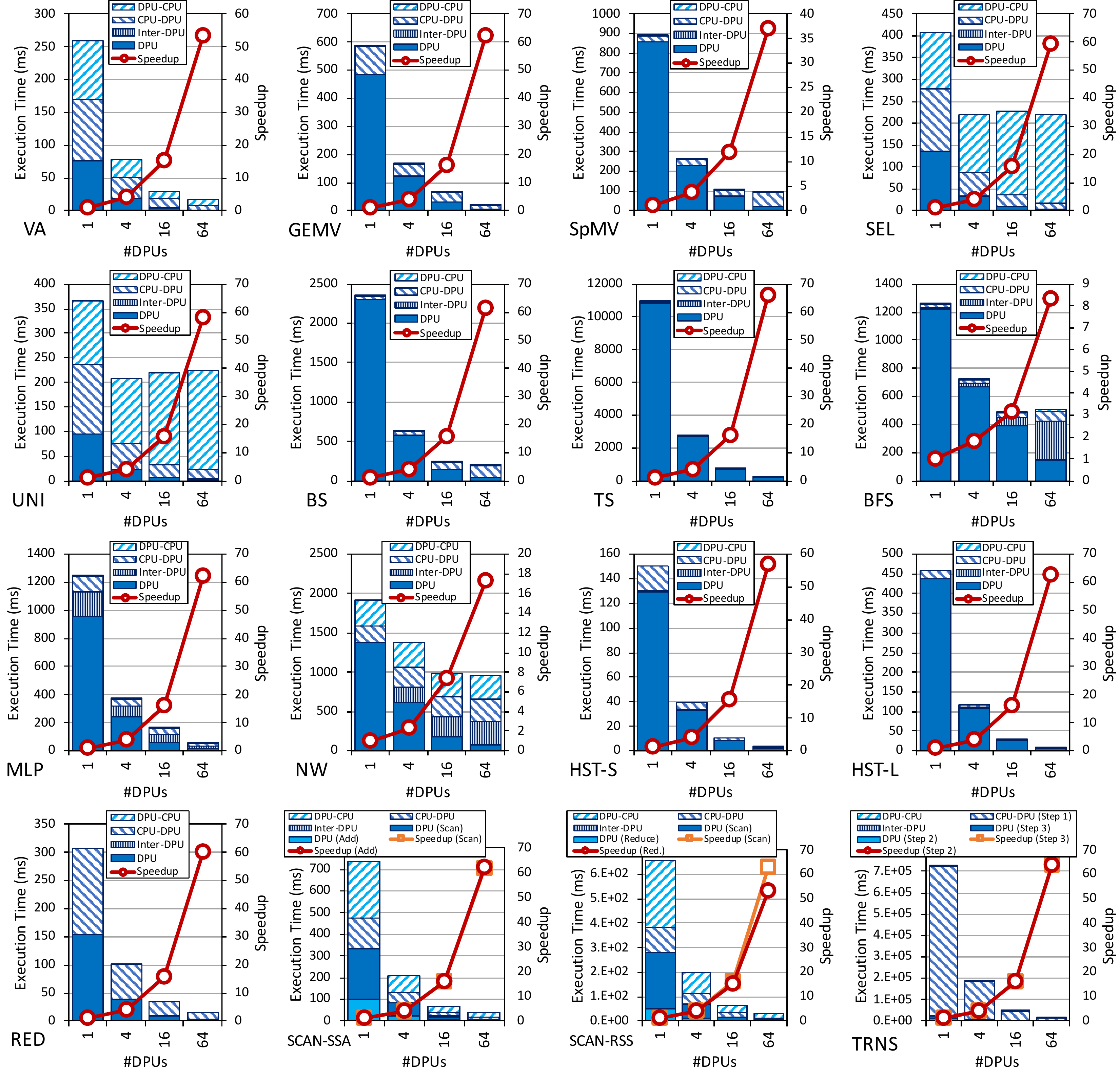}
\vspace{-2mm}
\caption{Execution time (ms) of 16 benchmarks on one rank (1, 4, 16, and 64 DPUs, using strong scaling$^3$) (left y-axis), and speedup ({considering only the portion of execution time spent} on the DPU) provided by more DPUs normalized to the performance of 1 DPU (right y-axis). Inside a DPU, we use the best-performing number of tasklets from Figure~\ref{fig:1dpu_strong}.}
\label{fig:64dpu_strong}
\end{figure*}

\noindent\paragraph{\textbf{One Rank (1-64 DPUs).}} We evaluate strong scaling with 1 to 64 DPUs. The size of the input is the dataset size we can fit in a single DPU (see Table~\ref{tab:datasets}). We especially examine {CPU-DPU transfer} and {DPU-CPU transfer times}, in order to assess how they change as we increase the number of DPUs. 
Figure~\ref{fig:64dpu_strong} shows execution time {and speedup scaling (versus DPU count)} results for 1, 4, 16, and 64 DPUs. 
The speedup results (right y-axis of each plot) correspond to only the execution time {portion spent} on the DPU ({i.e., the "DPU" portion of each bar} in Figure~\ref{fig:64dpu_strong}).
The breakdown of execution time is the same as that done in Figure~\ref{fig:1dpu_strong} for {the single-DPU} results.

We make the following seven observations from Figure~\ref{fig:64dpu_strong}.

First, we observe that DPU performance scaling is linear {with DPU count} (i.e., the execution times on the DPU reduce linearly as we increase the number of DPUs) for 
VA, GEMV, SpMV, SEL, UNI, \jieee{BS,} TS, MLP, HST-S, HST-L, RED, SCAN-SSA (both kernels), SCAN-RSS (both kernels), and TRNS (both kernels) (speedups between 
$3.1\times$ and $4.0\times$ when increasing the number of DPUs by 4). 
{As a result, increasing the DPU count from 1 to 64 for these benchmarks produces speedups between 37$\times$ (SpMV) and 64$\times$ (TS, TRNS).}

Second, scaling of DPU performance is sublinear for \jieee{two benchmarks (BFS, NW)}. Increasing the DPU count from 1 to 64 for these \jieee{two} benchmarks produces speedups between 8.3$\times$ (BFS) and \jieee{17.2$\times$ (NW)}.
For BFS, the speedups are \jieee{sublinear} ($1.7-2.7\times$ when increasing the number of DPUs by 4) 
due to load imbalance across DPUs produced by the irregular topology of the \emph{loc-gowalla} graph~\cite{graphgowalla}. 
In NW, the speedups are between \jieee{$2.2\times$ and $3.3\times$} when multiplying the DPU count by 4. 
In this benchmark, the parallelization factor in each iteration (i.e., number of active DPUs) depends on the size of the diagonal of the 2D score matrix that is processed, {and the number of large 2D blocks in the diagonal}. When we increase the number of DPUs by 4, the parallelization factor in smaller diagonals is low ({i.e., equal to only the number of blocks in these diagonals}), and only increases up to 4$\times$ in the larger diagonals {(i.e., when there are enough blocks to use all available DPUs)}. As a result, the scalability of NW is sublinear. 


Third, the overhead (if any) of inter-DPU synchronization ({as depicted by the "Inter-DPU" portion of each bar} in Figure~\ref{fig:64dpu_strong}) is low in 14 of the benchmarks (VA, GEMV, SpMV, SEL, UNI, BS, TS, 
HST-S, HST-L, RED, SCAN-SSA, SCAN-RSS, TRNS). As a result, these benchmarks achieve higher performance when {we increase} the number of DPUs (even including the inter-DPU synchronization time). 
There is no inter-DPU synchronization in 
VA, GEMV, SpMV, BS, TS, 
and TRNS. 
There is inter-DPU synchronization in HST-S and HST-L (for final histogram reduction), but its overhead is negligible. 
{The inter-DPU synchronization time is noticeable} in SEL, UNI, and RED (for final result merging) and in SCAN-SSA and SCAN-RSS (for intermediate scan step in the host). {For 64 DPUs, the inter-DPU synchronization times of SEL, UNI, RED, SCAN-SSA, and SCAN-RSS account for 53\%, 91\%, 48\%, 42\%, and 17\% the execution times on the DPUs (not visible in Figure~\ref{fig:64dpu_strong}), respectively.}
{Despite that,} we still obtain the best performance (including {portions of the} execution time {spent} on the DPUs, {i.e., "DPU",} and inter-DPU synchronization, {i.e., "Inter-DPU"}) with 64 DPUs {for SEL, UNI, RED, SCAN-SSA, and SCAN-RSS}. 

Fourth, we observe significantly higher overhead of inter-DPU synchronization for BFS, \jieee{MLP}, and NW. 
\jieee{In MLP, the inter-DPU synchronization overhead (due to the distribution of weights matrix and input vector to each layer) reduces as the number of DPUs increases. 
The reason is that the distribution of the weights matrix (i.e., copying assigned matrix rows to the corresponding DPUs) takes advantage of parallel CPU-DPU transfers, while the overhead of transferring the input vector is negligible}. 
\jieee{However, the trend is different for BFS and NW.} 
The overall performance (including portions of the execution time on the DPUs, i.e., "DPU", and inter-DPU synchronization, i.e., "Inter-DPU") of 64 DPUs is only 5\% and \jieee{17\%} higher than that of 16 DPUs for BFS and NW, respectively.
The reason in BFS is that, after each iteration, the CPU has to compute the union of the next frontier from all DPUs sequentially and redistribute it across the DPUs. Thus, the inter-DPU synchronization cost increases linearly with the number of DPUs. 
In NW, the inter-DPU synchronization overhead is substantial due to a similar reason. 
For each diagonal of the 2D score matrix, the host CPU (1) retrieves the results of the sub-blocks produced by all DPUs, and (2) sends the cells of the last row and the last column of each sub-block as input to the next diagonal (processed in the next iteration). 

Fifth, we observe the {CPU-DPU transfer} and {DPU-CPU transfer} times decrease for 
VA, GEMV, TS, MLP, HST-S, HST-L, RED, SCAN-SSA, SCAN-RSS, and TRNS, when we increase the number of DPUs in the strong scaling experiment for 1 rank. 
These benchmarks use parallel {CPU-DPU} and {DPU-CPU} transfers between the main memory of the host CPU and the MRAM banks.

Sixth, the {CPU-DPU} and {DPU-CPU transfer times} do not decrease for BS and NW with increasing number of DPUs, even though {BS and NW} use parallel transfers.
BS distributes the values to search in a sorted array across the available DPUs, but the sorted array is replicated in each DPU. As a result, the total {CPU-DPU} time increases with the number of DPUs. 
NW processes a diagonal in each iteration. For shorter diagonals, the algorithm does not need to use all available DPUs. Thus, more available DPUs does not always mean more parallelism in {CPU-DPU} and {DPU-CPU} transfers. 

Seventh, the remaining benchmarks (SpMV, SEL, UNI, BFS) cannot use parallel transfers to {copy input data and/or retrieve results}. 
In SEL and UNI, {DPU-CPU transfer times} {increase} when we increase the number of DPUs because we cannot use parallel transfers for retrieving results. In these two benchmarks, the size of the output in each DPU may differ as it depends on the element values of the input array. 
SpMV and BFS cannot use parallel {CPU-DPU and DPU-CPU} transfers because the size of the inputs assigned to each DPU may be different ({e.g., different number of nonzero elements of different sparse rows in SpMV, different numbers of edges for different} vertices in BFS). As a result, we observe that {CPU-DPU} and {DPU-CPU transfer times} do not reduce in SpMV and BFS when increasing the number of DPUs.


\pboxbegin{\ptask{pr}}
\textbf{Parallel CPU-DPU/DPU-CPU transfers inside a rank of UPMEM DRAM Processing Units are recommended for real-world workloads} when all transferred buffers {are of the same size}.
\pboxend

\begin{figure*}[h!]
\includegraphics[width=1.0\linewidth]{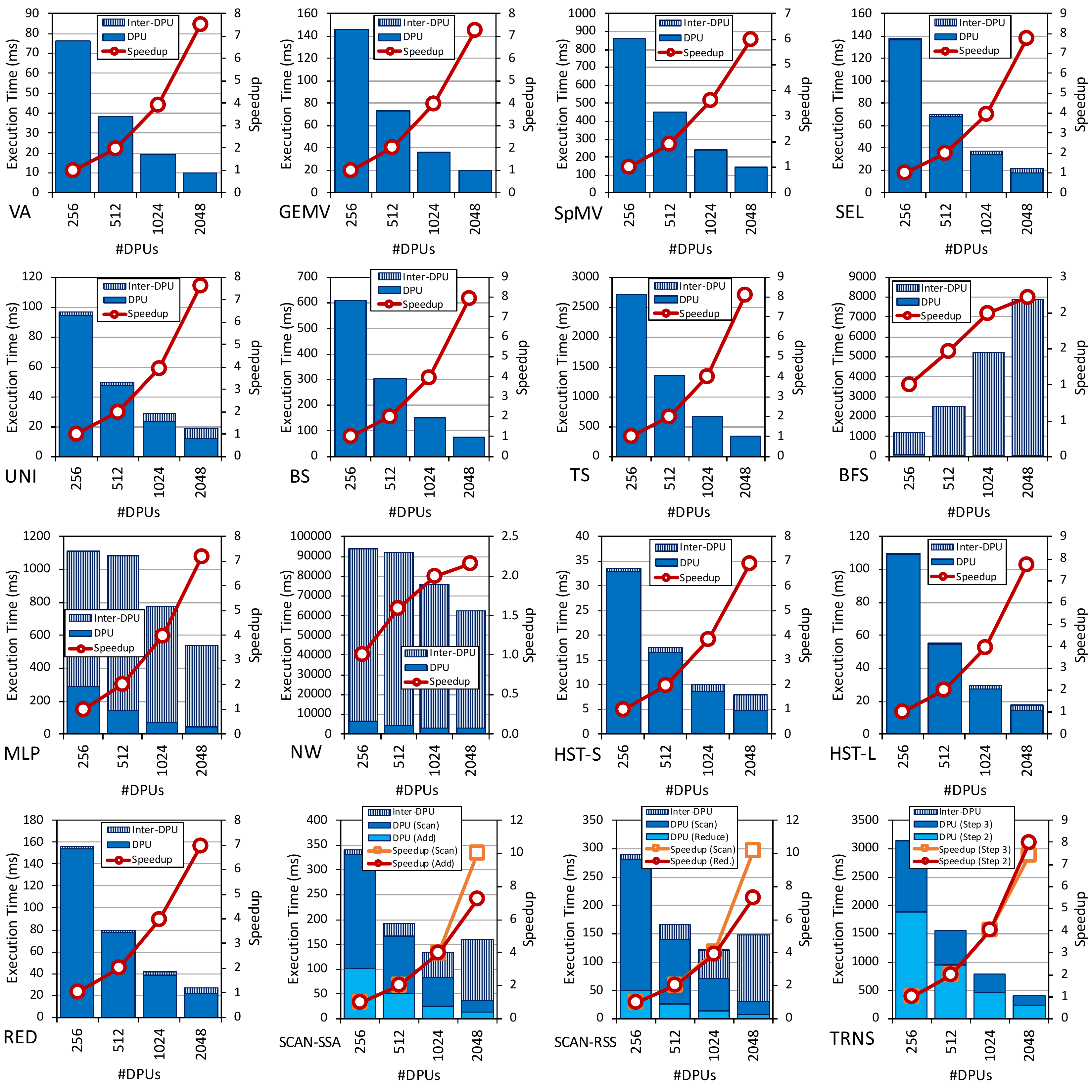}
\vspace{-3mm}
\caption{Execution time (ms) of 16 benchmarks on 4, 8, 16, and 32 ranks (256, 512, 1,024, and 2,048 DPUs, {using strong scaling$^3$}) (left y-axis), and speedup ({considering only the portion of execution time spent} on the DPU) provided by more DPUs normalized to the performance of 4 ranks (256 DPUs) (right y-axis). Inside a DPU, we use the best-performing number of tasklets from Figure~\ref{fig:1dpu_strong}.} 
\label{fig:640dpu_strong}
\vspace{-3mm}
\end{figure*}

\noindent\paragraph{\textbf{32 Ranks (256-2,048 DPUs).}} We evaluate strong scaling with 4, 8, 16, and 32 ranks. The size of the input is the maximum dataset size we can fit in four ranks (i.e., 256 DPUs), as shown in Table~\ref{tab:datasets}. We do not include {CPU-DPU} and {DPU-CPU transfer times} in our performance analysis, because {these} transfers are \emph{not} simultaneous across ranks, as we mentioned in Section~\ref{sec:cpu-dpu}. Figure~\ref{fig:640dpu_strong} shows execution time {and speedup scaling (versus DPU count)} results for 256, 512, 1,024, and 2,048 DPUs, corresponding to 4, 8, 16, and 32 ranks.
The speedup results (right y-axis of each plot) correspond to only the execution time {portion spent} on the DPU ({i.e., the "DPU" portion of each bar} in Figure~\ref{fig:640dpu_strong}).

We make the following observations from Figure~\ref{fig:640dpu_strong}.

First, as in the experiments on one rank, 
we observe that the execution times on the DPU ({i.e., the "DPU" portion of each bar} in Figure~\ref{fig:640dpu_strong}) reduce linearly with the number of DPUs (i.e., $\sim$2$\times$ when {we double} the number of DPUs, and $\sim$8$\times$ from 256 to 2,048 DPUs) for VA, GEMV, SEL, UNI, BS, TS, MLP, HST-S, HST-L, RED, SCAN-SSA (both kernels), SCAN-RSS (both kernels), and TRNS (both kernels).
For SCAN-SSA (Scan) and SCAN-RSS (Scan), we observe more than $8\times$ speedup {when we scale from 256 to 2,048 DPUs}. The reason is that the amount of synchronization across tasklets (i.e., handshakes in Scan) reduces when we distribute the input array across more DPUs. However, the downside is that the inter-DPU communication cost increases, as we explain below. 

Second, DPU performance scaling ({i.e., the "DPU" portion of each bar} in Figure~\ref{fig:640dpu_strong}) is sublinear for SpMV, BFS, and NW. 
For {SpMV and BFS}, there is load imbalance across DPUs due to the irregular nature of graphs and sparse matrices.
For NW, we observe small speedups when {we double} the number of DPUs (\jieee{$1.60\times$} from 256 to 512 DPUs, and \jieee{$1.25\times$} from 512 to 1,024 DPUs), and almost no speedup (only \jieee{8\%}) from 1,024 to 2,048 DPUs. As explained above, NW does not use all DPUs in all iterations, but only the number that is needed for the diagonal that is processed in each iteration. As a result, doubling the number of DPUs does not reduce the execution time {spent} on the DPU at the same rate. 

\gboxbegin{\rtask{ko}}
\textbf{Load balancing across DRAM Processing Units (DPUs) ensures linear reduction of the execution time {spent} on the DPUs} for a given problem size, when all available DPUs are used (as observed in strong scaling experiments).
\gboxend

Third, inter-DPU synchronization (as depicted {by the "Inter-DPU" portion of each bar} in Figure~\ref{fig:640dpu_strong}) imposes a small overhead (if any) for 12 of the benchmarks (VA, GEMV, SpMV, SEL, UNI, BS, TS, 
HST-S, HST-L, RED, and TRNS). 
VA, GEMV, SpMV, BS, TS, 
and TRNS do not require inter-DPU synchronization. 
For SEL, UNI, HST-S, HST-L, and RED, the inter-DPU synchronization involves {only} {DPU-CPU} transfers, since it is only used to merge final results at the end of execution. 
The inter-DPU synchronization overhead increases with the number of DPUs, since the amount of partial results to merge increases. However, the inter-DPU synchronization cost is small, and a larger number of DPUs results in larger overall performance. 

\gboxbegin{\rtask{ko}}
\textbf{The overhead of merging partial results from DRAM Processing Units in the host CPU is tolerable} across all {PrIM} benchmarks that need it.
\gboxend

Fourth, the inter-DPU synchronization imposes significant overhead when it requires more complex patterns (involving {both {CPU-DPU} and {DPU-CPU}} transfers). We observe this for \jieee{five} benchmarks (BFS, \jieee{MLP}, NW, SCAN-SSA, and SCAN-RSS). 
For NW \jieee{and MLP}, we observe that inter-DPU synchronization times are significantly higher than DPU times. If we compare these results to the results in Figure~\ref{fig:64dpu_strong}, we conclude that \jieee{these benchmarks'} overall performance is greatly burdened by inter-DPU synchronization when using more than one rank. 
SCAN-SSA and SCAN-RSS perform a more complex intermediate step in the CPU: (1) the CPU gathers partial results from the first kernel (Scan in SCAN-SSA, Reduce in SCAN-RSS) from the DPUs (via {DPU-CPU} transfers), (2) the CPU performs a scan operation, and (3) the CPU returns a value to be used by the second kernel (Add in SCAN-SSA, Scan in SCAN-RSS) to each DPU (via {CPU-DPU} transfers). 
The significant increase in "{Inter-DPU}" from 1,024 to 2,048 DPUs is due to the dual-socket system configuration (Section~\ref{sec:sys-org}), since the CPU in one socket obtains lower memory bandwidth from remote MRAM banks (i.e., in the other socket).
For BFS, the trend is even worse. We observe that the huge increase in the inter-DPU synchronization time makes 256 DPUs the best choice for executing BFS.
Our observations for BFS, SCAN-SSA, and SCAN-RSS are \emph{against} the general programming recommendation of using as many working DPUs as possible (Section~\ref{sec:general-recommendations}). These three benchmarks show that the best-performing number of DPUs is {limited} by the inter-DPU synchronization overhead.

\vspace{-2mm}
\gboxbegin{\rtask{ko}}
\vspace{-2mm}
\textbf{Complex synchronization across DRAM Processing Units (i.e., inter-DPU synchronization involving two-way communication with the host CPU) 
imposes significant overhead, which limits scalability {to more DPUs}.}
This is more noticeable when DPUs involved in the synchronization span multiple ranks. 
\vspace{-2mm}
\gboxend


\vspace{-3mm}
\subsubsection{{\textbf{Weak Scaling Results}}}
\label{sec:weak}
\begin{figure*}[t!]
\includegraphics[width=1.0\linewidth]{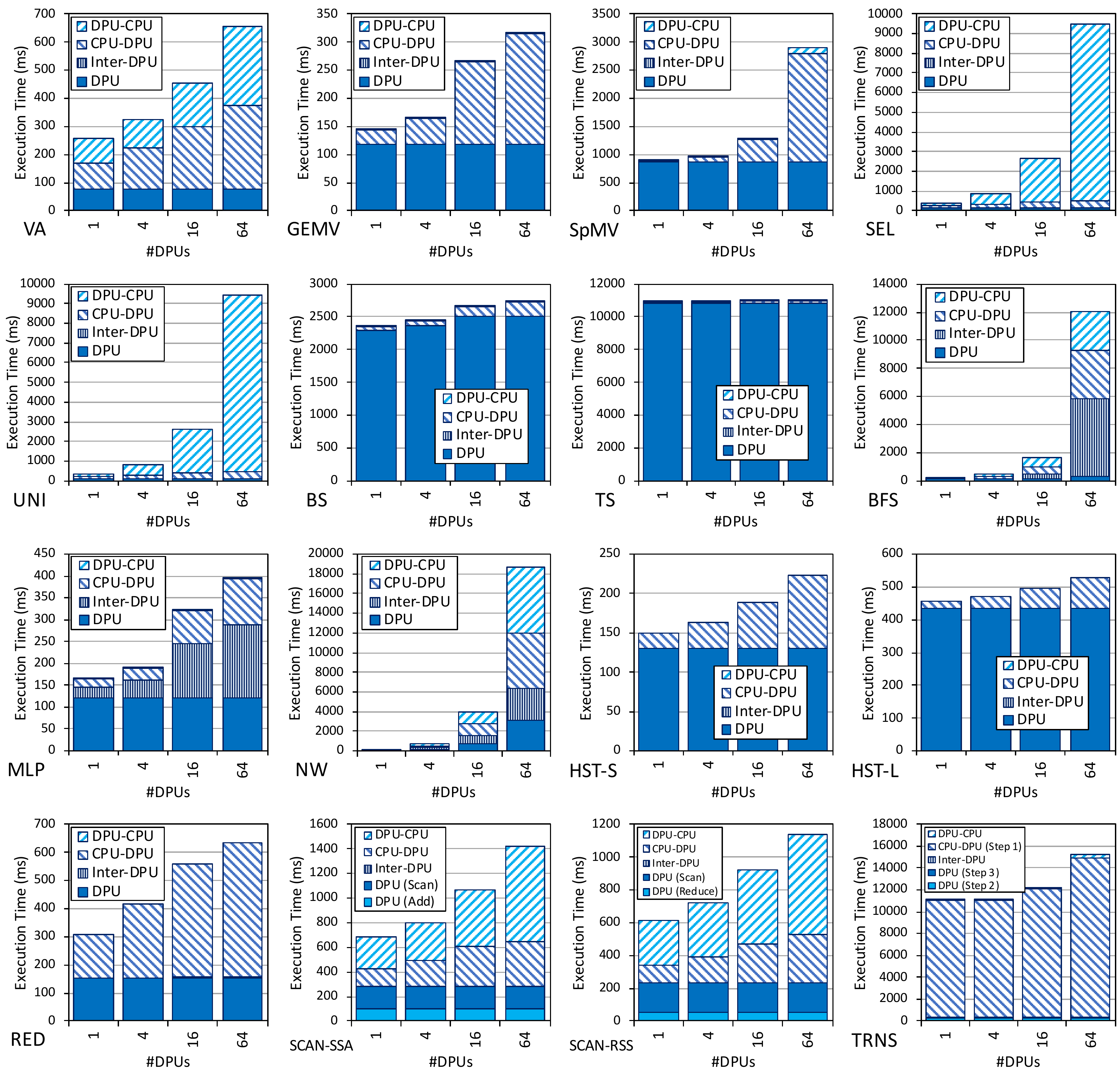}
\vspace{-2mm}
\caption{Execution time (ms) of 16 benchmarks on one rank (1, 4, 16, and 64 DPUs, {using weak scaling$^4$}). {Inside a DPU, we use the best-performing number of tasklets from Figure~\ref{fig:1dpu_strong}.}} \label{fig:64dpu_weak}
\end{figure*}

Figure~\ref{fig:64dpu_weak} shows the weak scaling results inside a single rank for 1, 4, 16, and 64 DPUs.
In each DPU, we run the number of tasklets that produces the best performance in Section~\ref{sec:1dpu}. 
The size of the dataset per DPU is the size shown in Table~\ref{tab:datasets}. 
The time is broken down into execution time on the DPU ("DPU"), inter-DPU synchronization time ("{Inter-DPU}"), and {CPU-DPU} and {DPU-CPU transfer times} ("{CPU-DPU}", "{DPU-CPU}"), similarly to the strong scaling results presented in Figures~\ref{fig:1dpu_strong} to~\ref{fig:640dpu_strong} in Section~\ref{sec:strong}.

We make the following five observations from Figure~\ref{fig:64dpu_weak}.

First, we observe perfect {(i.e., linear)} weak scaling on the DPU for 14 benchmarks: 
the execution time on the DPU ({i.e., the "DPU" portion of each bar} in Figure~\ref{fig:64dpu_weak}) remains constant for VA, GEMV, SpMV, SEL, UNI, BS, TS, MLP, 
HST-S, HST-L, RED, SCAN-SSA, SCAN-RSS, and TRNS, as we increase the number of DPUs (and the dataset size). 
Since there is no direct communication between DPUs {in these} kernels, the even distribution of workload {(i.e., load balance) among DPUs leads to} performance scaling. 
We observe a similar trend {of perfect weak scaling for BFS} even though there is some load imbalance across DPUs {in BFS}.

\vspace{-2mm}
\gboxbegin{\rtask{ko}}
\vspace{-1mm}
\textbf{Equally-sized problems assigned to different {DRAM Processing Units (DPUs)} {and little/no inter-DPU synchronization lead to} linear weak scaling} of the execution time {spent} on the DPUs (i.e., constant execution time when we increase the number of DPUs and the dataset size accordingly).
\vspace{-2mm}
\gboxend

Second, NW does \emph{not} scale linearly (i.e., the execution time {spent} on the DPU is \emph{not} constant) because the size of the problem does \emph{not} increase linearly with the number of DPUs. 
We {increase the lengths of the input sequences to NW} \emph{linearly} with the number of DPUs (see Table~\ref{tab:datasets}, weak scaling dataset). Thus, the size of the 2D score matrix increases \emph{quadratically} with the number of DPUs. 
As a result, the execution times on the DPUs increase when we increase the number of DPUs. 
However, the longest diagonal of the 2D score matrix increases linearly with the number of DPUs. The processing time of this diagonal shows linear weak scaling as we increase the number of DPUs. We show these experimental results in the Appendix (Section~\ref{app:nw}).

Third, among the benchmarks that require inter-DPU synchronization (SEL, UNI, BFS, \jieee{MLP}, NW, HST-S, HST-L, RED, SCAN-SSA, and SCAN-RSS, the synchronization overhead ({i.e., the "Inter-DPU" portion of each bar in Figure~\ref{fig:64dpu_weak}}) is negligible for SEL, UNI, 
HST-S, HST-L, RED, SCAN-SSA, and SCAN-RSS. 
For \jieee{MLP and} NW, the inter-DPU synchronization time takes a significant fraction of the overall execution time, 
and it increases with the number of DPUs because the total problem size (and thus, \jieee{the size of weight matrices in MLP and} the number of iterations \jieee{in NW})
increases, as indicated above. 
In BFS, the inter-DPU synchronization time increases linearly, as we explain {in Section~\ref{sec:strong} (Figures~\ref{fig:64dpu_strong} and~\ref{fig:640dpu_strong})} for strong scaling experiments. As a result, BFS obtains the best tradeoff between overall {execution time} (including {portions of the execution time spent on the DPUs, i.e., "DPU", and inter-DPU synchronization, i.e., "Inter-DPU"}) and number of DPUs {at} 16 DPUs (i.e., the ratio of overall execution time, {the "DPU" portions + the "Inter-DPU" portions}, over number of DPUs is lower for 16 DPUs).

Fourth, {CPU-DPU} and {DPU-CPU transfer times} increase slowly with the number of DPUs for the 13 benchmarks that use parallel transfers between main memory and MRAM banks (VA, GEMV, SEL (only {CPU-DPU}), UNI (only {CPU-DPU}), BS, TS, MLP, 
HST-S, HST-L, RED, SCAN-SSA, SCAN-RSS, and TRNS). 
As observed from Figure~\ref{fig:cpudpu}, the sustained {CPU-DPU} and {DPU-CPU} bandwidths increase sublinearly with the number of DPUs. On average, the increase in sustained {CPU-DPU}/{DPU-CPU} bandwidth for these 13 benchmarks from 1 DPU to 64 DPUs is 20.95$\times$/23.16$\times$.
NW uses parallel {CPU-DPU and DPU-CPU} transfers, but the {CPU-DPU transfer} and {DPU-CPU transfer times} increase with the number of DPUs because the amount of transferred data increases (i.e., the total problem size increases, as described above {in the second observation from Figure~\ref{fig:64dpu_weak}}).

Fifth, {CPU-DPU transfer} and {DPU-CPU transfer times} increase linearly with the number of DPUs for the benchmarks that cannot use parallel transfers.
SEL and UNI employ serial {DPU-CPU} transfers, as we discuss above. This makes the {DPU-CPU transfer times} in these two benchmarks increase dramatically with the number of DPUs, dominating the entire execution time.  
In SpMV and BFS, where we cannot use parallel transfers due to the irregular nature of datasets, {CPU-DPU transfer} and {DPU-CPU transfer times} also increase significantly.
In {full-blown} real-world applications, where SEL, UNI, SpMV, or BFS may be just one of {the multiple or many kernels executed by the application}, the CPU-DPU transfer and DPU-CPU transfer times {can} be amortized and their overhead alleviated.


\gboxbegin{\rtask{ko}}
\textbf{Sustained bandwidth of parallel {CPU-DPU}/{DPU-CPU} transfers inside a rank {of DRAM Processing Units (DPUs)} increases sublinearly} with the number of DPUs. 
\gboxend

\subsection{Comparison to CPU and GPU}
\label{sec:comparison}
We compare the UPMEM PIM architecture to a modern CPU and a modern GPU in terms of performance and energy consumption. 
Our {goal} is to quantify the potential of the UPMEM PIM architecture as a general-purpose accelerator. 
We use state-of-the-art CPU and GPU versions of PrIM benchmarks for comparison to our PIM implementations. The sources of the CPU and GPU versions of the benchmarks are listed in the Appendix (Table~\ref{tab:comparison}).

We compare the UPMEM-based PIM systems with 640 and 2,556 DPUs (Table~\ref{tab:pim-setups}) to an Intel Xeon E3-1225 v6 CPU~\cite{xeon-e3-1225} and an NVIDIA Titan V GPU~\cite{titanv} based on the Volta architecture~\cite{volta} for all our benchmarks. 
Table~\ref{tab:pim-cpugpu} summarizes key characteristics of the CPU, the GPU, and the two UPMEM-based PIM systems.

\begin{table*}[h]
        \begin{center}
        \captionof{table}{{Evaluated} CPU, GPU, and UPMEM-based PIM Systems.}
        \label{tab:pim-cpugpu}
        \resizebox{1.0\linewidth}{!}{
        \begin{tabular}{|l|c|ccc|cc|c|}
    \hline
    \multirow{2}{*}{\textbf{System}} & \textbf{Process} & \multicolumn{3}{c|}{\textbf{Processor Cores}} & \multicolumn{2}{c|}{\textbf{Memory}} & \multirow{2}{*}{\textbf{TDP}} \\
    \cline{3-7}
     & \textbf{Node} & \textbf{Total Cores} & \textbf{Frequency} & \textbf{Peak Performance} & \textbf{Capacity} & \textbf{Total Bandwidth} & \\
    \hline
    \hline
\textbf{Intel Xeon E3-1225 v6 CPU}~\cite{xeon-e3-1225} & 14 nm & 4 (8 threads) & 3.3 GHz & 26.4 GFLOPS$^\star$ & 32 GB & 37.5 GB/s & 73 W \\ \hline
  \textbf{NVIDIA Titan V GPU}~\cite{titanv} & 14 nm & 80 (5,120 SIMD lanes) & 1.2 GHz & 12,288.0 GFLOPS & 12 GB & 652.8 GB/s & 250 W \\ \hline
 \textbf{2,556-DPU PIM System} & 2x nm & 2,556$^9$ & 350 MHz & 894.6 GOPS & 159.75 GB & 1.7 TB/s & 383 W$^\dagger$ \\ \hline
 \textbf{640-DPU PIM System} & 2x nm & 640 & 267 MHz & 170.9 GOPS & 40 GB & 333.75 GB/s & 96 W$^\dagger$ \\ \hline
\end{tabular}

        }
        \end{center}
\begin{flushleft}
\resizebox{0.4\linewidth}{!}{
$\begin{tabular}{l}
 $^\star Estimated\ GFLOPS = 3.3\ GHz \times 4\ cores \times 2\ instructions\ per\ cycle$. \\
 $^\dagger Estimated\ TDP = \frac{Total\ DPUs}{DPUs/chip} \times 1.2\ W/chip$~\cite{devaux2019}.
\end{tabular}$
}
\end{flushleft}
\end{table*}

For our UPMEM-based PIM system performance measurements, we include the time spent in the DPU and the time spent for inter-DPU synchronization on the UPMEM-based PIM systems. 
For our CPU and GPU performance measurements, we include only the kernel times (i.e., we do not include data transfers between the host CPU and the GPU in the GPU versions).
For energy measurements, we use Intel RAPL~\cite{rapl} on the CPU and NVIDIA SMI~\cite{smi} on the GPU.
In the UPMEM PIM systems, we obtain the energy consumed by the DIMMs connected to the memory controllers, which can be done in x86 sockets~\cite{guide2011intel}. 
The measurements include only the energy of the PIM chips.

\subsubsection{Performance Comparison}
\label{sec:comparison-perf}

\begin{figure}[h]
        \includegraphics[width=\linewidth]{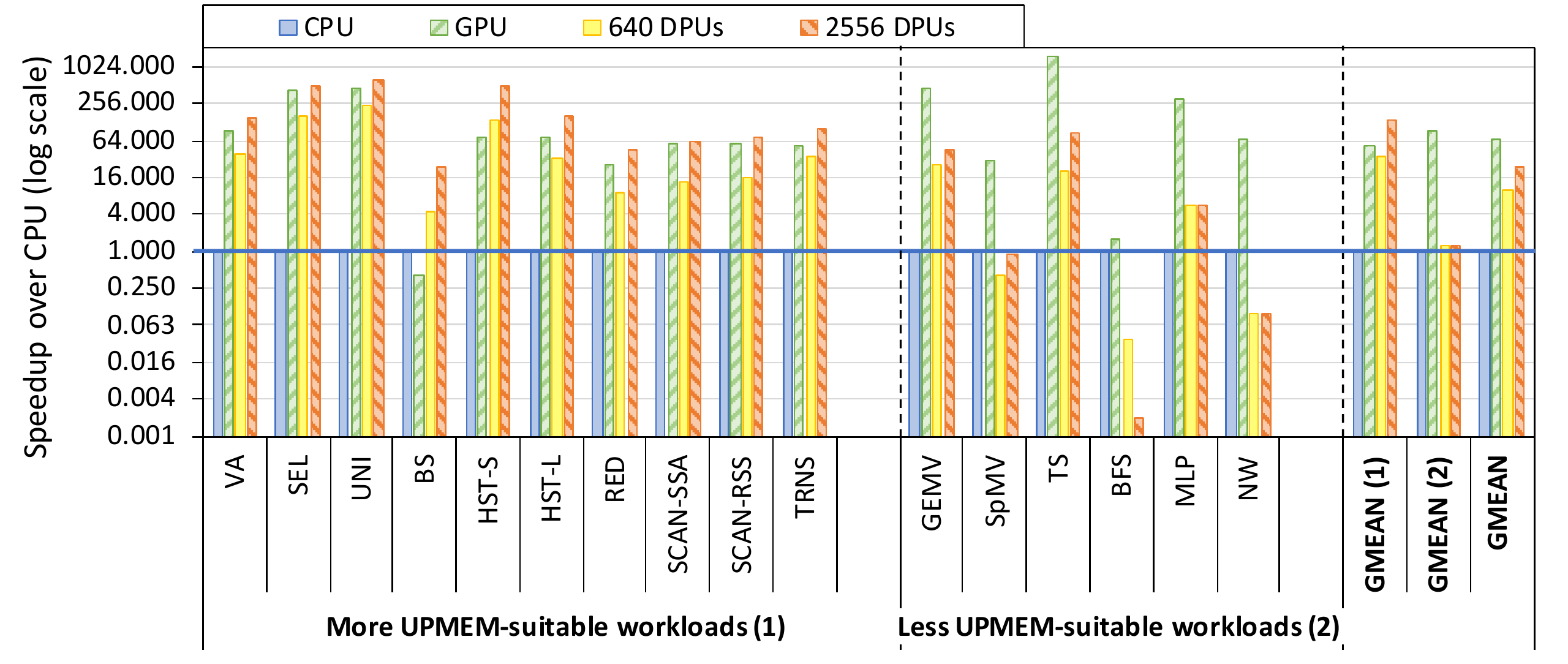}
        \caption{Performance comparison between the UPMEM-based PIM systems with 640 and 2,556 DPUs, a Titan V GPU, and an Intel Xeon E3-1240 CPU. Results are normalized to the CPU performance {(y-axis is log scale)}. 
        {There are two groups of benchmarks: (1) benchmarks that are more suitable to the UPMEM PIM architecture, and (2) benchmarks that are less suitable to the UPMEM PIM architecture.}} \label{fig:comparison-perf}
\end{figure}

Figure~\ref{fig:comparison-perf} shows the speedup of the UPMEM-based PIM systems with 640 and 2,556 DPUs and the Titan V GPU over the Intel Xeon CPU. 

We make four key observations from Figure~\ref{fig:comparison-perf}.

First, the 2,556-DPU system and the 640-DPU system are on average \jieee{23.2$\times$ and 10.1$\times$} faster than the CPU. The highest speedup is for UNI: the 2,556-DPU system is 629.5$\times$ and the 640-DPU system is 234.4$\times$ faster than the CPU. 
Even benchmarks that make heavy use of integer multiplication (GEMV, TS, and MLP) are much faster on the UPMEM-based PIM systems (\jieee{5.8-86.6$\times$} faster on the 2,556-DPU system, and \jieee{5.6-25.2$\times$} faster on the 640-DPU system). 
This observation reflects the large performance improvements that workloads running on a conventional system with a CPU can experience if we expand the system with DIMMs of PIM-enabled memory (see Figure~\ref{fig:scheme}).

Second, the UPMEM-based PIM systems outperform the CPU for all of the benchmarks except SpMV, BFS, and NW. 
SpMV has three characteristics that make it less suitable for UPMEM-based PIM systems: (1) it operates on floating point data, (2) it uses multiplication heavily, and (3) it suffers from load imbalance due to the irregular nature of sparse matrices. Regarding the first two characteristics, we know from our analyses in Sections~\ref{sec:arith-throughput} and~\ref{sec:throughput-oi} that floating point multiplication is very costly because of the lack of native support. Regarding the third characteristic, we know from our strong scaling evaluation in Section~\ref{sec:strong} that load imbalance across DPUs causes sublinear scaling for SpMV. 
BFS performs much worse than CPU on the UPMEM-based PIM systems because of the large overhead of inter-DPU synchronization via the host CPU, as we discuss in Section~\ref{sec:performance}. 
Since the inter-DPU synchronization overhead of BFS increases linearly with the number of DPUs, the 2,556-DPU system is significantly slower than the 640-DPU system.\footnote{BFS can obtain better performance by running it using much fewer DPUs. The reason is that BFS performance does not scale with many DPUs, as shown in Sections~\ref{sec:strong} and~\ref{sec:weak} (Figures~\ref{fig:64dpu_strong}-\ref{fig:64dpu_weak}). However, we do not {run BFS using much fewer DPUs} as we study the full-blown system performance utilizing \emph{all} DPUs in this experiment.} 
Note that the goal of these experiments is \emph{not} to show the performance of the best-performing number of DPUs for a given workload, but the performance of the full-blown systems with {all} 2,556 DPUs and 640 DPUs {active} for each workload. 
NW is one order of magnitude slower on both UPMEM-based PIM systems than on the CPU {due to the inter-DPU synchronization overhead}. The inter-DPU synchronization overhead of NW is not as dependent on the number of DPUs. As a result, the 2,556-DPU system has the same performance as the 640-DPU system for this benchmark.

Third, the 2,556-DPU system is faster than the GPU for 10 benchmarks: VA, SEL, UNI, BS, HST-S, HST-L, RED, SCAN-SSA, SCAN-RSS, and TRNS. 
{These 10 benchmarks are more suitable to the UPMEM PIM architecture due to three key characteristics:}
(1) streaming memory accesses, (2) no or little inter-DPU communication, and (3) no or little use of integer multiplication, integer division, or floating point operations. 
The speedups of the 2,556-DPU system over the GPU for these benchmarks range between 6\% (for SCAN-SSA) and \jieee{57.5$\times$} (for BS), with an average of \jieee{2.54$\times$}. 
It is especially interesting that the 2,556-DPU system outperforms the Titan V for some of these benchmarks, which are traditionally considered GPU-friendly and are subject of GPU optimization studies, libraries, and reference implementations, such as VA~\cite{cudasamples}, SEL and UNI~\cite{gomezluna2015ds,bell2012thrust}, HST-S and HST-L~\cite{gomez2013atomics,gomez2013optimized,van2013simulation}, RED~\cite{degonzalo2019automatic,harris2007optimizing}, SCAN-SSA~\cite{hwukirk2016.scan, sengupta2008efficient,bell2012thrust}, SCAN-RSS~\cite{yan2013streamscan, hwukirk2016.scan, dotsenko2008fast,merrill2015cuda}, and TRNS~\cite{sung2014matrix, gomez2016matrix, catanzaro2014decomposition}. 
In summary, the UPMEM PIM architecture outperforms {the modern GPU} for workloads {that exhibit} the three key characteristics {that make them potentially suitable for execution on the UPMEM-based PIM system}.

Fourth, the 640-DPU system is generally slower than the GPU, but for the 10 benchmarks where the 2,556-DPU system is faster than the GPU (VA, SEL, UNI, BS, HST-S, HST-L, RED, SCAN-SSA, SCAN-RSS, and TRNS) the average performance of the 640-DPU system is within \jieee{65\%} the performance of the GPU. 
Among these benchmarks, the 640-DPU system is faster than the GPU for two benchmarks: HST-S and BS.
The GPU version of histogram~\cite{gomez2013optimized,gomezluna2017chai} (the same one for HST-S and HST-L) uses atomic operations that burden the performance heavily~\cite{gomez2013atomics}. 
As a result, the 640-DPU system is 1.89$\times$ faster than the GPU for HST-S.
For BS, the GPU version suffers from many random memory accesses, which greatly reduce the achievable memory bandwidth. 
The 640-DPU system is \jieee{11.0$\times$} faster than the GPU for BS.

\gboxbegin{\rtask{ko}}
\textbf{The UPMEM-based PIM system {can outperform a modern GPU}} on workloads with \textbf{three key characteristics}:
\begin{enumerate}[1.]
\item Streaming memory accesses
\item No or little inter-DPU synchronization
\item No or little use of integer multiplication, integer division, or floating point operations
\end{enumerate}
{These three key characteristics {make} a \textbf{workload {potentially suitable} to the UPMEM PIM architecture}.}
\gboxend

\subsubsection{Energy Comparison}
\label{sec:comparison-ener}

\begin{figure}[h]
        \includegraphics[width=\linewidth]{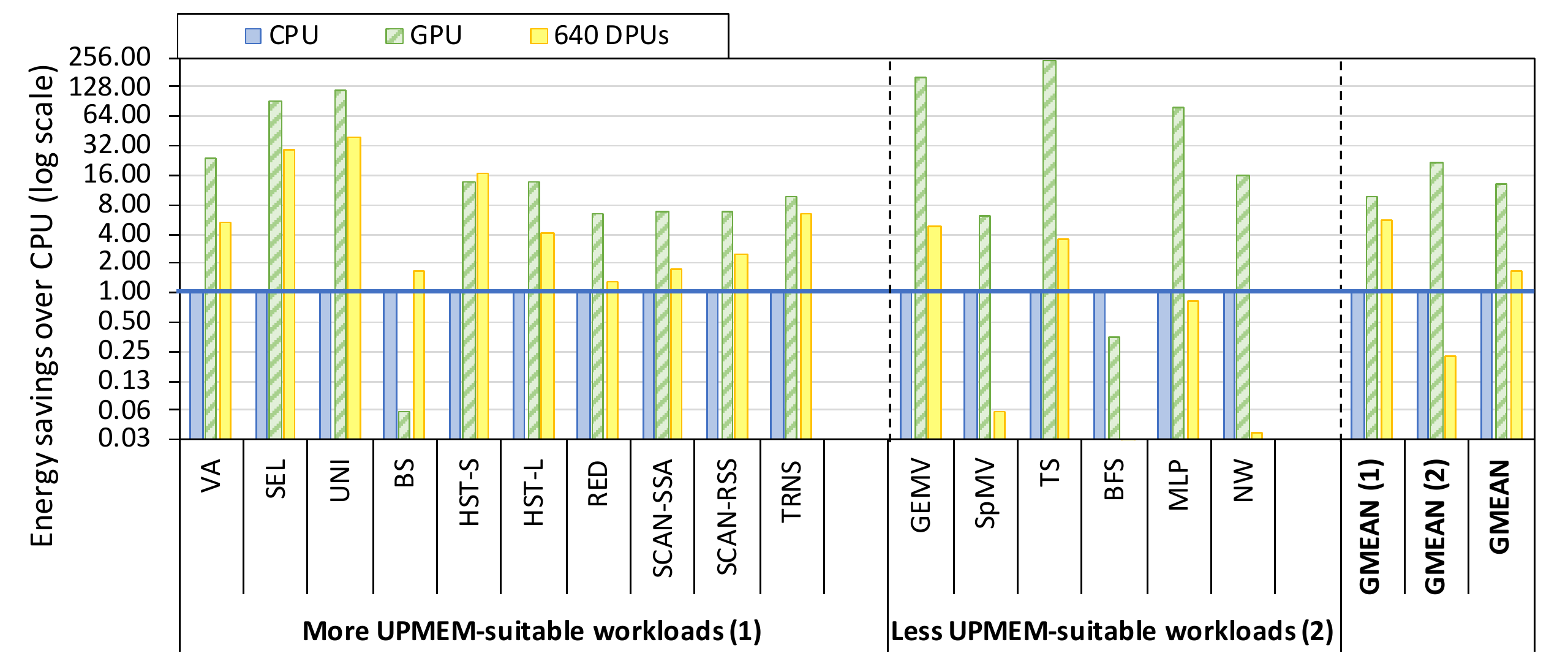}
        \vspace{-2mm}
        \caption{Energy comparison between the UPMEM-based PIM system with 640 DPUs, a Titan V GPU, and an Intel Xeon E3-1240 CPU. Results are normalized to the CPU performance {(y-axis is log scale)}.
        {There are two groups of benchmarks: (1) benchmarks that are more suitable to the UPMEM PIM architecture, and (2) benchmarks that are less suitable to the UPMEM PIM architecture.}} \label{fig:comparison-ener}
\end{figure}

Figure~\ref{fig:comparison-ener} shows the energy savings of the UPMEM-based PIM system with 640 DPUs and the Titan V GPU over the Intel Xeon CPU. 
At the time of writing, the 2,556-DPU system is not enabled to perform energy measurements, but we will aim to include them in an extended version of our work.

We make three key observations from Figure~\ref{fig:comparison-ener}.

First, the 640-DPU system consumes, on average, \jieee{1.64$\times$} less energy than the CPU for all 16 benchmarks. 
For \jieee{12} benchmarks (VA, GEMV, SEL, UNI, BS, TS, 
HST-S, HST-L, RED, SCAN-SSA, SCAN-RSS, and TRNS), the 640-DPU system provides an energy savings of \jieee{5.23$\times$} over the CPU. 
The maximum energy savings is 39.14$\times$ for UNI. 
Our experiments show that the 640-DPU system, featuring PIM-enabled memory with a capacity of 40 GB, provides outstanding energy savings over a modern Intel Xeon CPU (with memory capacity of 32 GB) for \jieee{12} out of 16 benchmarks.
This energy savings comes from the lower execution times of these \jieee{12} benchmarks on the 640-DPU system (Figure~\ref{fig:comparison-perf}).
We expect that the energy savings of the 2,556-DPU system, with $\sim$6$\times$ more DPUs, 160 GB of PIM-enabled memory, and higher frequency (350 vs. 267 MHz), over the CPU will be even higher due to higher performance (thus, lower static energy) and less data movement.

Second, the 640-DPU system is only less energy efficient than the CPU for SpMV, BFS, and NW, {which is} in line with our observations about performance (Section~\ref{sec:comparison-perf}).

Third, compared to the GPU, the 640-DPUs system consumes less energy for BS and HST-S, {since these are the two benchmarks for which the 640-DPU system outperforms the GPU (see Section~\ref{sec:comparison-perf}).} 
{For the 2,556-DPU system, we expect energy results to follow the performance results in Section~\ref{sec:comparison-perf}. 
The 10 benchmarks (VA, SEL, UNI, BS, HST-S, HST-L, RED, SCAN-SSA, SCAN-RSS, and TRNS) that run faster on the 2,556-DPU system than on the GPU will also likely consume less energy. 
This is because the major cause of performance improvement and energy reduction is the same: the reduction in data movement between memory and processors that the UPMEM-based PIM systems provide.}

\gboxbegin{\rtask{ko}}
\textbf{The UPMEM-based PIM system {provides large} energy savings over {a modern} CPU} due to higher performance (thus, {lower} static energy) and less data movement between memory and processors. 

{\textbf{The UPMEM-based PIM system provides energy savings over a modern CPU/GPU on workloads where it outperforms the CPU/GPU}. 
This is because the source of both performance improvement and energy savings is the same: \textbf{the significant reduction in data movement between the memory and the processor cores}, which the UPMEM-based PIM system can provide for PIM-suitable workloads.}
\gboxend

\subsubsection{Discussion}
\label{sec:comparison-discussion}

These observations are useful for programmers to anticipate how much performance and energy savings they can get from the UPMEM hardware compared to traditional {CPU and GPU systems} for different types of workloads.

One limitation of this comparison is the difficulty of establishing a common control factor across all three types of systems (CPU, GPU, and UPMEM-based PIM system) to ensure a fair comparison.
To this end, the 640-DPU PIM system has comparable memory capacity to the CPU (40 GB vs. 32 GB). 
However, the 2,556-DPU system has much higher memory capacity ($\sim$160 GB).
On the other hand, the {640-DPU} UPMEM-based PIM system and the GPU have comparable cost (the 640-DPU system being a little cheaper). 
Other {hardware} characteristics, such as fabrication technology, process node, number of cores, or frequency (Table~\ref{tab:comparison}), are very different {across the four systems that we evaluate in Section~\ref{sec:comparison}}.

{We note} that the UPMEM hardware is still maturing and is expected to run at a higher frequency in the near future (400-450 MHz instead of 350 or 267 MHz) and potentially be manufactured with a smaller technology node~\cite{comm-upmem}. 
Hence, the results we report in this comparison may underestimate the full potential of the UPMEM-based PIM architecture. 
CPU and GPU systems have been heavily optimized for decades in terms of architecture, software, and manufacturing. 
We believe the architecture, software, and manufacturing of PIM systems will continue to improve (see our suggestions for future improvement in Section~\ref{sec:discussion}).





\ignore{
{Our first observation is that the UPMEM PIM system consumes on average $6.32\times$ less energy than the CPU. 
We also observe that the GPU can achieve more energy savings, but we expect that this difference reduces as the technology node of the UPMEM chip decreases.}
\todo{Revise.}

\ie{Include comment about fairness and that UPMEM results are underestimated (see Comment 15)}
}

\section{Key Takeaways}\label{sec:discussion}

In this section, we reiterate several key empirical observations {in the form of} four key takeaways we {provide} throughout this paper. {We also provide} implications on workload suitability and good programming practices for the UPMEM PIM architecture, and suggestions for hardware and architecture designers of future PIM systems.


\noindent\paragraph{\textbf{Key Takeaway \#1}.}
\textbf{The UPMEM PIM architecture is fundamentally compute bound}.
Section~\ref{sec:mram-bandwidth} shows that workloads with more complex operations than integer addition fully utilize the instruction pipeline before they can potentially saturate the memory bandwidth.
Section~\ref{sec:throughput-oi} shows that even workloads with as simple operations as integer addition saturate the compute throughput with an operational intensity as low as 0.25 operations/byte (1 addition per integer accessed).
This {key takeaway} shows that \textbf{the most suitable workloads for the UPMEM PIM architecture are memory-bound workloads}. 
From a programmer's perspective, the architecture requires a shift in how we think about computation and data access, since the relative cost of computation vs. data access in the PIM system is very different from that in {the dominant processor-centric architectures of today}.

\tboxbegin{\ttask{kt}}
\textbf{The UPMEM PIM architecture is fundamentally compute bound}. 
As a result, \textbf{the most suitable workloads are memory-bound}.
\tboxend

\noindent\paragraph{\textbf{Key Takeaway \#2}.}
\textbf{The workloads most well-suited for the UPMEM PIM architecture are those with simple or no arithmetic operations}. {This is because} DPUs include native support for \emph{only} integer addition/subtraction and bitwise operations. More complex integer (e.g., multiplication, division) and floating point operations are implemented using software library routines.
Section~\ref{sec:arith-throughput} shows that the arithmetic throughput of more complex integer operations and floating point operations are an order of magnitude lower than that of simple addition and subtraction.
Section~\ref{sec:comparison} shows that benchmarks with little amount of computation and no use of multiplication, division, or floating point operations (10 out of 16 benchmarks) 
run faster (\jieee{2.54$\times$} on average) on a 2,556-DPU system than on a {modern NVIDIA} Titan V GPU. 
These observations show that \textbf{the workloads most well-suited for the UPMEM PIM architecture are those with no arithmetic operations or simple operations (e.g., bitwise operations and integer addition/subtraction)}.
Based on this {key takeaway}, we recommend devising {much} more efficient software library routines or, {more importantly, specialized and fast} in-memory hardware for complex operations in future PIM architecture generations {to improve the general-purpose performance of PIM systems}.

\tboxbegin{\ttask{kt}}
\textbf{The most well-suited workloads for the UPMEM PIM architecture use no arithmetic operations or use only simple operations (e.g., bitwise operations and integer addition/subtraction).}
\tboxend

\noindent\paragraph{\textbf{Key Takeaway \#3}.}
\textbf{The workloads most well-suited for the UPMEM PIM architecture are those with little global communication}, 
because there is no direct communication channel among DPUs. 
As a result, there is a huge disparity in performance scalability of benchmarks that do \emph{not} require inter-DPU communication and benchmarks that do (especially if parallel transfers across MRAM banks cannot be used), as Section~\ref{sec:performance} shows. 
This {key takeaway} shows that \textbf{the workloads most well-suited for the UPMEM PIM architecture are those with little or no inter-DPU communication}.
Based on this {takeaway}, we recommend that the {hardware} architecture and the software stack be enhanced with support for inter-DPU communication (e.g., by leveraging new in-DRAM data copy techniques~\cite{seshadri2018rowclone,seshadri2013rowclone, wang2020figaro, rezaei2020nom, chang.hpca16, seshadri2020indram, seshadri.bookchapter17} {and providing better connectivity inside DRAM~\cite{chang.hpca16, rezaei2020nom}}).

\tboxbegin{\ttask{kt}}
\textbf{The most well-suited workloads for the UPMEM PIM architecture require little or no communication {across DRAM Processing Units (inter-DPU communication)}.}
\tboxend

\noindent\paragraph{{\textbf{Summary}.}} 
We find that the workloads most suitable for the UPMEM PIM architecture in its current form are (1) memory-bound workloads with (2) simple or no arithmetic operations and (3) little or no inter-DPU communication. 

\noindent\paragraph{\textbf{Key Takeaway \#4}.}
We observe that the existing UPMEM-based PIM systems greatly improve energy efficiency and performance over modern CPU and GPU systems across many workloads we examine. Section~\ref{sec:comparison} shows that the 2,556-DPU and the 640-DPU systems are \jieee{23.2$\times$ and 10.1$\times$} faster, respectively, than a modern Intel Xeon CPU, {averaged across the entire set of 16 PrIM benchmarks}. 
The 640-DPU system is \jieee{1.64$\times$} more energy efficient than the CPU, averaged across the entire set of 16 PrIM benchmarks, and \jieee{5.23$\times$} more energy efficient for \jieee{12} of the PrIM benchmarks. 

The 2,556-DPU system is faster (on average by \jieee{2.54$\times$}) than the modern GPU in 10 out of 16 PrIM benchmarks, which have three key characteristics that define a workload's PIM suitability: (1) streaming memory accesses, (2) {little or no} inter-DPU communication, and (3) {little or no} use of multiplication, division, or floating point operations.\footnote{{Note that these three characteristics are not exactly the same three characteristics highlighted by key takeaways \#1 to \#3, but more specific. The difference is that the 2,556-DPU system outperforms the GPU for memory-bound workloads with streaming memory accesses. {These workloads do not need to have \emph{only} streaming memory accesses}, since BS, HST-S, HST-L, and TRNS, {for which the 2,556-DPU system outperforms the GPU,} have also random accesses. Since all PrIM workloads (see Table~\ref{tab:benchmarks}) contain some streaming memory accesses, we cannot say that the 2,556-DPU system outperforms the GPU for workloads with \emph{only} strided and/or random accesses.}}

We expect that the 2,556-DPU system will {provide even higher} performance and energy {benefits}, and that future PIM systems {will be even better (especially after implementing our recommendations for future PIM hardware).} 
If the architecture is improved based on our recommendations under Key Takeaways 1-3, we believe the {future} PIM system will be even more attractive, leading to much higher performance and energy benefits {versus modern CPUs and GPUs} over potentially all workloads.

\tboxbegin{\ttask{kt}}
\begin{itemize}[wide, labelsep=0.5em]
\item UPMEM-based PIM systems \textbf{outperform modern CPUs in terms of performance and energy efficiency on most of PrIM benchmarks}. 
\item UPMEM-based PIM systems \textbf{outperform modern GPUs on a majority of PrIM benchmarks}, and the outlook is even more positive for future PIM systems.
\item {UPMEM-based PIM systems are \textbf{more energy-efficient than modern CPUs and GPUs on workloads that they provide performance improvements} over the CPUs and the GPUs.}
\end{itemize}
\tboxend

\section{Related Work}

To our knowledge, {this paper} provides the first comprehensive {characterization and} analysis of the first publicly-available real-world PIM architecture {along with} the first {open-source} benchmark suite for a real-world PIM architecture.

We briefly review related work on PIM architectures.

There are two main approaches to PIM~\cite{mutlu2019,mutlu2020modern,ghoseibm2019,ghose2019arxiv}: (1) \emph{processing-using-memory} (\emph{PUM}) and (2) \emph{processing-near-memory} (\emph{PNM}). 
No prior work on PUM or PNM provides results from real commercial systems or a benchmark suite to evaluate PIM architectures.

\textbf{Processing using memory (PUM)} exploits the existing memory architecture and the operational principles of the memory cells and circuitry to perform operations within each memory chip at low cost. 
Prior works propose PUM mechanisms using SRAM~\cite{aga.hpca17,eckert2018neural,fujiki2019duality,kang.icassp14}, DRAM~\cite{seshadri.micro17,seshadri.arxiv16,seshadri2018rowclone,seshadri2013rowclone,angizi2019graphide,kim.hpca18,kim.hpca19,gao2020computedram,chang.hpca16,xin2020elp2im,li.micro17,deng.dac2018,hajinazarsimdram,rezaei2020nom,wang2020figaro,ali2019memory,seshadri2020indram, seshadri.bookchapter17,seshadri.thesis16,Seshadri:2015:ANDOR,seshadri.bookchapter17.arxiv,ferreira2021pluto,olgun2021quactrng, olgun2021pidram}, 
PCM~\cite{li.dac16}, 
MRAM~\cite{angizi2018pima,angizi2018cmp,angizi2019dna}, 
or RRAM/memristive~\cite{levy.microelec14,kvatinsky.tcasii14,shafiee2016isaac,kvatinsky.iccd11,kvatinsky.tvlsi14,gaillardon2016plim,bhattacharjee2017revamp,hamdioui2015memristor,xie2015fast,hamdioui2017myth,yu2018memristive,puma-asplos2019, ankit2020panther,chi2016prime,song2018graphr,zheng2016tcam,xi2020memory, yavits2021giraf} memories. 
PUM mechanisms enable different types of operations such as data copy and initialization~\cite{chang.hpca16,seshadri2018rowclone,seshadri2013rowclone,aga.hpca17,rezaei2020nom,wang2020figaro, seshadri.bookchapter17,olgun2021pidram}, 
bulk bitwise operations (e.g., a functionally-complete set of Boolean logic operations)~\cite{seshadri.micro17,seshadri.arxiv16,li.dac16,angizi2018pima,angizi2018cmp,angizi2019dna,aga.hpca17,li.micro17,mandelman.ibmjrd02,xin2020elp2im,gao2020computedram, seshadri2020indram, Seshadri:2015:ANDOR}, 
and simple arithmetic operations (e.g., addition, multiplication, implication)~\cite{levy.microelec14,kvatinsky.tcasii14,aga.hpca17,kang.icassp14,li.micro17,shafiee2016isaac,eckert2018neural,fujiki2019duality,kvatinsky.iccd11,kvatinsky.tvlsi14,gaillardon2016plim,bhattacharjee2017revamp,hamdioui2015memristor,xie2015fast,hamdioui2017myth,yu2018memristive,deng.dac2018,angizi2019graphide,ferreira2021pluto}. 
A recent work, called SIMDRAM~\cite{hajinazarsimdram}, designs a framework for implementing and executing arbitrary operations in a {bit-serial} SIMD fashion inside DRAM arrays, {building on the Ambit substrate~\cite{seshadri.micro17,seshadri.arxiv16}}.

\textbf{Processing near memory (PNM)} integrates processing elements (e.g., functional units, accelerators, simple processing cores, reconfigurable logic) near or inside the memory (e.g.,~\cite{syncron,fernandez2020natsa,cali2020genasm,alser2020accelerating,kim.arxiv17,kim.bmc18,ahn.pei.isca15,ahn.tesseract.isca15,boroumand.asplos18,boroumand2019conda,boroumand.arxiv17,boroumand2016pim,singh2019napel,asghari-moghaddam.micro16,DBLP:conf/sigmod/BabarinsaI15,farmahini2015nda,gao.pact15,DBLP:conf/hpca/GaoK16,gu.isca16,guo2014wondp,hashemi.isca16,cont-runahead,hsieh.isca16,kim.isca16,kim.sc17,DBLP:conf/IEEEpact/LeeSK15,liu-spaa17,morad.taco15,nai2017graphpim,pattnaik.pact16,pugsley2014ndc,zhang.hpdc14,zhu2013accelerating,DBLP:conf/isca/AkinFH15,gao2017tetris,drumond2017mondrian,dai2018graphh,zhang2018graphp,huang2020heterogeneous,zhuo2019graphq,santos2017operand,boroumand2021polynesia,boroumand2021mitigating,besta2021sisa,lloyd2015memory,landgraf2021combining,rodrigues2016scattergather,lloyd2018dse,lloyd2017keyvalue,gokhale2015rearr,nair2015active,jacob2016compiling,balasubramonian2014near,lee2022isscc, ke2021near, kwon202125, lee2021hardware, asgarifafnir, herruzo2021enabling, singh2021fpga, singh2021accelerating, oliveira2021pimbench, boroumand2021google, boroumand2021google_arxiv}). 
Many of these PNM works place PIM logic inside the logic layer of 3D-stacked memories~\cite{syncron,cali2020genasm,alser2020accelerating,kim.arxiv17,kim.bmc18,ahn.pei.isca15,ahn.tesseract.isca15,boroumand.asplos18,boroumand2019conda,boroumand.arxiv17,boroumand2016pim,singh2019napel,guo2014wondp,hsieh.isca16,kim.isca16,kim.sc17,liu-spaa17,nai2017graphpim,pattnaik.pact16,pugsley2014ndc,zhang.hpdc14,DBLP:conf/isca/AkinFH15,gao2017tetris,drumond2017mondrian,dai2018graphh,zhang2018graphp,huang2020heterogeneous,zhuo2019graphq,boroumand2021polynesia,boroumand2021mitigating,lloyd2015memory,lloyd2018dse,gokhale2015rearr,nair2015active,jacob2016compiling,sura2015data,balasubramonian2014near,oliveira2021pimbench, boroumand2021google, boroumand2021google_arxiv}, at the memory controller~\cite{hashemi.isca16,cont-runahead}, on the DDRX DIMMs~\cite{ke2021near,asghari-moghaddam.micro16,alves2015opportunities,medal2019}, or in the same package as the CPU connected via silicon interposers~\cite{fernandez2020natsa,singh2020nero,singh2021fpga, singh2021accelerating}.

Another body of recent works study and propose solutions to system integration challenges in PIM-enabled systems, such as memory coherence~\cite{boroumand.arxiv17,boroumand2016pim, boroumand2019conda}, virtual memory~\cite{impica,hajinazar2020virtual}, synchronization~\cite{syncron}, or PIM suitability of workloads~\cite{deoliveira2021,oliveira2021pimbench}.

Several works explore the acceleration opportunities offered by the UPMEM PIM architecture for bioinformatics~\cite{lavenier2016,lavenier2020}, skyline computation~\cite{Zois2018}, compression~\cite{nider2020}, or sparse linear algebra~\cite{giannoula2022sparsep}. 
Readers can refer to these works for in-depth analysis of specific applications on the UPMEM PIM architecture. 
Our work is the first one that performs a comprehensive architecture characterization of the UPMEM PIM architecture and studies the PIM suitability of a large number of workloads. 
We are also the first to openly {and freely} provide the first benchmark suite for real PIM systems.

A recent work~\cite{kwon202125,lee2021hardware} presents a real-world PIM system with programmable near-bank {computation} units, called FIMDRAM, based on HBM technology~\cite{jedec.hbm.spec,lee.taco16}. The FIMDRAM architecture, {designed specifically for} machine learning applications, implements a SIMD pipeline with simple multiply-and-accumulate units~\cite{shin2018mcdram,cho2020mcdram}. 
More recently presented, Accelerator-in-Memory~\cite{lee2022isscc} is a GDDR6-based PIM architecture with specialized units for multiply-and-accumulate and activation functions for deep learning applications. 
AxDIMM~\cite{ke2021near} is a DIMM-based solution which places an FPGA fabric in the buffer chip of the DIMM. It has been tested for recommendation inference. 
Compared to the {more general-purpose} UPMEM PIM architecture, 
these architectures focus 
on a specific domain of applications (i.e., machine learning), and thus it may lack flexibility {to support} a wider range of applications. 
A comprehensive characterization and analysis of these architectures, 
along the lines of our work, can greatly help researchers, programmers, and architects to understand their potential.

\section{Summary \& Conclusion}

We present the first comprehensive {characterization and} analysis of a real commercial PIM architecture. 
{Through this analysis, we develop a} rigorous, thorough understanding of the UPMEM PIM architecture, the first publicly-available PIM architecture, and its suitability to various {types of} workloads.

{First,} we conduct a characterization of the UPMEM-based PIM system using microbenchmarks to assess various architecture limits such as compute throughput and memory bandwidth, yielding new insights. 
Second, we present PrIM, a benchmark suite of 16 memory-bound workloads from different application domains (e.g., {dense/sparse linear algebra, databases, data analytics, graph processing, neural networks, bioinformatics, image processing}).

Our extensive evaluation of PrIM benchmarks conducted on two real systems with UPMEM memory modules provides new insights about suitability of different workloads to the PIM system, programming recommendations for software designers, and suggestions and hints for hardware and architecture designers of future PIM systems. 
We compare the performance and energy consumption of the UPMEM-based PIM systems for PrIM benchmarks to those of a modern CPU and a modern GPU, and identify key workload characteristics that can successfully leverage the key strengths of a real PIM system over conventional processor-centric architectures.
We note that we compare the first ever commercial PIM system to CPU and GPU systems that have been heavily optimized for decades in terms of architecture, software, and manufacturing. As the architecture, software, and manufacturing of PIM systems continue to improve, it will be possible to do more fair comparisons to CPU and GPU systems, which reveal even higher benefits for PIM systems in the future.

We believe and hope that our work will provide valuable insights to programmers, users and architects of this PIM architecture as well as of future PIM systems, and will represent an enabling milestone in the development of memory-centric computing systems.


\begin{acks}
    We thank UPMEM's Fabrice Devaux, Rémy Cimadomo, Romaric Jodin, and Vincent Palatin for their valuable support.
    We acknowledge the support of SAFARI Research Group's industrial partners, especially ASML, Facebook, Google, Huawei, Intel, Microsoft, VMware, and the Semiconductor Research Corporation.
    Izzat El Hajj acknowledges the support of the University Research Board of the American University of Beirut (URB-AUB-103951-25960).

This article is a greatly extended version of~\cite{gomez2021cut}, a summary of our work presented at the 2021 12th International Green and Sustainable Computing Conference (IGSC). 
Talk videos for this work are available on YouTube, including a 3-minute talk video (\url{https://youtu.be/SrFD_u46EDA}), a 20-minute talk video (\url{https://youtu.be/Pp9jSU2b9oM}), a 1-hour talk video (\url{https://youtu.be/6Ws3h_CQO_Q}), and a long talk video (\url{https://youtu.be/D8Hjy2iU9l4}).
    
\end{acks}


{
  \bstctlcite{IEEEexample:BSTcontrol} 
  \let\OLDthebibliography\thebibliography
  \renewcommand\thebibliography[1]{
    \OLDthebibliography{#1}
    \setlength{\parskip}{0pt}
    \setlength{\itemsep}{0pt}
  }
  \bibliographystyle{IEEEtran}
  \bibliography{references}
}

\section{{\Large APPENDIX}}

This appendix presents some additional results for one of our microbenchmarks (Section~\ref{app:throughput-oi}) and four of the PrIM benchmarks (Section~\ref{sec:appendix-results}). 
Section~\ref{sec:appendix-comparison} shows the sources of the CPU and GPU versions of PrIM benchmarks.

\subsection{Arithmetic Throughput versus Number of Tasklets}
\label{app:throughput-oi}
\begin{figure*}[h]
        \centering
        \centering
        \includegraphics[width=1.0\linewidth]{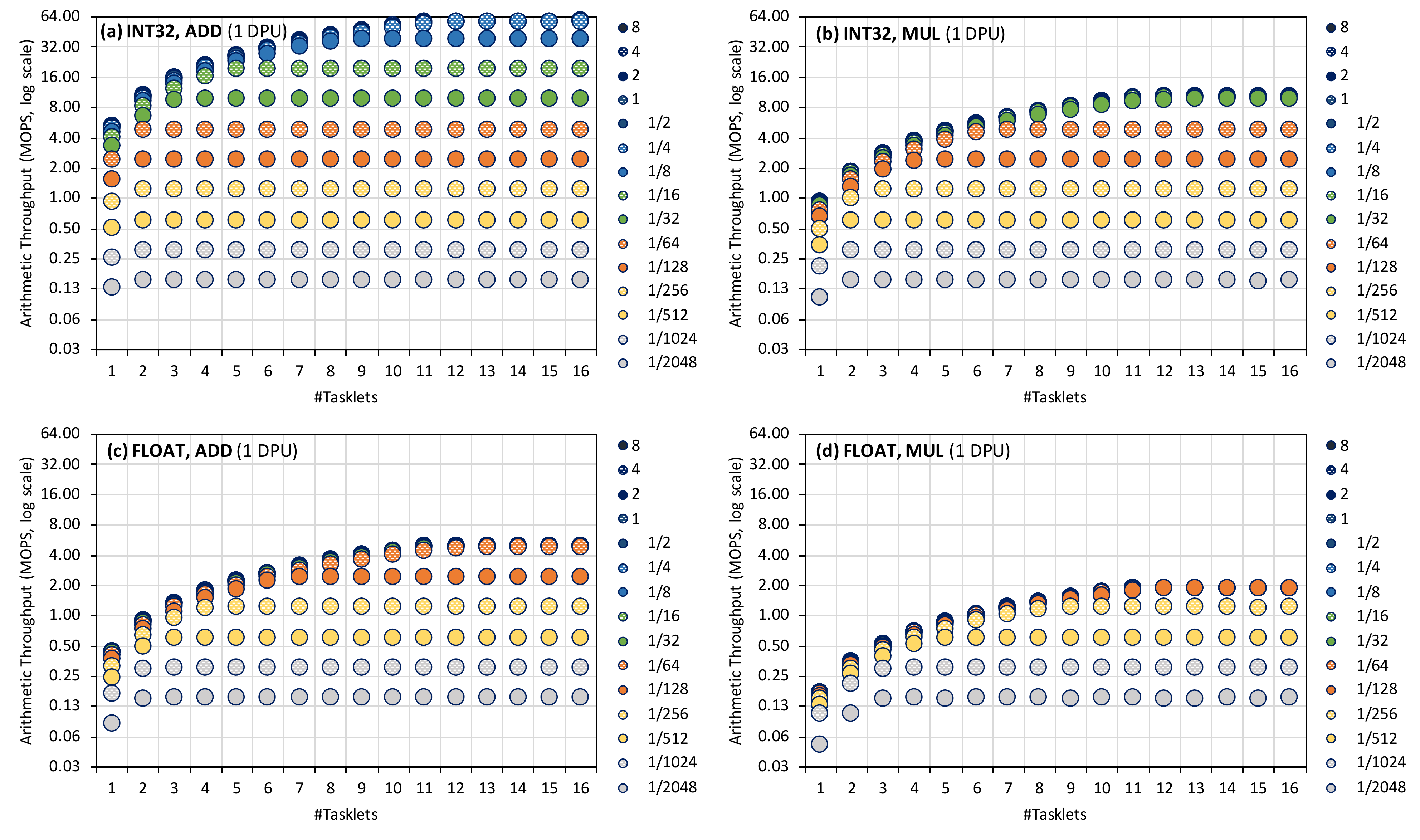}
        \vspace{-2mm}
        \captionof{figure}{Arithmetic throughput versus number of tasklets for different operational intensities of (a) 32-bit integer addition, (b) 32-bit integer multiplication, (c) 32-bit floating point addition, and (d) 32-bit floating point multiplication. The legend shows the operational intensity values (in OP/B). The y-axis is log scale.}
        \label{fig:ai-dpu-tasklets}
\end{figure*}

Figure~\ref{fig:ai-dpu-tasklets} presents arithmetic throughput results for different numbers of tasklets at different operational intensities. 
This figure shows a different view of the experimental results presented in Figure~\ref{fig:ai-dpu}, with the {goal} of showing the variation in arithmetic throughput for different operational intensities.

We make two key observations from Figure~\ref{fig:ai-dpu-tasklets}.

First, for any data type and operation, the highest possible throughput is achieved at 11 tasklets, i.e., the number of tasklets to fully utilize the pipeline. 
However, the operational intensity at which the highest throughput value is reached depends on the actual data type and operation. 
For example, the highest throughput of 32-bit integer addition is achieved at $\frac{1}{4}$~OP/B, i.e., 1 addition per 32-bit element. For floating point multiplication, {the highest throughput} is achieved at $\frac{1}{128}$~OP/B, i.e., 1 multiplication every 32 32-bit elements.

Second, for lower operational intensities, the number of tasklets necessary to reach the saturation throughput is less than 11. 
This happens in the memory-bound regions, where the MRAM access latency dominates the overall latency. This observation is in line with our observations for COPY and ADD benchmarks in Section~\ref{sec:mram-streaming}.

\subsection{Extended results for Needleman-Wunsch, image histogram, reduction, and scan 
}
\label{sec:appendix-results}

This section presents some additional results for four of the PrIM benchmarks.
First, we present an extended evaluation of NW (Section~\ref{app:nw}).
Second, we compare HST-S and HST-L for different histogram sizes (Section~\ref{app:histogram}).
Third, we show an evaluation of RED with three different mechanisms to perform local intra-DPU reduction (Section~\ref{app:reduction}). 
Fourth, we compare SCAN-SSA and SCAN-RSS for different array sizes (Section~\ref{app:scan}). 

\subsubsection{Needleman-Wunsch}
\label{app:nw}

\begin{figure}[h]
\includegraphics[width=\linewidth]{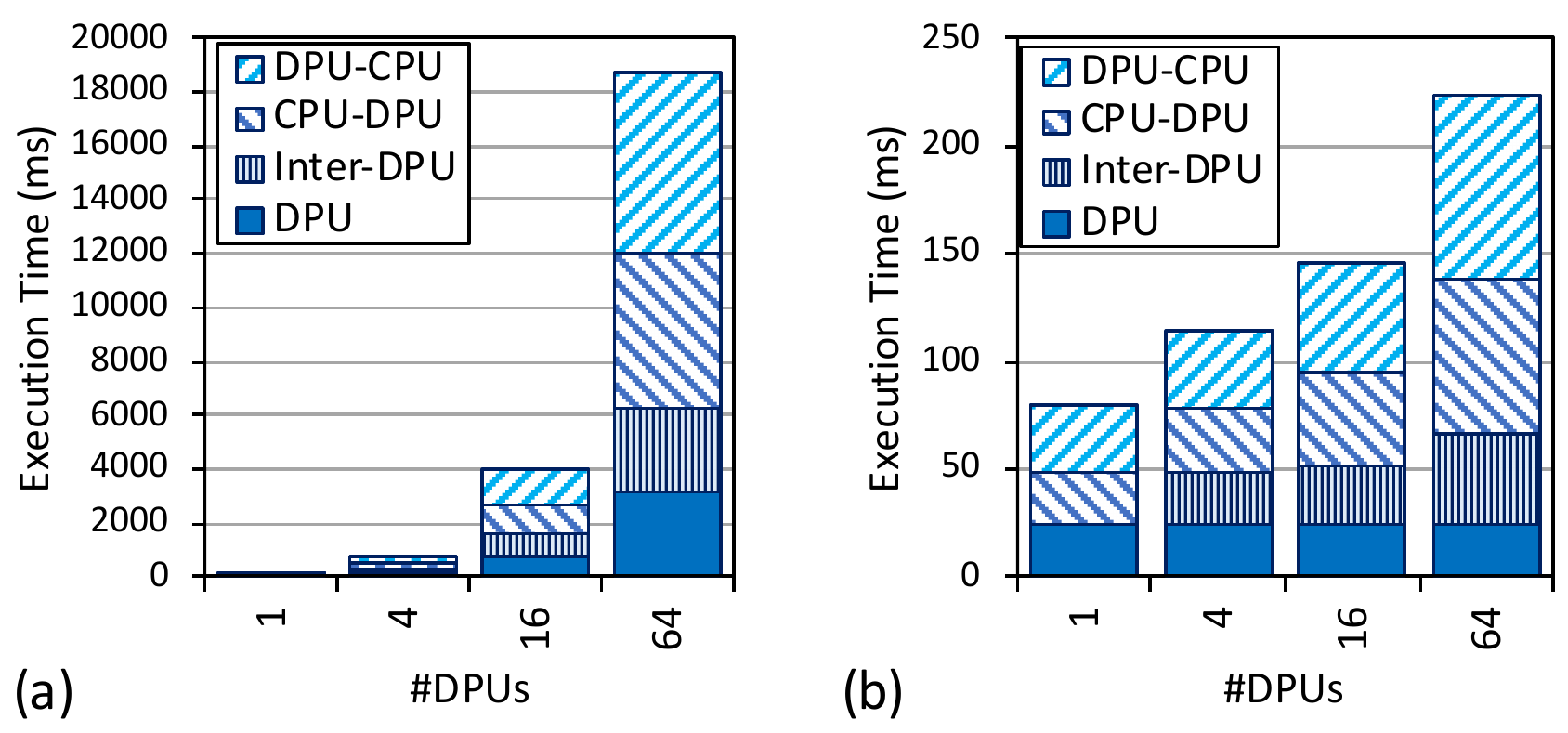}
\vspace{-2mm}
\caption{Weak scaling evaluation of NW: (a) complete execution of NW, (b) execution of the longest diagonal.} \label{fig:appendix_nw}
\end{figure}

We present additional results for the weak scaling experiment of NW. 
In this experiment, we increase the length of the sequences to align proportionally to the number of DPUs. Thus, the size of the 2D score matrix increases quadratically with the number of DPUs.
Figure~\ref{fig:appendix_nw} shows weak scaling results of (a) the complete execution of NW (including all iterations) and (b) the execution of only the longest diagonal.

We make two observations from Figure~\ref{fig:appendix_nw}. 
First, the execution times on the DPUs for the complete execution (Figure~\ref{fig:appendix_nw}a) increase with the number of DPUs, since the size of the problem (the 2D score matrix) increases quadratically. We make the same observation in Section~\ref{sec:weak}.
Second, the execution times on the DPUs for the longest diagonal (Figure~\ref{fig:appendix_nw}b) remain flat as the number of DPUs increases. The reason is that the length of the longest diagonal increases linearly with the length of the sequences and the number of DPUs. As a result, we observe linear weak scaling for the longest diagonal.

These results show (1) {that} a larger number of active DPUs is {more} beneficial for NW {in the computation of the longest diagonals of the 2D score matrix}, and (2) why we do not observe linear scaling for the complete NW.

\subsubsection{Image Histogram}
\label{app:histogram}

\begin{figure}[h]
\includegraphics[width=\linewidth]{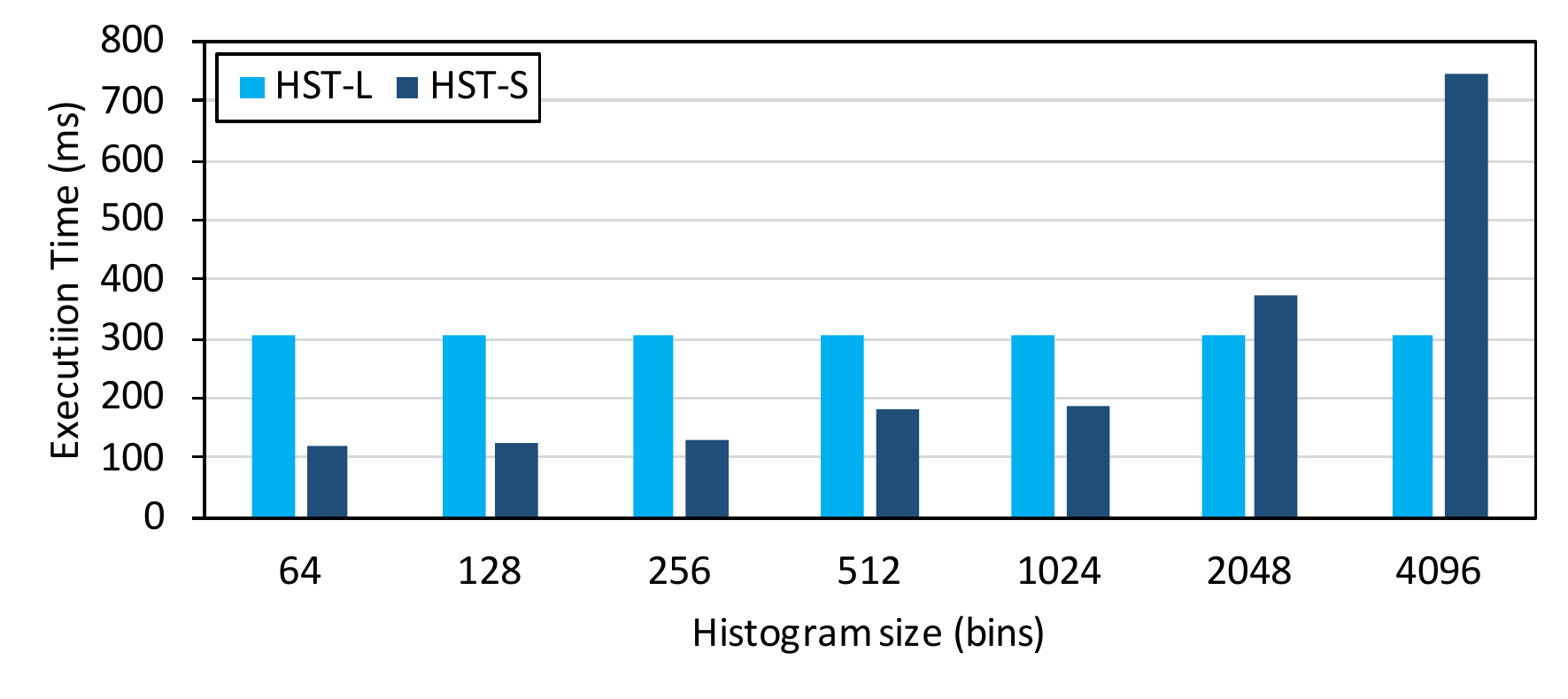}
\vspace{-2mm}
\caption{{Execution times (ms) of} two versions of histogram (HST-L, HST-S) on 1 DPU.} \label{fig:appendix_histo}
\end{figure}

We present results for different histogram sizes for our two versions of histogram (HST-S, HST-L). 
Figure~\ref{fig:appendix_histo} shows the execution time results for histogram sizes between 64 and 4096. The input is the one specified in Table~\ref{tab:datasets}, which is an image of 12-bit depth (thus, maximum histogram size is 4096).

The results show that HST-S is 
$1.6-2.5\times$ faster than HST-L for histograms between 64 and 1024 bins. 
The performance of HST-S gets worse when increasing the histogram size because the number of tasklets that it is possible to {run on a DPU} reduces. For example, for 512 bins, only 8 tasklets can be launched because of the limited amount of WRAM (each tasklet has its own local histogram). 
For 4046 bins, HST-S can only launch 2 tasklets. 
After 2048 bins, HST-L performs faster, as its execution time is independent of the histogram size.

\subsubsection{Reduction}
\label{app:reduction}
\begin{figure}[h]
\includegraphics[width=\linewidth]{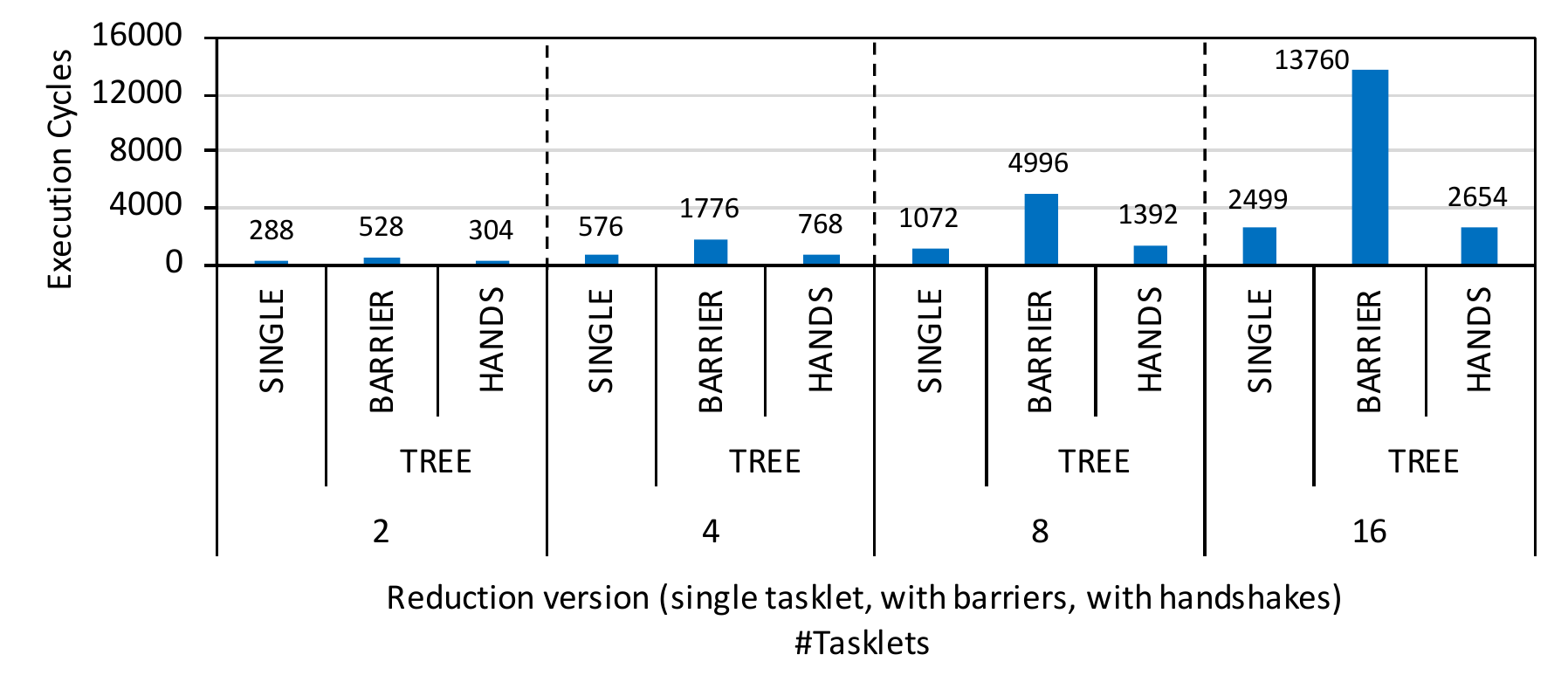}
\vspace{-2mm}
\caption{{Effect of} sequential reduction (SINGLE) vs. parallel tree-based reductions (BARRIER, HANDS), {in the second step of the RED benchmark}.} \label{fig:appendix_red_sync}
\end{figure}

\begin{figure}[h]
    \centering
    \includegraphics[width=\linewidth]{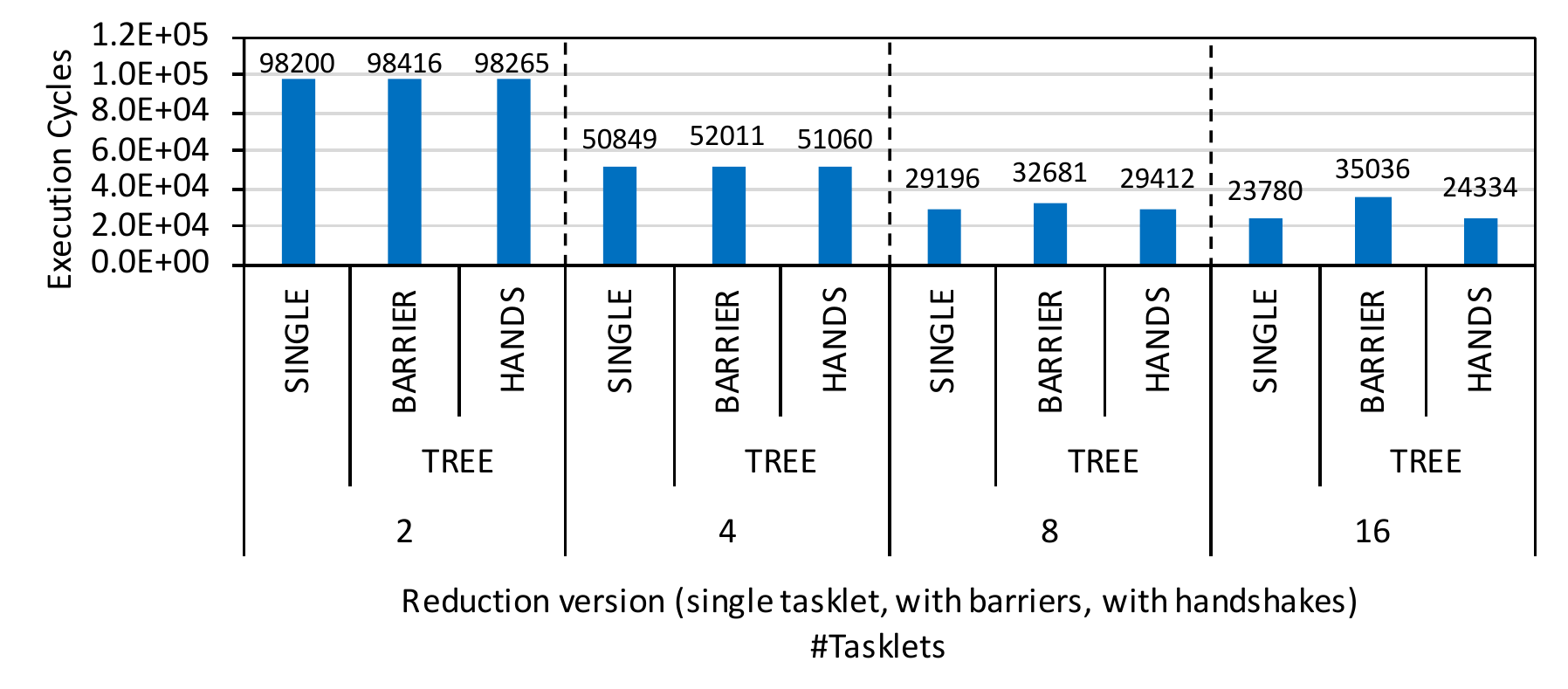}
    \vspace{-2mm}
    \caption{{Execution cycles of} three versions of reduction of 2K 64-bit elements on 1 DPU.} \label{fig:appendix_red_2K}
\end{figure}

\begin{figure}[h]
    \centering
    \includegraphics[width=\linewidth]{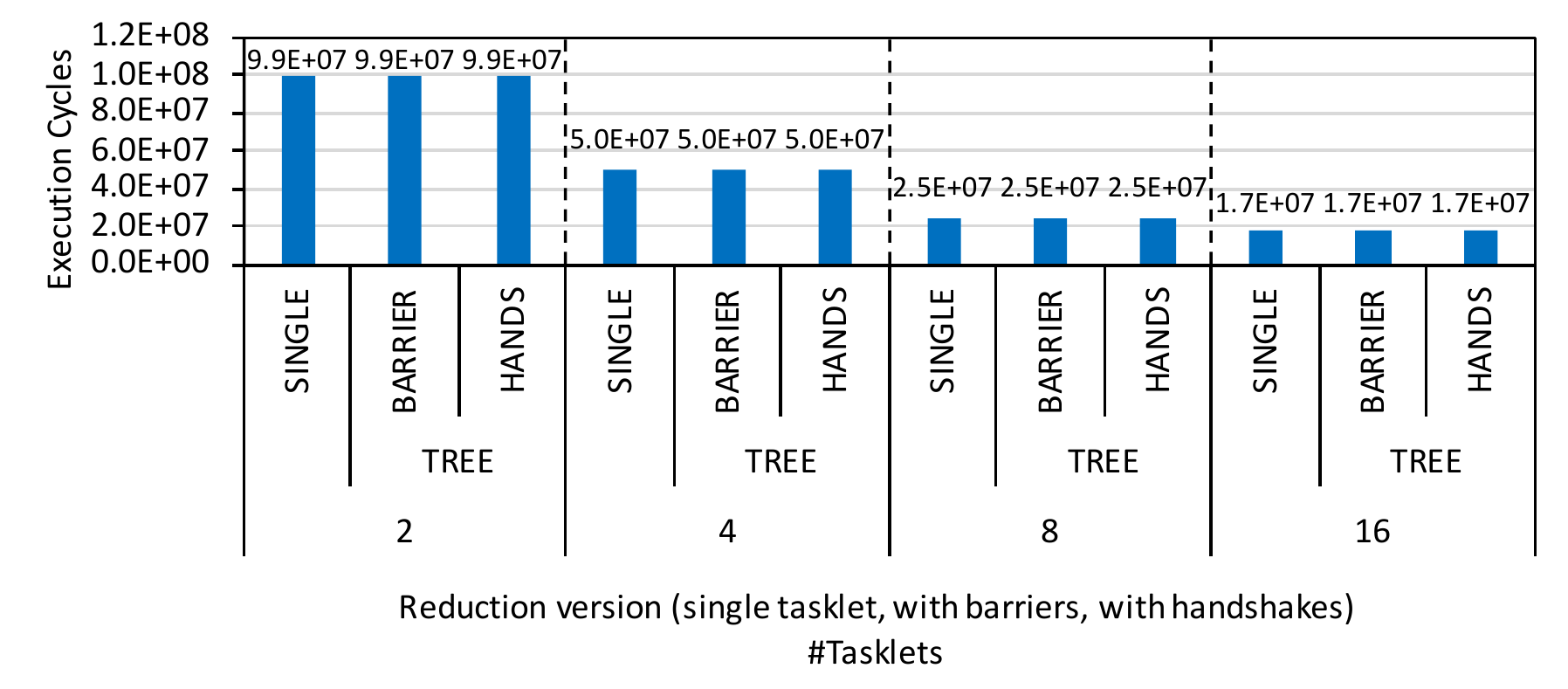}
    \vspace{-2mm}
    \caption{{Execution cycles of} three versions of reduction of 2M 64-bit elements on 1 DPU.} \label{fig:appendix_red_2M}
\end{figure}

We compare three versions of RED that we introduce in Section~\ref{sec:reduction}. 
Recall that RED has two steps. In the first step, each tasklet accumulates the values of an assigned chunk of the input array. In the second step, RED performs the final reduction of the local sums of all tasklets. 
The difference between the three versions {is in how the second step is implemented}. 
The first version uses a single tasklet to perform a sequential reduction in the second step {(SINGLE in Figures~\ref{fig:appendix_red_sync} to~\ref{fig:appendix_red_2M})}. 
The other two versions implement a parallel tree-based reduction in the second step {(see Section~\ref{sec:reduction})}. The only difference {between the other two versions} is the synchronization primitive used for synchronization at the end of each tree level: (1) barriers for all tasklets {(BARRIER in Figures~\ref{fig:appendix_red_sync} to~\ref{fig:appendix_red_2M})}, or (2) handshakes between pairs of tasklets {(HANDS in Figures~\ref{fig:appendix_red_sync} to~\ref{fig:appendix_red_2M})}. 
Figure~\ref{fig:appendix_red_sync} shows the number of execution cycles needed to perform sequential (SINGLE) or the parallel tree-based (BARRIER, HANDS) reduction for 2 to 16 tasklets on one DPU.

We observe that the most efficient of the three versions is the sequential reduction (SINGLE). 
However, it is only a few cycles faster {(6\% faster with 16 tasklets)} that the tree-based version with handshakes (HANDS).
We also observe the high cost of barriers when the number of tasklets increases. 
These results indicate that synchronization primitives impose high overhead in the current implementation of the UPMEM PIM architecture.
Nevertheless, the relative weight of the final reduction is negligible when the input array is large. 
Figure~\ref{fig:appendix_red_2K} shows the execution cycles of the three versions for an input array of 2K 64-bit elements with 2-16 tasklets on one DPU. The difference between the three versions is very small, but we still observe that SINGLE is {slightly} faster {(i.e., 2\% over HANDS, and 47\% over BARRIER)}.

For an array of 2M 64-bit elements (Figure~\ref{fig:appendix_red_2M}), the difference in performance {of the three versions} is completely negligible, since most of the execution cycles are spent in the first step of RED.

\subsubsection{Prefix Sum (Scan)}
\label{app:scan}
\begin{figure}[h]
    \centering
    \includegraphics[width=\linewidth]{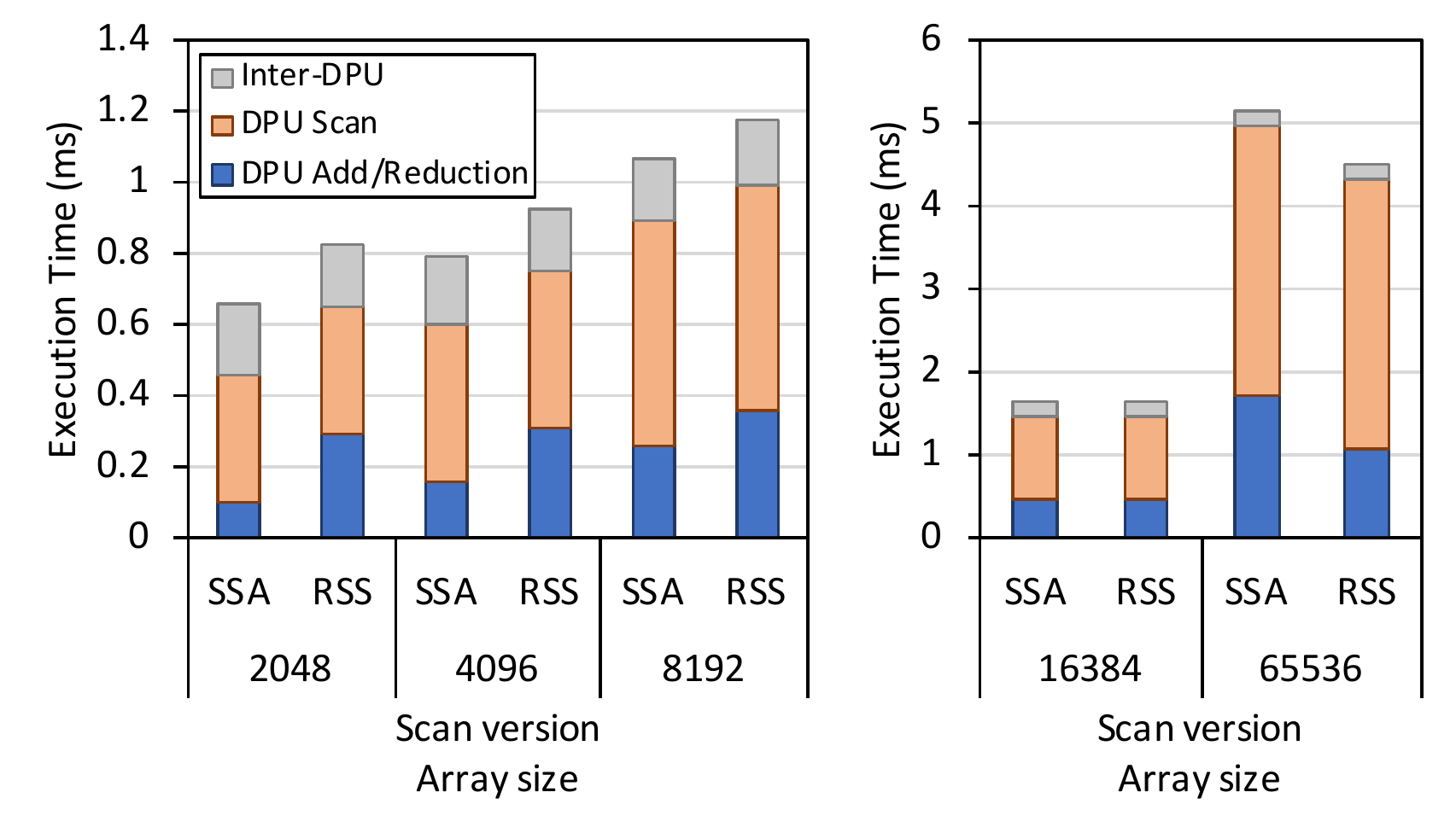}
    \vspace{-2mm}
    \caption{Two versions of scan (SCAN-SSA, SCAN-RSS) on 1 DPU.} \label{fig:appendix_scan}
\end{figure}

We compare the execution time of our two versions of scan, SCAN-SSA and SCAN-RSS, for different array sizes (2048, 4096, 8192, 16384, 65536 elements) {on the DPU}. 
Figure~\ref{fig:appendix_scan} shows the execution time results. For both versions, the figure shows the breakdown of DPU kernel times ({"DPU Scan" + "DPU Add"} in SCAN-SSA, and {"DPU Reduction" + "DPU Scan"} in SCAN-RSS) and the intermediate scan in the host CPU ("{Inter-DPU}").

The main observation from these results is that SCAN-SSA runs faster for small arrays (2048-8192). 
Scan kernel time and {Inter-DPU} time are very similar in both SCAN-SSA and SCAN-RSS, but the Add kernel is faster than the Reduction kernel for small sizes. 
The reason is that the Reduction kernel is burdened by the overhead of intra-DPU synchronization (barrier) and the final reduction, where {only} a single tasklet works. This overhead {becomes} negligible for larger arrays. As a result, SCAN-RSS is faster for large arrays (more than 16384 elements).


\subsection{CPU and GPU versions of the benchmarks}
\label{sec:appendix-comparison}

Table~\ref{tab:comparison} shows the sources of the CPU and GPU versions of PrIM benchmarks, which we use for comparison purposes in Section~\ref{sec:comparison}. 
{We provide these CPU and GPU versions as part of our PrIM benchmark suite~\cite{gomezluna2021repo}.}

\begin{table}[h]
            \begin{center}
                \caption{CPU and GPU versions of PrIM benchmarks.}
                \label{tab:comparison}
                \resizebox{\linewidth}{!}{
                    \begin{tabular}{|l|l|l|}
    \hline
    \textbf{Benchmark} & \textbf{CPU version} & \textbf{GPU version} \\
    \hline
    \hline
 VA & OpenMP (custom) & CUDA SDK~\cite{cudasamples} \\ \hline
 GEMV & OpenMP (custom) & CUDA (custom)  \\ \hline
 SpMV & OpenMP (custom) & CUDA (custom) \\ \hline
 SEL & DS algorithms~\cite{gomezluna2015ds} & DS algorithms~\cite{gomezluna2015ds} \\ \hline
 UNI & DS algorithms~\cite{gomezluna2015ds} & DS algorithms~\cite{gomezluna2015ds} \\ \hline
 BS & OpenMP (custom) & CUDA (custom) \\ \hline
 TS & \cite{zhu2018matrix} & \cite{zhu2018matrix}  \\ \hline
 BFS & OpenMP (custom) & CUDA~\cite{hwukirk2016.bfs} \\ \hline
 NW & Rodinia~\cite{che2009rodinia} & Rodinia~\cite{che2009rodinia} \\ \hline
 MLP & OpenMP (custom) & CUDA (custom) \\ \hline
 HST & Chai~\cite{gomezluna2017chai} & Chai~\cite{gomezluna2017chai} \\ \hline
 RED & Thrust~\cite{bell2012thrust} & Thrust~\cite{bell2012thrust} \\ \hline
 SCAN & Thrust~\cite{bell2012thrust} & Thrust~\cite{bell2012thrust} \\ \hline
 TRNS & \cite{gomez2016matrix} & \cite{gomez2016matrix} \\\hline
\end{tabular}

                }
            \end{center}

\end{table}








\nobalance

\begin{wrapfigure}{l}{25mm} 
    \includegraphics[width=1in,height=1.25in,clip,keepaspectratio]{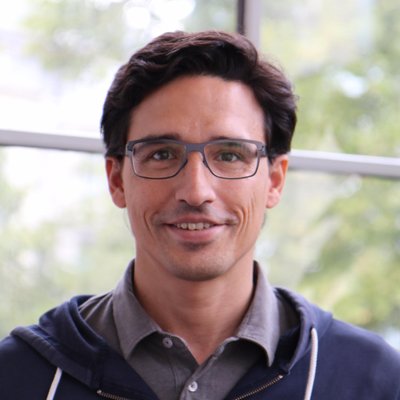}
\end{wrapfigure}\par
\textbf{Juan Gómez-Luna} is a senior researcher and lecturer at SAFARI Research Group @ ETH Zürich. He received the BS and MS degrees in Telecommunication Engineering from the University of Sevilla, Spain, in 2001, and the PhD degree in Computer Science from the University of Córdoba, Spain, in 2012. Between 2005 and 2017, he was a faculty member of the University of Córdoba. His research interests focus on processing-in-memory, memory systems, heterogeneous computing, and hardware and software acceleration of medical imaging and bioinformatics. He is the lead author of PrIM (https://github.com/CMU-SAFARI/prim-benchmarks), the first publicly-available benchmark suite for a real-world processing-in-memory architecture, and Chai (https://github.com/chai-benchmarks/chai), a benchmark suite for heterogeneous systems with CPU/GPU/FPGA.\par

\vspace{2mm}
\begin{wrapfigure}{l}{25mm} 
    \includegraphics[width=1in,height=1.25in,clip,keepaspectratio]{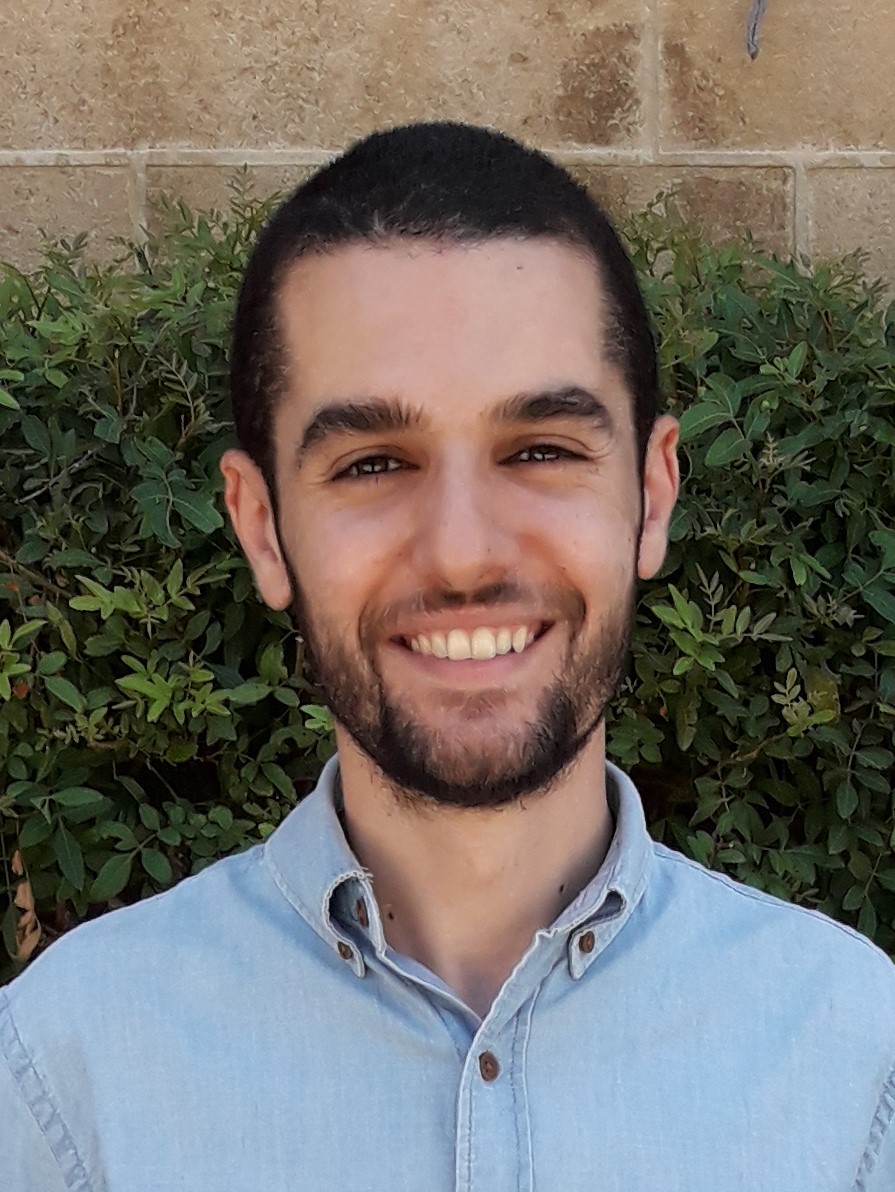}
\end{wrapfigure}\par
\textbf{Izzat El Hajj} is an Assistant Professor in the Department of Computer Science at the American University of Beirut. He received his M.S. and Ph.D. in Electrical and Computer Engineering in 2018 from the University of Illinois at Urbana-Champaign, where he received the Dan Vivoli Endowed Fellowship. Prior to that, he received his B.E. in Electrical and Computer Engineering in 2011 from the American University of Beirut, where he received the Distinguished Graduate Award. His research interests are in application acceleration and programming support for emerging accelerators and memories, with a particular interest in GPUs and processing-in-memory.\par

\vspace{2mm}
\begin{wrapfigure}{l}{25mm} 
    \includegraphics[width=1in,height=1.25in,clip,keepaspectratio]{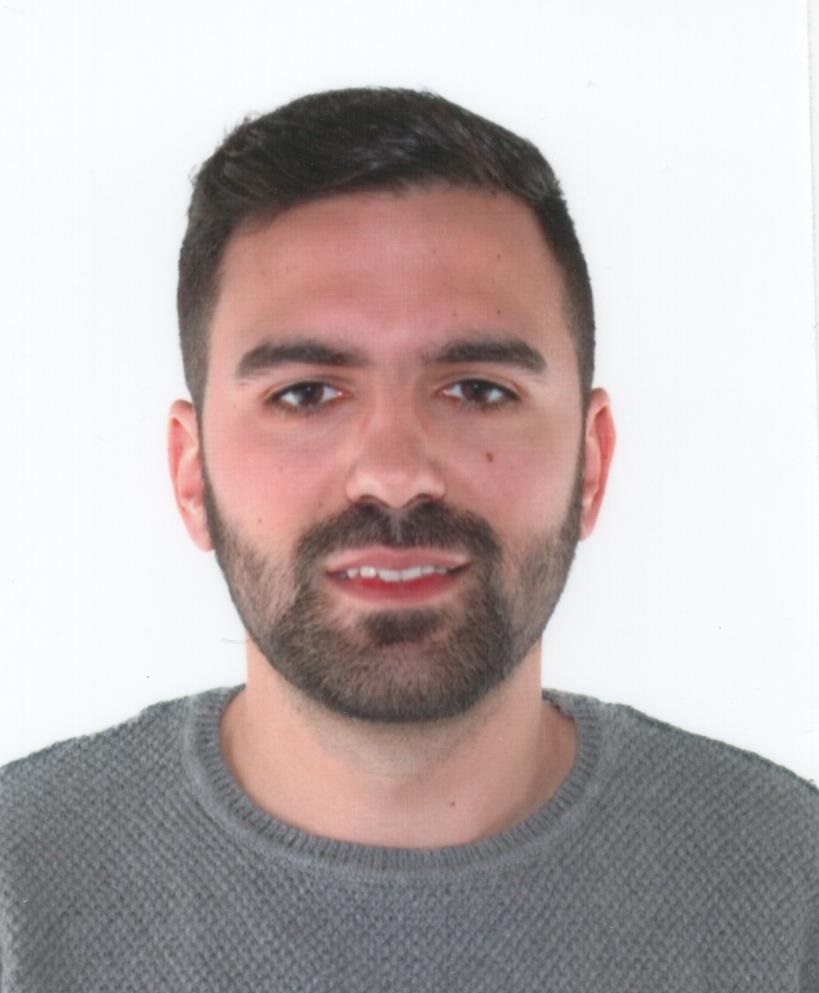}
\end{wrapfigure}\par
\textbf{Ivan Fernandez} received his B.S. degree in computer engineering and his M.S. degree in mechatronics engineering from University of Malaga in 2017 and 2018, respectively. He is currently working toward the Ph.D. degree at the University of Malaga. His current research interests include processing in memory, near-data processing, stacked memory architectures, high-performance computing, transprecision computing, and time series analysis.\par

\vspace{2mm}
\begin{wrapfigure}{l}{25mm} 
    \includegraphics[width=1in,height=1.25in,clip,keepaspectratio]{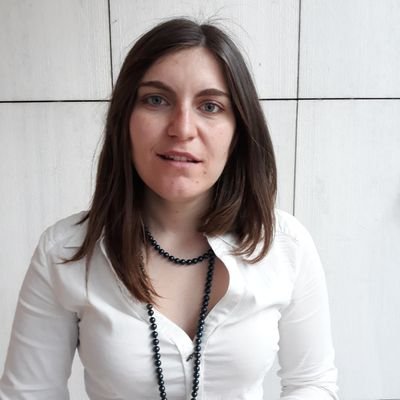}
\end{wrapfigure}\par
\textbf{Christina Giannoula} Christina Giannoula is a Ph.D. student at the School of Electrical and Computer Engineering of the National Technical University of Athens. She received a Diploma in Electrical and Computer Engineering from NTUA in 2016, graduating in the top 2\% of her class. Her research interests lie in the intersection of computer architecture and high-performance computing. Specifically, her research focuses on the hardware/software co-design of emerging applications, including graph processing, pointer-chasing data structures, machine learning workloads, and sparse linear algebra, with modern computing paradigms, such as large-scale multicore systems and near-data processing architectures. She is a member of ACM, ACM-W, and of the Technical Chamber of Greece.\par

\vspace{2mm}
\begin{wrapfigure}{l}{25mm} 
    \includegraphics[width=1in,height=1.25in,clip,keepaspectratio]{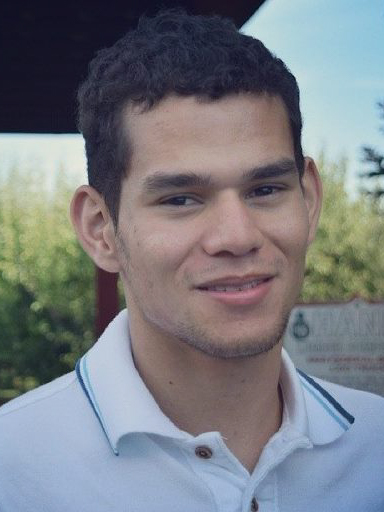}
\end{wrapfigure}\par
\textbf{Geraldo F. Oliveira} received a B.S. degree in computer science from the Federal University of Viçosa, Viçosa, Brazil, in 2015, and an M.S. degree in computer science from the Federal University of Rio Grande do Sul, Porto Alegre, Brazil, in 2017. Since 2018, he has been working toward a Ph.D. degree with Onur Mutlu at ETH Zürich, Zürich, Switzerland. His current research interests include system support for processing-in-memory and processing-using-memory architectures, data-centric accelerators for emerging applications, approximate computing, and emerging memory systems for consumer devices. He has several publications on these topics.\par

\vspace{2mm}
\begin{wrapfigure}{l}{25mm} 
    \includegraphics[width=1in,height=1.25in,clip,keepaspectratio]{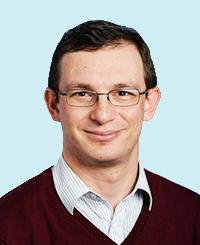}
\end{wrapfigure}\par
\textbf{Onur Mutlu} is a Professor of Computer Science at ETH Zurich. He is also a faculty member at Carnegie Mellon University, where he previously held the Strecker Early Career Professorship.  His current broader research interests are in computer architecture, systems, hardware security, and bioinformatics. A variety of techniques he, along with his group and collaborators, has invented over the years have influenced industry and have been employed in commercial microprocessors and memory/storage systems. He obtained his PhD and MS in ECE from the University of Texas at Austin and BS degrees in Computer Engineering and Psychology from the University of Michigan, Ann Arbor. He started the Computer Architecture Group at Microsoft Research (2006-2009), and held various product and research positions at Intel Corporation, Advanced Micro Devices, VMware, and Google.  He received the IEEE High Performance Computer Architecture Test of Time Award, the IEEE Computer Society Edward J. McCluskey Technical Achievement Award, ACM SIGARCH Maurice Wilkes Award, the inaugural IEEE Computer Society Young Computer Architect Award, the inaugural Intel Early Career Faculty Award, US National Science Foundation CAREER Award, Carnegie Mellon University Ladd Research Award, faculty partnership awards from various companies, and a healthy number of best paper or "Top Pick" paper recognitions at various computer systems, architecture, and security venues. He is an ACM Fellow, IEEE Fellow for, and an elected member of the Academy of Europe (Academia Europaea). His computer architecture and digital logic design course lectures and materials are freely available on YouTube (https://www.youtube.com/OnurMutluLectures), and his research group makes a wide variety of software and hardware artifacts freely available online (https://safari.ethz.ch/). For more information, please see his webpage at https://people.inf.ethz.ch/omutlu/.\par

\end{document}